\documentclass[12pt,a4paper]{article}

\usepackage{geometry}
\usepackage{hyperref}
\usepackage{cite}
\usepackage{url}
\usepackage{amsmath}
\usepackage{amssymb}
\usepackage[dvips]{graphicx}
\usepackage{latexsym}
\usepackage{pstricks}
\usepackage{bm}
\usepackage{pbox}
\usepackage{placeins}
\usepackage{graphicx}
\usepackage{caption}
\usepackage{subcaption}
\usepackage[T1]{fontenc}
\usepackage{footnote}
\usepackage{pdfpages}
\usepackage{hhline}
\usepackage{multirow}
\usepackage{slashbox}
\usepackage[toc,page]{appendix}
\usepackage{tikz}

\DeclareMathOperator*{\argmin}{argmin}

\newcommand{\be}{\begin{equation}}
\newcommand{\ee}{\end{equation}}

\newcommand{\beq}[1] {\begin{equation}\label{#1} }
\newcommand{\eeq} {\end{equation} }
\newcommand{\bea}[1]{\begin{eqnarray}\label{#1} }
\newcommand{\eea}{\end{eqnarray}}

\def\beqn{\begin{eqnarray}}
\def\eeqn{\end{eqnarray}}

\def\beq{\begin{equation}}
\def\eeq{\end{equation}}
\def\bea{\begin{equation}}
\def\eea{\end{equation}}

\def\vs{\vspace}

\begin{document}
\vspace*{-0.2in}
\begin{flushright}
OSU-HEP-16-10
\end{flushright}
\vs{0.5cm}
\renewcommand{\thefootnote}{\fnsymbol{footnote}}
\begin{center}
{\Large\bf Anarchy with Hierarchy: A Probabilistic Appraisal}\\
\end{center}
\vspace{0.5cm}
\begin{center}
{\large
{}~K.S. Babu\footnote{E-mail: babu@okstate.edu},{}~
Alexander Khanov\footnote{E-mail: khanov@okstate.edu} and
{}~Shaikh Saad\footnote{E-mail: shaikh.saad@okstate.edu}
}
\vspace{0.5cm}

{\em Department of Physics \\
Oklahoma State University\\
Stillwater, OK 74078, USA }
\end{center}
\renewcommand{\thefootnote}{\arabic{footnote}}
\setcounter{footnote}{0}
\thispagestyle{empty}

\begin{abstract}
The masses of the charged fermion and the mixing angles among quarks are observed to be strongly hierarchical, while analogous parameters in the neutrino sector appear to be structure-less or anarchical.  We develop a class of unified models based on $SU(5)$ symmetry that explains these differing features probabilistically. With the aid of three input parameters that are hierarchical, and with the assumption that all the Yukawa couplings are uncorrelated random variables described by Gaussian distributions, we show by Monte Carlo simulations that the observed features of the entire fermion spectrum can be nicely reproduced. We extend our analysis to an $SU(5)$-based flavor $U(1)$ model making use of the Froggatt-Nielsen mechanism where the order one Yukawa couplings are modeled as random variables, which also shows good agreement with observations.
\end{abstract}

\newpage
\section{Introduction}

Although the Standard Model (SM) of particle physics has been highly successful, it does not address some of the observed phenomena. For example, neutrinos in the SM are strictly massless.  Non-zero masses for the neutrinos have been firmly established through oscillations experiments conducted with atmospheric~\cite{Fukuda:1998mi}, solar~\cite{Ahmad:2002jz}, accelerator~\cite{Abe:2011sj} and  reactor~\cite{Abe:2011fz} neutrinos, requiring modification of the minimal model. An aesthetic shortcoming of the SM, arising from the enormous freedom available in the Yukawa Lagrangian, is that it provides very little insight into the masses and mixings of quarks and leptons.  This shortcoming is often dubbed as the ``flavor puzzle'' and many extensions of the SM are constructed to address this issue.  The purpose of this paper is to interpret the apparently diverse set of flavor parameters -- quark masses, quark mixing angles, charged fermion masses, neutrino masses and leptonic mixing angles -- in a unified fashion probabilistically.

The observed masses in the charged fermion sector show a hierarchical structure, with the strongest hierarchy seen in the up-type quark sector,  and a somewhat milder hierarchy seen in the down-type quark and charged lepton sectors. These mass parameters, at the momentum scale $\mu=\rm{M_{Z}}$, are approximately given by (in units of $m_{t}=1$):
\begin{eqnarray}\label{eq:mass}
\begin{aligned}
m_{u} \sim 7.5\times 10^{-6};\; m_{c} \sim 3.6\times 10^{-3};\; m_{t} \sim 1; \\
m_{d} \sim 1.6\times 10^{-5};\; m_{s} \sim 3\times 10^{-4};\; m_{b} \sim 1.6\times 10^{-2}; \\
m_{e} \sim 3\times 10^{-6};\; m_{\mu} \sim 6\times 10^{-4};\; m_{\tau} \sim 1\times 10^{-2}.
\end{aligned}
\end{eqnarray}

\noindent
In contrast, the two neutrino squared-mass differences measured in oscillation experiments yield values given by~\cite{Olive:2016xmw}
\begin{eqnarray}\label{eq:Nmass}
\begin{aligned}
\Delta m^{2}_{\rm{sol}} \sim 7.5\times 10^{-5} \; \rm{eV^{2}} \; \rm{and} \;  \Delta \it{m}^{\rm{2}}_{\rm{atm}} \sim \rm{ 2.5\times 10^{-3}} \; \rm{eV^{2}}.
\end{aligned}
\end{eqnarray}
Adopting a normal ordering of the mass spectrum with $m_1 <  m_2 \ll m_3$ with $m_i$ being the neutrino masses, these values would indicate a mild or almost no hierarchy with $m_2/m_3 \sim 1/5$, quite different from the hierarchy seen in the other sectors (Cf: Eq. (\ref{eq:mass})).
Additionally, the inter-generational mixing angles in the quark sector are found to be small, while the leptonic mixing angles are measured to be large:
\begin{eqnarray}\label{eq:angle}
\begin{aligned}
\theta^{\rm{CKM}}_{12} \sim 13^{\circ};\; \theta^{\rm{CKM}}_{23} \sim 2.4^{\circ};\; \theta^{\rm{CKM}}_{13} \sim 0.2^{\circ}; \\
\theta^{\rm{PMNS}}_{12} \sim 34^{\circ};\; \theta^{\rm{PMNS}}_{23} \sim 38^{\circ};\; \theta^{\rm{PMNS}}_{13} \sim 9^{\circ}.
\end{aligned}
\end{eqnarray}

Understanding these patterns observed in the fermion spectrum is a fundamental unresolved problem in particle physics. Various attempts have been made to explain the hierarchy in the charged fermion masses and mixings, adopting highly regulated mass matrices supported by flavor symmetries (for a review see  Ref.~\cite{Babu:2009fd}). On the other hand, random structure-less matrices may be better suited to explain the non-hierarchical mass spectrum and the large mixing angles observed in the neutrino sector~\cite{Hall:1999sn}. The use of such random matrices to explain neutrino mixing angles has been termed ``anarchy hypothesis''. A probability measure should be specified for these random matrices such that the matrix elements remain random after a basis transformation.  For random unitary matrices this is achieved uniquely by the Haar measure~\cite{Haba:2000be}. Such matrices have been shown to be successful in explaining the observed large mixing angles in the neutrino sector~\cite{Hall:1999sn, Haba:2000be, Altarelli:2002sg,deGouvea:2003xe,Espinosa:2003qz,deGouvea:2012ac,
Altarelli:2012ia,Brdar:2015jwo,Fortin:2016zyf}. When basis independence of the random matrix is combined with the requirement that each entry of the matrix has a distribution independent of other entries,  the measure gets determined uniquely to be Gaussian~\cite{mehta,Bai:2012zn,Bergstrom:2014owa,Lu:2014cla}.  Anarchical neutrino mixing angles as well as mass ratios have been analyzed with the Gaussian measure in Ref. \cite{Lu:2014cla}.

In this paper we unify the anarchy hypothesis in the neutrino sector with the hierarchy observed in the quark and charged lepton sectors~
\cite{Babu:1995hr}, \cite{Strassler:1995ia}, \cite{Nelson:1996km}, \cite{Haba:2000be} and analyze the resulting models from a probabilistic perspective.  Such a unification is achieved in the framework of $SU(5)$ grand unified theories, which treat quarks and leptons on similar footing.  For concreteness we adopt a supersymmetric framework, which admits a one step symmetry breaking of $SU(5)$ down to the MSSM.  These models have at most three parameters which are  hierarchical and determined from a fit to data. They also contain five complex Yukawa coupling matrices which are taken to be structure-less or anarchical.  Elements of these Yukawa coupling matrices are treated as uncorrelated random variables obeying Gaussian distributions.  We perform Monte Carlo simulations of this framework and compare theoretical expectations with experimental data, which show good agreement.

Our main analysis is focused on the Yukawa coupling structure obtained in SUSY $SU(5)$ unified theories where the three families of $10_i$ fermions mix with vector-like fermions belonging to $10_\alpha + \overline{10}_\alpha$ representations that have GUT scale masses \cite{Babu:1995hr}.  A variant of this model using the Froggatt-Nielsen mechanism \cite{Froggatt:1978nt}, where the three families of $10_i$ fermions are distinguished by a flavor $U(1)$ symmetry while the three families of $\overline{5}_i$ are universal, is also analyzed allowing for effective non-renormalizable operators \cite{Haba:2000be}.  This class of models is a special case of the general class, with only two hierarchical input parameters.  A second variant, also using a similar $U(1)$ flavor symmetry, which now distinguishes the first family $\overline{5}_1$ from the $\overline{5}_{2,3}$ fields is also analyzed, with a single hierarchy parameter as input \cite{Babu:2003zz,Babu:2004th}.  Good fit to the entire fermion spectrum is obtained in all cases with the Yukawa couplings taking on uncorrelated Gaussian distributions.

It should be noted that ways to understand the neutrino mass anarchy along with charged fermion mass hierarchy has been explored in extra dimensional models with some success \cite{Agashe:2008fe,Feruglio:2014jla,Brummer:2011cp,Yoshioka:1999ds}. These models have not yet been subject to a detailed Monte Carlo analysis for testing quantitatively the goodness of the fit.  The (renormalizable) models we discuss here share some common qualitative features with these extra dimensional models.

We also develop a constrained Monte Carlo simulation method to evaluate the figure of merit of the uncorrelated Gaussian distributions adopted for the random variables. In this method we calculate a specific projection of the probability density distribution of the original random parameters onto a surface that corresponds to random parameters that satisfy the experimental constraints.
The figure of merit that is optimized in this simulation is the distortion of the distributions of the random parameters with respect to their original (unconstrained) distributions.  This constrained Monte Carlo result can be thought of as a multi-dimensional analog of the Kolmogorov-Smirnov statistical test for a single variable.  Our analysis shows that the distortions from the original Gaussian distributions are not much, suggesting a good quality fit.

While the class of models studied here cannot be tested in their precise predictions, they may become strongly favored or disfavored once we know more about the neutrino mass and mixing parameters.  With an anarchical structure the CP-violating parameter $\sin\delta$ in the neutrino sector is found to be peaked at maximal values ($\pm 1$), although variations from these peak values are not excluded. The probability distribution of the neutrino mass ratio $m_{1}/m_{2}$ is peaked around 0.3, with the probability of measuring it below 1/100 found to be about 4\%.

This paper is organized as follows.  In Sec. 2 we present our unified SUSY $SU(5)$ model which allows for the mixing of the three families of $10_i$ with vector-like fermions in the $10_\alpha + \overline{10}_\alpha$ representations.  Here we also present special cases of this general framework making use of flavor $U(1)$ symmetries.  In Sec. 3 we present the results of our Monte Carlo simulations for the fermion mass and mixing parameters for the main model as well as for its variants.  In Sec. 4 we develop a new constrained Monte Carlo method to evaluate the goodness of the fits and compare the distortions of these new distributions from the original Gaussian distributions.  In Sec. 5 we conclude.  Two Appendices contain further details of our analysis. In Appendix A we present the distributions of the various flavor observables for the special cases with flavor $U(1)$ symmetries with either two or one parameter(s).  In Appendix B we present the distributions of the flavor observables obtained from our constrained Monte Carlo simulation for the main model.

\section{Unifying Anarchy with Hierarchy in \boldmath{$SU(5)$}}

As noted in the introduction, grand unified theories based on $SU(5)$ allow for a unified description of anarchy in the neutrino sector and hierarchy in the  quark sector.  We work in the context of SUSY $SU(5)$. The GUT symmetry breaks spontaneously down to the MSSM at an energy scale of $2 \times 10^{16}$ GeV.  The effective low energy theory is the MSSM.  Our focus is the Yukawa couplings of the quarks and leptons in these theories.  At the MSSM level, the Yukawa coupling matrices for the up quarks, down quarks, charged leptons, Dirac neutrinos and the right-handed Majorana neutrinos derived from these models will take the form \cite{Babu:1995hr}:

\vspace{-20pt}
\begin{align}
Y_U&=  H^T Y^0_U H, \label{U4} \\
Y_D&= \epsilon_4 \;  Y^0_D H, \label{D4}\\
Y_L&= \epsilon_4 \; H^T Y^0_L, \label{UL4}\\
Y_N&=   Y^0_N, \label{N4} \\
Y_R&= Y^0_R.  \label{L4}
\end{align}

\noindent Here the superpotential couplings are written as $(f^c_i (Y_f)_{ij} f_j)\,H_f$ with $H_u$ and $H_d$ denoting the two Higgs fields of MSSM.
The fermion mass matrices obtained from Eqs. (\ref{U4})-(\ref{L4}) have the form
\begin{equation}
M_U = Y_U v_u, \,\,M_D = Y_D v_d,\,\, M_L = Y_L v_d,\,\, {\rm and}\,\, M_N = Y_N v_u,\,\,M_R = Y_R v_R
\end{equation}
with $v_u$ and $v_d$ being the VEVs of $H_u$ and $H_d$.
We have assumed the right-handed Majorana neutrino masses arise through the vacuum expectation value (VEV) $v_R$ of a SM singlet field.  In $SU(5)$ unified theories, bare Majorana masses for the gauge singlet right-handed neutrinos may be written down.  If such bare masses are adopted, the scale $v_R$ should be treated as an overall scale in the Majorana mass matrix. The light neutrino mass matrix, obtained via the seesaw mechanism \cite{Minkowski:1977sc}, has the form:
\begin{equation}\label{nu}
M_\nu = \left(Y_N^T Y_R^{-1} Y_N\right) \frac{v_u^2}{v_R}~.
\end{equation}
An explicit derivation of the Yukawa matrices of Eqs. (\ref{U4})-(\ref{L4}) based on $SU(5)$ will be given in the next subsection.  Here we note their salient features which enable the unification of hierarchy and anarchy.

The matrix $H$ in Eqs. (\ref{U4})-(\ref{UL4}) is Hermitian, which may be chosen to be diagonal, real and positive:
\begin{equation}
H = {\rm diag}(\epsilon_1,\, \epsilon_2,\,\epsilon_3).
\end{equation}
Here $\epsilon_1 \ll \epsilon_2 \ll \epsilon_3 \sim 1$ are input parameters of the model which take hierarchical values \cite{Babu:1995hr}.  $\epsilon_3 =1$ can be chosen by redefining other parameters of the model. These parameters arise in the model by virtue of mixing between the three chiral $10_i$-plets of fermions with vector-like $10_\alpha+\overline{10}_\alpha$ of fermions with GUT scale masses.  $Y_f^0$ in Eqs. (\ref{U4})-(\ref{L4}) are the ``bare'' Yukawa coupling matrices -- coupling matrices in the absence of mixing with the vector-like $10_\alpha+\overline{10}_\alpha$ fermions -- which will be assumed to have no specific structure.  $SU(5)$ invariance implies that the same $H$ multiplies all the bare Yukawa coupling matrices in Eqs. (\ref{U4})-(\ref{UL4}). Note that $H$ appears on the right of $Y_D^0$, while it appears on the left of $Y_L^0$.  This occurs in $SU(5)$ since the $d^c$ field -- the $SU(2)_L$ singlet down-type anti-quark -- is unified with the left-handed lepton doublet in a $\overline{5}$ representation.  As a consequence, the left-handed lepton mixing angles will be of order unity, simultaneously with order one mixing in the right-handed down quark sector (which are unobservable). Note also that the mass matrices for down quarks and charged leptons are ``lopsided'' \cite{Babu:1995hr,Albright:1998vf,Sato:1997hv,Irges:1998ax,Maekawa:2001uk}.  Furthermore, $H$ appears on both sides of $Y_U^0$ in Eq. (\ref{U4}) (while it appears only on one side of $Y_D^0$ and $Y_E^0$ in Eqs. (\ref{D4})-(\ref{UL4})), which is due to the presence of $u$ and $u^c$ fields in the same $10$-plet of $SU(5)$.  As a result, the mass hierarchy in the up-quark sector would be stronger compared to the hierarchy in the down-quark and charged lepton sectors:
\begin{eqnarray}
m_d: m_s: m_b \sim \epsilon_1: \epsilon_2: 1 \\
m_e: m_\mu: m_\tau \sim \epsilon_1: \epsilon_2: 1\\
m_u: m_c: m_t \sim \epsilon_1^2: \epsilon_2^2: 1
\end{eqnarray}
Such a pattern  is consistent with observations.

As for the mixing angles, Eqs. (\ref{U4})-(\ref{L4}) will lead to
\begin{eqnarray}\label{eq:04}
\begin{aligned}
V^{CKM}_{i j} &\sim  \frac{\epsilon_{i}}{\epsilon_{j}} ,\; i<j ; \\
V^{lepton}_{i j} &\sim 1 ,\; i<j.
\end{aligned}
\end{eqnarray}
That is, small quark mixings are realized along with large leptonic mixings in these models.

The parameter $\epsilon_4$ in Eqs. (\ref{D4})-(\ref{UL4}) is a third hierarchy parameter, corresponding to an overall suppression of $Y_D$ and $Y_L$ compared to $Y_U$, which has its origin in the mixing of Higgs doublets at the GUT scale.  (In certain minimal models such mixings may be absent, in which case $\epsilon_4 = 1$.  We have investigated this scenario and found that the goodness of the fit to data is poor.) Since there is no hierarchy parameter in $Y_N$ and $Y_R$ in Eqs. (\ref{N4})-(\ref{L4}), the light neutrino masses do not exhibit any hierarchy in this construction, see Eq. (\ref{nu})).

The form of the Yukawa matrices given in Eqs. (\ref{U4})-(\ref{L4}) may also be obtained in other ways in the context of $SU(5)$ unification.  It has been suggested that these forms may follow if the $10$-plet fermions are composite, while the $\overline{5}$-plet fermions are elementary \cite{Strassler:1995ia}.  Alternatively, if there is a flavor symmetry that distinguishes the three families of $10$-plets, with the $\overline{5}$-plets being indistinguishable by this symmetry \cite{Haba:2000be}, the forms of Eqs. (\ref{U4})-(\ref{L4}) may follow with the restriction that $\epsilon_1 \simeq \epsilon_2^2$.  A flavor-dependent $U(1)$ symmetry that distinguishes $\overline{5}_1$ from $\overline{5}_{2,3}$ can lead to yet another constrained model, which may have only a single hierarchy parameter \cite{Babu:2003zz,Babu:2004th}.  We shall analyze these special cases as well.

\subsection{Anarchy and hierarchy via mixing with vector-like fermions}
\label{SU5M}
In this subsection we provide an explicit construction of the fermion Yukawa matrices of Eqs. (\ref{U4})-(\ref{L4}) based on $SU(5)$ symmetry. The setup that we present here is quite general, we will discuss some of its special cases in subsequent subsections.  The construction involves mixing of the chiral families in the $10_i$ representations of $SU(5)$ with vector-like $10_\alpha+\overline{10}_\alpha$ fermions which have GUT scale masses.
Such mixings provide the needed hierarchy factors to explain the charged fermion masses and quark mixing angles. All the Yukawa couplings of the model will be assumed to be structure-less or anarchical.  This applies to the Yukawa couplings in the quark sector, charged lepton sector, and the neutrino sector universally. Thus, in the spirit of anarchy, these Yukawa coupling matrix elements will all be taken as uncorrelated random variables with Gaussian distributions.

The three families of fermions belong to the $10_i + \overline{5}_i$ multiplets of $SU(5)$ ($i=1-3$ is the generation index).  Quarks and leptons are unified in these multiplets as  $10_i=\{e^c_i, u^c_i, Q_i  \}$ and $\overline{5}_i=\{L_i, d^c_i   \}$, where $Q_i=(u_i \; d_i)^T$ and $L_i=(\nu_i \; e_i)^T$. To generate small neutrino masses via the seesaw mechanism three $SU(5)$ singlet fermions $1_i$ ($\nu^c_i$) are introduced.
If only a $5_H+\overline{5}_H$ Higgs pair is involved in the Yukawa couplings as usually assumed in minimal SUSY $SU(5)$, the relation $M_L=M^T_D$ will result among the down-type quark and charged lepton mass matrices, which is unacceptable.  To correct for this at the renormalizable level, we extend the Higgs sector by introducing a $45_H+\overline{45}_H$ pair \cite{Georgi:1979df}. Then the Yukawa superpotential is given by (assuming the usual $R$-parity)
\begin{eqnarray}\label{45}
\mathcal{W}_Y &=&  10_i Y^5_{ij} 10_j 5_H +10_i Y^{45}_{ij} 10_j 45_H + \overline{5}_i Y^{\overline{5}}_{ij} 10_j \overline{5}_H +\overline{5}_i Y^{\overline{45}}_{ij} 10_j \overline{45}_H  \nonumber \\
&+& \overline{5}_i Y^1_{ij} 1_j 5_H + \frac{1}{2} (M_R)_{ij} 1_i 1_j  ,
\end{eqnarray}

\noindent where $Y^{\overline{5}}$, $Y^{\overline{45}}$ and $Y^1$ are general complex matrices, while $Y^5$  and $Y^{45}$ are complex symmetric  and antisymmetric matrices.  These ``bare'' Yukawa coupling matrix elements (as well as the Majorana mass terms $M_R$ for the right-handed neutrinos, up to an overall scale) will all be taken to be random variables obeying Gaussian distributions.

The model also contains a set of vector-like fermions belonging to $10_\alpha + \overline{10}_\alpha$ representations, where $\alpha = 1, 2,.. n$ where $n$ is the number of copies used.  The choice of $n=3$ is natural, in which case there would be 3 pairs of such fields.  The superpotential now admits additional mass terms given by
 \vspace{-5pt}
\begin{align}\label{vector}
\mathcal{W}_Y \supset m_{\alpha j} \overline{10}_{\alpha} 10_j + M_{\alpha \beta}\overline{10}_{\alpha} 10_{\beta},
\end{align}

\noindent where the first term represents the mixing of the ordinary fermions with the vector-like fermions and the second term generates bare masses for these vector-like fermions.  Other possible gauge invariant couplings are assumed to be absent due to additional symmetries. An example of such a symmetry is a $Z_2\times Z_2$ with the vector-like fermions $\overline{10}_{\alpha}$ being odd under the first $Z_2$, and the rest of the fields being even. This choice will prevent unwanted terms of the type $\overline{10}_{\alpha} 10_{\beta} 24_H$ and $\overline{10}_{\alpha} 10_i 24_H$, involving the $SU(5)$ breaking Higgs field $24_H$.  Such a $Z_2$ is broken by the terms in Eq. \eqref{vector}, but only softly.  Under the second $Z_2$, both $10_\alpha$ and $\overline{10}_\alpha$ fields are odd, while the remaining fields are even.  This $Z_2$, which is also broken softly by the first term in Eq. (\ref{vector}), will prevent mixed Yukawa coupling of the type $10_i 10_\alpha 5_H$.  (This second $Z_2$ is optional, since the presence of mixed Yukawa couplings of the type $10_i 10_\alpha 5_H$ do not have any effect on our analysis.)

In Eq. \eqref{vector} the mass terms $m$ and $M$ are SM singlets, and will be assumed to be of order the GUT scale.  The presence of these terms in the Yukawa Lagrangian modifies the structure of the mass matrices of the SM  fermions.  From Eq. \eqref{vector}, the heavy states are found to be $10^{H}_{\alpha} \propto m_{\alpha i} 10_i + M_{\alpha \beta} 10_{\beta}$, with the light states $10^{L}_i$ being orthogonal to the $10^H_\alpha$ states. This system can be inverted to express $10_i$ and $10_{\alpha}$ in terms of $10^{L,H}$ states: $10_i= (H\; 10^L+H^{\prime}\; 10^H)_i$ with

\vspace{-20pt}
\begin{align}
H= (I+m M^{-1} {M^{-1}}^{\dagger} m^{\dagger})^{-\frac{1}{2}} .
\label{H}
\end{align}

\noindent Substituting this form of $10_i$ in Eq. \eqref{45}, one can write down the light quark and light lepton mass matrices as \cite{Babu:1995hr}:
\vspace{-8pt}
\begin{align}
M_U&= H^T M^0_U H, \label{U2} \\
M_D&=  M^0_D H, \\
M_L&= H^T M^0_L ,  \label{K1}\\
M_N&= M^0_N,    \\
M_R&= M^0_R,  \label{L2}
\end{align}

\noindent where $M_{U,D}$ are the up-type and down-type quark mass matrices, $M_L$ is the charged lepton mass matrix, $M_N$ is the Dirac type neutrino mass matrix and $M_R$ is the right-handed neutrino Majorana mass matrix. In writing these mass matrices we have defined \cite{Dorsner:2006dj}
\begin{align}
M^0_U &=  \langle 5_H \rangle Y^5 + \langle 45_H \rangle Y^{45}, \label{U1}\\
M^0_D &= \langle \overline{5}_H \rangle Y^{\overline{5}} + \langle \overline{45}_H \rangle Y^{\overline{45}},\\
M^0_L &=\langle \overline{5}_H \rangle {Y^{\overline{5}}}^T -3 \langle \overline{45}_H \rangle {Y^{\overline{45}}}^T  , \\
M^0_N&= \langle 5_H \rangle\; Y^1 , \label{N1}\\
M^0_R&= v_R\; Y^0_R \label{L1}.
\end{align}
Note that all matrices in Eqs. (\ref{U1})-(\ref{N1}) are general complex, while $M_R^0$ in Eq. (\ref{L1}) is complex symmetric.  ($M^0_U$ has symmetric contributions from $\langle 5_H \rangle$ as well as antisymmetric contributions from $\langle 45_H \rangle$, with the sum being neither symmetric nor antisymmetric.)

The Hermitian matrix $H$ in Eq. (\ref{H}) can be written as $H = U^{\dagger} {\rm diag} \{ \epsilon_{1},\epsilon_{2},\epsilon_{3}  \} U$, with $U$ being a unitary matrix and  $\epsilon_{i}$'s being real and positive ($i=1,2,3$). Substituting this form of $H$ in Eqs. \eqref{U2}-\eqref{K1} and redefining the quark and lepton fields, one can absorb the unitary matrix $U$ into the non-hierarchical matrices $M_{U,D,L}^0$ without affecting the numerical results.  Thus, we choose
$H = {\rm diag}\{ \epsilon_{1}, \epsilon_{2}, \epsilon_3 \}$. A hierarchy $\epsilon_1 \ll \epsilon_2 \ll \epsilon_3 \sim 1$ can be generated within the model by arranging for unequal mixings between the $10_i$ and $10_\alpha$ for different families. For example, for the third family, we may take $M_3 \gg m_3$ (ignoring generation mixing for simplicity of explaining) while for the second and first families we may take $M_2 \ll m_2$ and $M_1 \lll m_1$, see Eq. (\ref{H}) \cite{Babu:1995hr}.  We shall set $\epsilon_3=1$, since this parameter is of order one, and redefining other parameters of the theory enables this choice. Consequently, we will choose
\begin{equation}
H = {\rm diag}\{ \epsilon_{1},\epsilon_{2},1 \}
\end{equation}
for our analysis.

The MSSM up-type Higgs doublet $H_u$ that remains light to low energies is a linear combination of up-type doublets from the $5_H$, $45_H$ and other possible up-type Higgs doublets present in the $SU(5)$ model. Similarly the light MSSM field $H_d$ is a linear combination of down-type Higgs doublets from $\overline{5}_H$, $\overline{45}_H$ and other possible down-type Higgs doublets in the model. An example of such additional up-type and down-type Higgs doublets is a pair of $5'_H+\overline{5}'_H$ fields with no Yukawa couplings to the fermions.  We then have

\vspace{-25pt}
\begin{align}
H_u&= \alpha_u \; h^u_{5} + \beta_u \; h^u_{45} + \sum_i \gamma_i^u h^{\prime u}_i \\
H_d&= \alpha_d \; h^d_{\overline{5}}+ \beta_d \;  h^d_{\overline{45}}+ \sum_i \gamma_i^d h^{\prime d}_i
\end{align}

\noindent with $|\alpha_u|^2 +|\beta_u|^2 + \sum_i|\gamma_i^u|^2 = 1 = |\alpha_d|^2 +|\beta_d|^2 + \sum_i|\gamma_i^d|^2$.
Here $h^u_{5}= (1,2,\frac{1}{2}) \subset 5_H$, $h^u_{45}= (1,2,\frac{1}{2}) \subset 45_H$, $h^d_{\overline{5}}= (1,2,-\frac{1}{2}) \subset \overline{5}_H$ and $h^d_{\overline{45}}= (1,2,-\frac{1}{2}) \subset \overline{45}_H$, where the quantum numbers under the SM gauge symmetry are indicated. The fields $h^{\prime u}_i$ and $h^{\prime d}_i$ are $(1,2,\frac{1}{2})$ and $(1,2,-\frac{1}{2})$ fields from additional Higgs multiplets, such as $5'_H+\overline{5}'_H$ pairs.  All fields orthogonal to $H_u$ and $H_d$ remain superheavy.  The VEVs of the doublet components of the various fields are related to the VEVs $v_u$ and $v_d$ of the MSSM fields $H_u$ and $H_d$ as

\vspace*{-20pt}
\begin{align}
v_5&= \alpha_u^* v_u ,\;\; v_{45}= \beta_u^* v_u, \\
v_{\overline{5}}&= \alpha_d^* v_d ,\;\; v_{\overline{45}}= \beta_d^* v_u.
\label{vev}
\end{align}

\noindent Substituting these relations, one can rewrite the effective mass matrices Eqs. \eqref{U2}-\eqref{L2} for the fermions  as:

\vspace{-25pt}
\begin{align}
M_U&= v_u\; H^T Y^0_U H \equiv v_u\;  Y_U, \label{U3} \\
M_D&= v_d\; \epsilon_4 \; Y^0_D H \equiv v_d\;  Y_D, \label{A1}\\
M_L&= v_d\; \epsilon_4 \; H^T Y^0_L \equiv v_d\;  Y_L,  \label{A2}\\
M_N&= v_u\;  Y^0_N \equiv v_u\; Y_N,    \\
M_R&= v_R\; Y^0_R \equiv v_R\; Y_R.  \label{L3}
\end{align}

\noindent Here $Y_U^0, \, Y_D^0$ etc are the bare Yukawa coupling matrices derived from Eqs. (\ref{U1})-(\ref{N1}), using the definitions given in
Eq. (\ref{vev}):

\vspace{-15pt}
\begin{align}
Y^0_U &=  \alpha_u^*  \,Y^5 + \beta_u^* \,Y^{45}, \\
Y^0_D &= \alpha_d^*\, Y^{\overline{5}} + \beta_d^* \, Y^{\overline{45}},\\
Y^0_L &=\alpha_d^* \,{Y^{\overline{5}}}^T -3 \beta_d^* \,{Y^{\overline{45}}}^T  , \\
Y^0_N&= \alpha_u^*\; Y^1 ~.
\end{align}

\noindent Thus, we see that the effective Yukawa coupling matrices of the quarks and leptons with the MSSM Higgs fields as given  in Eqs. (\ref{U4})-(\ref{L4}) are generated. The bare Yukawa couplings $Y^0_{U,D,L,N,R}$ in these equations will be treated as random variables obeying Gaussian distributions in our numerical analysis.  The parameter $\epsilon_4$ appearing in Eqs. (\ref{A1})-(\ref{A2}) arises from the Higgs doublet mixing expressed in terms of $(\alpha_{u,d},\,\beta_{u,d})$.  To realize values of $\epsilon_4$ in the range $\epsilon_4=(0.04-0.1)$ as our fits would prefer, it is sufficient to take $\alpha_d$ and $\beta_d$ somewhat smaller than one.  Unitarity of the Higgs mixing matrix is maintained due to the presence of additional Higgs doublets such as $5'_H+\overline{5}'_H$ in the model.  The model also has $\tan\beta = v_u/v_d$ as an input parameter.  A relation between the $\tan \beta= v_u/v_d$ and $\epsilon_4$ can be obtained from Eqs. (\ref{U3})-(\ref{A1}):

\vspace*{-15pt}
\begin{align}\label{tanbeta}
\epsilon_4 \simeq \frac{m_b}{m_t} \tan\beta  \frac{(Y^{0}_U)_{33}}{|(\overrightarrow{d_{0}})_{3}|}
\end{align}

\noindent where we have defined $(\overrightarrow{d_{0}})_{3}=\{(Y^{0}_D)_{13},(Y^{0}_D)_{23},(Y^{0}_D)_{33}\}$.  Note that to set $\epsilon_3 = 1$ which we have adopted, we redefine $\epsilon_4$ in Eqs.(\ref{A1})-(\ref{A2}), and also redefine $v_u$ in Eq. (\ref{U3}).

Since the masses of the vector-like fermions are of the order of  GUT scale, any effect of these particles at  low energies will be suppressed by a factor of 1/$M_{GUT}$, except for the dimension four fermion mass operators as discussed in the text. Hence their presence does not change the phenomenology of the MSSM or the Higgs boson mass. Even though the super-heavy vector-like fermions decouple, they may leave imprints on the SUSY flavor structure at low energies. However, SUSY models with large superpartner masses  or gauge-mediated SUSY breaking models can potentially suppress any such flavor violating effects.

As noted previously, there are other ways of generating the Yukawa structure shown in Eqs. (\ref{U4})-(\ref{L4}) by assuming $U(1)$ flavor symmetry that distinguishes the three families of $10_i$ \cite{Haba:2000be}, and/or the first family of $\overline{5}_1$ from $\overline{5}_{2,3}$ \cite{Babu:2003zz,Babu:2004th}, by hypothesizing that the $10_i$-plets are composite \cite{Strassler:1995ia,Haba:1998wf}, or postulating extra dimensions \cite{Yoshioka:1999ds,Agashe:2008fe,Feruglio:2014jla}. Another interesting class of models proposed recently in Ref. \cite{Barr:2012ma,Barr:2015rrm} has a very similar structure for the mass matrices, which we shall not investigate here. We do analyze the flavor $U(1)$ models as special cases of the general class of models described here, which are described next.

\subsection{\boldmath{$SU(5)$}-inspired  models with \boldmath{$U(1)$} flavor symmetry}
\label{SU5F}
In this subsection we briefly describe a class of $SU(5)$-inspired models with $U(1)$ flavor symmetry. Models of this type can explain the hierarchical structure in the fermion masses and mixings by using the Fraggatt-Nielsen mechanism~\cite{Froggatt:1978nt}.  Smaller entries in the mass matrices are induced as higher dimensional operators suppressed by differing inverse powers of a fundamental mass scale.  Assigning different charges to different families will lead to a hierarchy in masses and mixings.

The models we study here are inspired by SUSY $SU(5)$ unification -- in the sense that the flavor $U(1)$ charge assignment will be compatible with $SU(5)$ -- but we can work just within the framework of MSSM.  We shall use the language of $SU(5)$, however, for simplicity.  The three fermion families are assigned to $10_i+\overline{5}_i$, and we include three families of SM singlet $1_i$ ($\nu_i^c$) fields for the seesaw mechanism.  In order to reproduce the observed hierarchical structure in fermion masses, we make specific $U(1)$ charge assignment to the fermion fields as shown in Table   \ref{chargeSU5}.  The integer charges $q_1$, $q_2$ and $p$ are left unspecified in the table, two different choices will be presented below.

\begin{table}[th]
\centering
\footnotesize
\resizebox{0.35\textwidth}{!}{
\begin{tabular}{|c|c|} \hline
Field & $U_A(1)$ charge \\ [1ex] \hline
$10_1, 10_2, 10_3 $&2$q_1$, $q_1$, 0\\ \hline
$\overline{5}_1, \overline{5}_2, \overline{5}_3 $&$q_2+p$, $p$, $p$\\ \hline
$1_1, 1_2,1_3 $&$q_2$, 0, 0\\ \hline
\end{tabular}
}
\caption{The flavor $U(1)$ charge assignment of the fermion fields in $SU(5)$ notation. The Yukawa matrices of Eqs. (\ref{U4})-(\ref{L4}) will be induced with the choice $q_1 = 1$, $q_2=p=0$.  Yukawa couplings given in Eqs. (\ref{eq:100A})- (\ref{eq:100B}) will result with the choice $q_1 = 2$, $q_2 = 1$,
$p = 0, 1$ or 2, corresponding to large, medium and small $\tan\beta$. These models also contain a flavon field $S$ with $U(1)$ charge of $-1$ that acquires a VEV.  The Higgs doublets $H_u$ and $H_d$ of MSSM are neutral under this $U(1)$.}
\label{chargeSU5}
\end{table}

In these models, the $U(1)$ flavor symmetry is broken by a single parameter $\epsilon =\langle S \rangle /M_{\ast}$, where $\langle S \rangle$ is the VEV of an $SU(5)$ singlet flavon field $S$ with $U(1)$ charge $-1$ and $M_{\ast}>M_{GUT}$ is a fundamental scale such as the string scale. The Yukawa superpotential contains higher dimensional terms suppressed by inverse powers of $M_{\ast}$, with coefficients which are all of order one.  These couplings have the form
\begin{align}
{\cal W}_Y &\supset Y^u_{ij} Q_i u^c_j H_u \left(\frac{S}{M_\ast}\right)^{n_{ij}^u}
+ Y^d_{ij} Q_i d^c_j H_d \left(\frac{S}{M_\ast}\right)^{n_{ij}^d}
+
Y^\ell_{ij} L_i e^c_j H_d \left(\frac{S}{M_\ast}\right)^{n_{ij}^\ell}
\nonumber \\&
+
Y^{\nu}_{ij} L_i \nu^c_j H_u \left(\frac{S}{M_\ast}\right)^{n_{ij}^{\nu}}
+
v_R Y^{R}_{ij} \nu^c_i \nu^c_j \left(\frac{S}{M_\ast}\right)^{n_{ij}^{\nu^c}} .
\label{fn}
\end{align}
Here the integers $n^u_{ij}$ etc are chosen such that the corresponding Yukawa coupling $Y^u_{ij}$ is charge neutral. The couplings $Y_{ij}^u$ etc are all taken to be of order unity. Still hierarchical masses and mixings are induced since the $(ij)$ entry in the mass matrix has a suppression factor $\epsilon^{n_{ij}}$.

In our first flavor $U(1)$ model we choose the $U(1)$ charges of Table \ref{chargeSU5} to be  $\{q_1=1, q_2=0, p=0\}$ \cite{Haba:2000be}. In this case the Yukawa coupling matrices will have the same form as in Eqs. \eqref{U4}-\eqref{L4}.  Note that in this model the three families of $\overline{5}_i$ are neutral under $U(1)$, while the $10_i$ carry differing charges given as $(2,\,1,\,0)$.  Since the $U(1)$ symmetry is broken by a single parameter, the Hermitian matrix  $H$ appearing in Eqs. \eqref{U4}-\eqref{L4} is now given by

\vspace*{-20pt}
\begin{align}\label{HH}
H= \begin{pmatrix}
\epsilon^2&0&0\\
0&\epsilon&0\\
0&0&1
\end{pmatrix}.
\end{align}

\noindent The only difference from the general model of the previous subsection is that here $\epsilon_2 \equiv \epsilon$ and $\epsilon_1 = \epsilon^2$.\footnote{Strictly, $\epsilon_1 = O(1) \epsilon^2$, but this $O(1)$ coefficient may be absorbed into other $O(1)$ Yukawa couplings, which is what we shall do.}  This model will be analyzed separately, with the assumption that the Yukawa couplings entering Eq. (\ref{fn}) are random variables taking Gaussian distributions.  The light neutrino mass matrix retains exactly the same structure-less pattern as before, since the $\nu^c$ fields as well as the $L_i$ fields are all neutral under the $U(1)$. If the model is embedded in $SU(5)$ minimally, the wrong relation $Y_L=Y^T_D$ would result.  This would require the extension of the scalar sector by a $45_H+\overline{45}_H$ pair. As before, the parameter $\epsilon_4$ has the same definition as in Eqs. \eqref{U4}-\eqref{L4}, and such models have two hierarchical parameters $\{\epsilon, \epsilon_4 \}$.

A second flavor $U(1)$ model is obtained by the choice of $U(1)$ charges in Table \ref{chargeSU5} as $\{q_1=2, q_2=1, p=0,\,1,\,{\rm or}~2\}$ along with the charges of the scalar fields given by $\{H_u, H_d, S\}=\{0, 0, -1\}$.  Here the first family $\overline{5}_1$ has a shifted charge compared to $\overline{5}_{2,3}$.  This is the only difference of this model compared to the first flavor $U(1)$ model just discussed.  Such a model has been studied in Ref. ~\cite{Babu:2003zz,Babu:2004th}, where the Yukawa coupling matrices written in the basis $f^c_i(Y_f)_{ij} f_j$ are shown to take the form:

\vspace{-10pt}
\begin{align}
Y_U &\sim \begin{pmatrix}
\epsilon^{8} & \epsilon^{6} & \epsilon^{4} \\
\epsilon^{6} & \epsilon^{4} & \epsilon^{2} \\
\epsilon^{4} & \epsilon^{2} & 1
\end{pmatrix} , \;\;\;
Y_D \sim \epsilon^{p} \begin{pmatrix}
\epsilon^{5} & \epsilon^{3} & \epsilon \\
\epsilon^{4} & \epsilon^{2} & 1 \\
\epsilon^4 & \epsilon^{2} & 1
\end{pmatrix} , \label{eq:100A}\\
Y_L &\sim \epsilon^{p} \begin{pmatrix}
\epsilon^{5} & \epsilon^{4} & \epsilon^4 \\
\epsilon^{3} & \epsilon^{2} & \epsilon^2 \\
\epsilon & 1 & 1
\end{pmatrix} , \;
Y_N \sim \epsilon^{p} \begin{pmatrix}
\epsilon^{2} & \epsilon & \epsilon \\
\epsilon & 1 & 1 \\
\epsilon & 1 & 1
\end{pmatrix} , \\
Y_{R} &\sim \begin{pmatrix}
\epsilon^{2} & \epsilon & \epsilon \\
\epsilon & 1 & 1 \\
\epsilon & 1 & 1
\end{pmatrix} , \;\;\;\;\;\;
 \mathcal{Y}_{\nu} \sim \epsilon^{2 p} \begin{pmatrix}
\epsilon^{2} & \epsilon & \epsilon \\
\epsilon & 1 & 1 \\
\epsilon & 1 & 1
\end{pmatrix} .\label{eq:100B}
\end{align}

\noindent Here $ \mathcal{Y}_{\nu}$ determines the light neutrino mass matrix via the seesaw relation $M_\nu =  \mathcal{Y}_{\nu} v_u^2/v_R$. The integer $p$ is allowed to take three different values, $p = 0,1$ or 2, corresponding to large, medium, and small values of $\tan\beta$.
In Eqs. (\ref{eq:100A})-(\ref{eq:100B}), each matrix element has an $O(1)$ coefficient $c^f_{ij}$ that is not explicitly shown.  These entries are taken to be of order unity.  For our statistical analysis of the model, we shall take these $c^f_{ij}$ to be random variables obeying uncorrelated Gaussian distributions. One clearly sees that although the charged fermion mass matrices here are quite similar to the previously discussed models, the light neutrino mass matrix is significantly different. Unlike the previous cases, it is no longer given by a matrix with order unity entries everywhere; rather it has somewhat of a hierarchical structure.  In this model, it is possible to correct the $SU(5)$ relation $M_L = M_D^T$ via higher dimensional operators involving the $24_H$ field, and therefore, a parameter analogous to $\epsilon_4$ is not required.  As we shall see, a good fit to all data is obtained in this model with a single hierarchy parameter $\epsilon$.

\section{Statistical Analysis of Flavor Parameters in \boldmath{$SU(5)$}-based Models}
\label{SU5A}

In this section we perform a statistical analysis of the general class of unified theories based on $SU(5)$. The general model described in Sec. \ref{SU5M} contains three hierarchical input parameters $\{\epsilon_1,\,\epsilon_2,\,\epsilon_4\}$ as well as $\tan\beta$ in the flavor sector.
In addition, these models have five complex Yukawa coupling matrices, see Eqs. ({\ref{U4})-(\ref{L4}), the elements of which are treated as uncorrelated random variables with Gaussian distributions. After a detailed analysis of this general setup, we repeat the analysis for the two
$SU(5)$-inspired flavor $U(1)$ variants.  These variants have either two set of hierarchical parameters $\{\epsilon,\,\epsilon_4\}$, or
a single parameter $\epsilon$.

The primary goal of this section is to investigate how well the theoretical predictions of this class of models agree with the  experimentally  observed quantities on average. We perform a Monte Carlo simulation and derive the theoretical expectations for these models. We start with the MSSM Yukawa coupling matrices given in Eqs. \eqref{U4}-\eqref{L4}.  As noted before, the matrices
$Y^{0}_{F}$  in Eqs. \eqref{U4}-\eqref{L4} are random matrices with all elements of order $O(1)$. The matrices $Y^{0}_{F}$ for $F=U, D, L, N$ are of the Dirac-type and in general complex matrices. The right-handed neutrino Yukawa coupling matrix $Y^{0}_{R}$ in Eq. (\ref{L4}) is of the Majorana-type which is complex symmetric. We assume that each of these matrix elements is a random variable independent of other elements. The probability distributions of the matrix elements are assumed to be completely independent of the hierarchical model parameters $\{\epsilon_1, \,\epsilon_2,\, \epsilon_4\}$. Basis independence as well as absence of correlation between various matrix elements determine uniquely the probability measures for these random variables to be Gaussian~\cite{Bai:2012zn,Lu:2014cla}:

\vspace{-10pt}
\begin{eqnarray}\label{eq:77}
\begin{aligned}
dY^{0}_{\rm{D}} &= \prod\limits_{ij} dY^{0}_{ij}\; e^{-|Y^{0}_{ij}|^{2} }, \\
dY^{0}_{\rm{M}} &= \prod\limits_{i} dY^{0}_{ii}\; e^{-|Y^{0}_{ii}|^{2} } \; \prod\limits_{i<j} dY^{0}_{ij}\; e^{-2 |Y^{0}_{ij}|^{2} },
\end{aligned}
\end{eqnarray}

\noindent
Here the subscripts D and M represent Dirac-type and Majorana-type respectively. These measures are defined up to a scale factor $e^{-c}$, which has been set equal to 1.  (When Gaussian distributions are applied to mass matrices, this scale factor can be used to fix the overall scale of the VEV, see Ref.~\cite{Lu:2014cla} for details). From Eq. \eqref{eq:77}, all the elements of a general complex random matrix are independently generated with Gaussian distribution of variance 0.5 for both the real and imaginary parts separately. Similarly, for the complex symmetric random matrix, the   real and imaginary parts  are generated independently with Gaussian distribution of variance 0.5 and 0.25 for diagonal and off-diagonal entries respectively.

\begin{table}[th]
\centering
\footnotesize
\resizebox{0.4\textwidth}{!}{
\begin{tabular}{|c|c|}
\hline
\pbox{10cm}{Yukawa Couplings \\ and CKM parameters} & $\mu= \rm{M_{Z}}$ \\ [1ex] \hline
$y_{u}/10^{-6}$ & $6.65 \pm 2.25$  \\ \hline
$y_{c}/10^{-3}$ & $3.60 \pm 0.11$  \\ \hline
$y_{t}$ & $0.9860 \pm 0.00865$  \\ \hline
$y_{d}/10^{-5}$ & $1.645 \pm 0.165$  \\ \hline
$y_{s}/10^{-4}$ & $3.125 \pm 0.165$  \\ \hline
$y_{b}/10^{-2}$ & $1.639 \pm 0.015$  \\ \hline
$y_{e}/10^{-6}$ & $2.79475 \pm 0.0000155$  \\ \hline
$y_{\mu}/10^{-4}$ & $5.89986 \pm 0.0000185$  \\ \hline
$y_{\tau}/10^{-2}$ & $1.00295 \pm 0.0000905$  \\ \hline
$\theta^{CKM}_{12}$ & $0.22735\pm 0.000072$ \\ \hline
$\theta^{CKM}_{23}/10^{-2}$ & $4.208\pm 0.064$ \\ \hline
$\theta^{CKM}_{13}/10^{-3}$ & $3.64\pm 0.13$ \\ \hline
$\delta^{CKM}$ & $1.208\pm0.054$ \\ [0.5ex] \hline
\end{tabular}
}
\caption{  Observables in the charged fermion sector at the $\rm{M_{Z}}$ scale taken from Ref.~\cite{Antusch:2013jca}. For quantities with asymmetrical error bars, we have symmetrized and presented the experimental central values with associated 1 $\sigma$ uncertainties. The fermion masses are given by the relations $m_{i}(\rm{M_{Z}})=\it{v} \; \it{y}^{\rm{SM}}_{i}(\rm{M_{Z}})$, with $\it{v}=\rm{174}$ GeV.}
\label{table:1}
\end{table}

The class of models with Yukawa matrices given in Eqs. \eqref{U4}-\eqref{L4} has three input parameters,  $\epsilon_{i}$ (i=1,2,4) and 84 random variables (72 in four general complex random matrices and 12 in one random complex symmetric matrix). In this section we present a Monte Carlo analysis of these models adopting Gaussian measure for the random matrix elements. The parameters $\epsilon_{i}$ are however not random, instead they are  fixed by $\chi^{2}$-function minimization.  We have seen previously that these parameters do not enter in the neutrino sector.  Thus, in order to fix the numerical values of these parameters we only include in the $\chi^2$-minimization the observables in the charged fermion sector.  The minimization is carried out at the GUT scale with 3 input parameters to fit 13 observables.

To perform the $\chi^2$-minimization at the GUT scale we take the experimentally observed values of the charged fermion observables at the $\rm{M_{Z}}$ scale from Ref.~\cite{Antusch:2013jca}. These values are quoted in Table \ref{table:1}. We use the renormalization group running factors corresponding to MSSM, $\eta_{i}=m_{i}(\rm{M_{GUT}})/ \it{m_{i}}(\rm{M_{Z}}) $,
taken from Ref.~\cite{Xing:2007fb} for the evolution of the Yukawa couplings from the
 $\rm{M_{Z}}$ scale to the GUT scale.  These running factors are listed in Table \ref{table:6}. We perform the Monte Carlo analysis for two values of the parameter $\tan\beta$, $10$ and $50$. The Yukawa couplings at the GUT scale are obtained from the couplings determined at  $\mu = \rm{M_{Z}}$ with the help of  these  renormalization running factors by using the relations $y^{\rm{MSSM}}_{\it{u_{i}}}(\rm{M_{GUT}})=\it{y}^{\rm{SM}}_{\it{u_{i}}}(\rm{M_{Z}}) \eta_{\it{u_{i}}}/ \sin\beta$ for up-type quarks and $\it{y}^{\rm{MSSM}}_{\it{d_{i},e_{i}}}(\rm{M_{GUT}})=\it{y}^{\rm{SM}}_{\it{d_{i},e_{i}}}(\rm{M_{Z}}) \eta_{\it{d_{i},e_{i}}}/ \cos\beta$ for down-type quarks and charged leptons. We also run the CKM mixing parameters from $\rm{M_{Z}}$ to the GUT scale using the MSSM renormalization group equations \cite{Babu:1987im,Barger:1992pk}.  The renormalization running factors of the CKM matrix elements are presented in Table \ref{table:6}.  The Yukawa couplings and the CKM mixing parameters at the GUT scale are presented in Table \ref{table:2}. For the associated one sigma uncertainties of these observables at the GUT scale, we take the same percentage uncertainty with respect to the central value of each quantity as that at the $\rm{M_{Z}}$ scale. For the charged lepton Yukawa couplings, a relative uncertainty of $1\%$ is assumed, instead of smaller experimental statistical errors, in order to take into account the theoretical uncertainties such as SUSY and GUT scale threshold effects.

\begin{table}[th]
\centering
\footnotesize
\resizebox{0.6\textwidth}{!}{
\begin{tabular}{|c|c|c|}
\hline
$\tan\beta$ & $10$ & $50$ \\ [1ex] \hline
$(\eta_{u},\eta_{c},\eta_{t})$ & (0.385, 0.381, 0.536) & (0.377, 0.382, 0.551)  \\ \hline
$(\eta_{d},\eta_{s},\eta_{b})$ & (0.241, 0.236, 0.273) & (0.175, 0.181, 0.211)  \\ \hline
$(\eta_{e},\eta_{\mu},\eta_{\tau})$ & (0.583,  0.583, 0.585) & (0.423,  0.423, 0.442)  \\ \hline
$(\eta^{CKM}_{us},\eta^{CKM}_{cb},\eta^{CKM}_{ub})$ & (0.999,  0.890, 0.890) & (0.999,  0.826, 0.826) \\ \hline
\end{tabular}
}
\caption{  Renormalization group running factors for the masses, $\eta_{i}=m_{i}(\rm{M_{GUT}})/ \it{m_{i}}(\rm{M_{Z}}) $ (taken from Ref.~\cite{Xing:2007fb}). These values are obtained with two-loop MSSM renormalization group evolution with appropriate one-loop matching conditions.  In the last row the renormalization group running factors $\eta^{CKM}_{ij}=V_{ij}(\rm{M_{GUT}})/ \it{V_{ij}}(\rm{M_{Z}})$ of the CKM matrix elements  are listed, which are obtained by evolving the RGEs for these parameters~\cite{Babu:1987im,Barger:1992pk} from low energy to $\rm{M_{GUT}}$.}
\label{table:6}
\end{table}

\begin{table}[th]
\centering
\footnotesize
\resizebox{0.6\textwidth}{!}{
\begin{tabular}{|c|c|c|}
\hline
\pbox{20cm}{Yukawa Couplings and \\  CKM mixing parameters} & \pbox{20cm}{$\tan\beta =10$ \\ (at $\mu= \rm{M_{GUT}}$)}  & \pbox{20cm}{$\tan\beta =50$ \\ (at $\mu= \rm{M_{GUT}}$)} \\ [1ex] \hline
$y_{u}/10^{-6}$ & $2.57 \pm 0.86 $ & $2.51 \pm 0.84 $  \\ \hline
$y_{c}/10^{-3}$ & $1.37 \pm 0.04$ & $1.37 \pm 0.04$   \\ \hline
$y_{t}/10^{-1}$ & $5.31 \pm 0.04$ & $5.43 \pm 0.04$ \\ \hline
$y_{d}/10^{-4}$ & $0.39 \pm 0.04$ & $1.44 \pm 0.14$  \\ \hline
$y_{s}/10^{-3}$ & $0.74 \pm 0.03$ & $2.84 \pm 0.14$ \\ \hline
$y_{b}/10^{-2}$ & $4.49 \pm 0.04$ & $17.29 \pm 0.15$ \\ \hline
$y_{e}/10^{-5}$ & $1.63 \pm 0.01$ & $5.91 \pm 0.05$ \\ \hline
$y_{\mu}/10^{-3}$ & $3.45  \pm 0.03$ & $12.49  \pm 0.12$ \\ \hline
$y_{\tau}/10^{-2}$ & $5.89  \pm 0.05$ & $22.21  \pm 0.22$ \\ \hline
$ \lvert V_{us} \rvert /10^{-2}$ & $22.53 \pm 0.07$ & $22.53 \pm 0.07$ \\ \hline
$\lvert V_{cb} \rvert /10^{-2}$ & $3.74 \pm 0.05$ & $3.47 \pm 0.05$ \\ \hline
$\lvert V_{ub} \rvert /10^{-3}$ & $3.24 \pm 0.11$  & $3.00 \pm 0.10$ \\ \hline
$\eta_{W}$ & $0.35 \pm 0.01$ & $0.35 \pm 0.01$ \\ [1ex] \hline
\end{tabular}
}
\caption{ Input values at $\rm{M_{GUT}}$ used in our fits. Central values and 1 $\sigma$ errors are quoted.
For Yukawa couplings, these numbers are found with the help of Tables \ref{table:1} and \ref{table:6} and by using the equations $y^{\rm{MSSM}}_{\it{u_{i}}}(\rm{M_{GUT}})=\it{y}^{\rm{SM}}_{\it{u_{i}}}(\rm{M_{Z}}) \eta_{\it{u_{i}}}/ \sin\beta$ for up-type quarks and $\it{y}^{\rm{MSSM}}_{\it{d_{i},e_{i}}}(\rm{M_{GUT}})=\it{y}^{\rm{SM}}_{\it{d_{i},e_{i}}}(\rm{M_{Z}}) \eta_{\it{d_{i},e_{i}}}/ \cos\beta$ for down-type quarks and charged leptons. For the charged lepton Yukawa couplings, a relative uncertainty of $1\%$ is assumed, instead of smaller experimental statistical errors, in order to take into account the theoretical uncertainties from threshold effects.  For the CKM mixing parameters, we evolve the quantities  from low scale to $\rm{M_{GUT}}$  by using the RGEs provided in Ref.~\cite{Babu:1987im,Barger:1992pk}.  }
\label{table:2}
\end{table}

With these GUT scale inputs, using the Eqs. \eqref{U4}-\eqref{L4}, we perform $\chi^{2}$ minimization by treating $\epsilon_{1}$, $\epsilon_{2}$ and $\epsilon_{4}$ as parameters and fit the data in the charged fermion sector.  Here $n_{obs}=13$ is the number of observables, with 3 parameters to fit them.  The elements of the random matrices pick up random values independently according to Gaussian distribution. For our analysis the error, pull and $\chi^{2}$-function are defined as follows:
\vspace{-5pt}
\begin{eqnarray}\label{eq:dis}
\begin{aligned}
\sigma_{i} &= \sqrt{ \sigma^{2}_{i\; \rm{th}} +  \sigma^{2}_{i\; \rm{exp}} }, \\
P_{i} &= \frac{O_{i\; \rm{th}}-E_{i\; \rm{exp}}}{\sigma_{i}}, \\
\chi^{2} &= \sum_{i} P_{i}^{2},
\end{aligned}
\end{eqnarray}

\noindent
where $\sigma_{i\; \rm{th}}$ and $\sigma_{i\; \rm{exp}}$ represent the theoretical standard deviation (TSD) and experimental 1$\sigma$ uncertainty respectively and $O_{i\; \rm{th}}$, $E_{i\; \rm{exp}}$ and $P_{i}$ represent the theoretical mean value (TMV), experimental central value (ECV) and pull of an observable $i$.

\begin{table}[th]
\centering
\footnotesize
\resizebox{0.43\textwidth}{!}{
\begin{tabular}{|c|c|c|}
\hline
$\tan\beta$ & $10$ & $50$ \\ [1ex] \hline
$\epsilon_{1}$ & 0.00181$\pm$0.00010 & 0.00169$\pm$0.00009  \\ \hline
$\epsilon_{2}$ & 0.0388$\pm$0.00222 & 0.03659$\pm$0.00215  \\ \hline
$\epsilon_{4}$ & 0.04055$\pm$0.00229 & 0.15716$\pm$0.00894 \\ [0.5ex] \hline
\end{tabular}
}
\caption{  Model parameters determined by $\chi^{2}$ minimization for the $SU(5)$-based GUTs  defined in Eqs. \eqref{U4}-\eqref{L4}.}
\label{table:10}
\end{table}

We find the minimum with $\chi^{2}/n_{obs} \sim 1$ along with the model parameters shown in Table \ref{table:10}.  The best fit values of the observables obtained with these fixed model parameters resulting from our Monte Carlo optimization are shown in Table \ref{table:4}. In Fig. \ref{fig:001} we plot the histogram distributions of the observables in the quark and the charged lepton sectors corresponding to the fixed model parameters given in Table~\ref{table:10} for the case where  $\tan\beta=10$ (plots for the case $\tan\beta=50$ are similar). In producing these distributions we have taken the sample size to be $10^{4}$ and chose the bin size (N bins) to be 50.

\begin{table}[!ht]
\centering
\footnotesize
\resizebox{0.9\textwidth}{!}{
\begin{tabular}{|c|c|c||c|c|c|c|}\hline
Observables & \multicolumn{2}{c|}{TMV$\pm$TSD} & \multicolumn{2}{c|}{$\frac{\rm{TMV}}{\rm{ECV}}$} & \multicolumn{2}{c|}{pull} \\\cline{2-7}
 & $\tan\beta=10$ & $\tan\beta=50$ & $\tan\beta=10$ & $\tan\beta=50$ & $\tan\beta=10$ & $\tan\beta=50$ \\\hline
$y_{u}/10^{-6}$ & 7.23$\pm$7.76 & 6.39$\pm$6.93    & 2.81 &  2.54 & 0.59 &  0.55  \\ \hline
$y_{c}/10^{-3}$ & 2.55$\pm$2.53 &  2.26$\pm$2.37   & 1.85 &  1.64 & 0.46 &  0.37   \\ \hline
$y_{t}$ & 0.88$\pm$0.46 &  0.89$\pm$0.46    & 1.67 &  1.63 & 0.77 &  0.74  \\ \hline
$y_{d}/10^{-4}$ & 0.64$\pm$0.33 &  2.3$\pm$1.23     & 1.61 &  1.62 & 0.73 &  0.73  \\ \hline
$y_{s}/10^{-3}$ & 2.10$\pm$0.77 &  7.59$\pm$2.79     & 2.83 &  2.67 & 1.75 &  1.69  \\ \hline
$y_{b}/10^{-1}$ & 0.67$\pm$0.19 &   2.61$\pm$0.76   & 1.50 &  1.51 & 1.13 &  1.15 \\ \hline
$y_{e}/10^{-4}$ & 0.64$\pm$0.34 &  2.34$\pm$1.22     & 3.96 &  3.96 & 1.42 &  1.42  \\ \hline
$y_{\mu}/10^{-3}$ & 2.10$\pm$0.75 &  7.63$\pm$2.74    & 0.60 &  0.61 & -1.79 &  -1.76   \\ \hline
$y_{\tau}/10^{-1}$ & 0.67$\pm$0.19 &  2.59$\pm$0.76    & 1.14 &  1.16 & 0.42 &   0.48  \\ \hline
$\lvert V_{us} \rvert /10^{-2}$ & 8.17$\pm$7.80 &  8.07$\pm$7.87     & 0.36 &  0.35 & -1.83 &  -1.83  \\ \hline
$\lvert V_{cb} \rvert /10^{-2}$ & 6.15$\pm$6.37 &  5.99$\pm$6.34   & 1.64 &  1.72 & 0.37 &  0.39  \\ \hline
$\lvert V_{ub} \rvert /10^{-3}$ & 3.42$\pm$3.67 &  3.23$\pm$3.75    & 1.05 &  1.07 & 0.04 &  0.06  \\ \hline
$\eta_{W}$ & 0.05$\pm$3.13 &  0.05$\pm$2.59     & 0.14 &  0.14 & -0.09 &  -0.11 \\ [1ex]  \hline
\end{tabular}
}
\caption{$\chi^{2}$ best fit values of the observables for the $SU(5)$-based GUTs   defined in Eqs. \eqref{U4}-\eqref{L4} with the fixed model parameters given in Table \ref{table:10}. The best fit values shown in this table correspond to $\chi^{2}/n_{obs}=$ 1.13 and 1.12 for $\tan\beta=$ 10 and 50 respectively. Here TMV$=$theoretical mean value, TSD$=$theoretical standard deviation, ECV$=$experimental central value and pull is defined in Eq. \eqref{eq:dis}. }\label{table:4}
\end{table}

\begin{figure}[th!]
\centering
\includegraphics[scale=0.3]{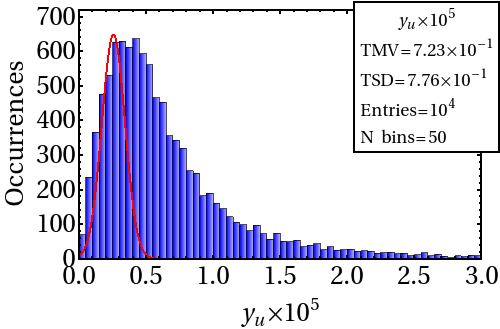}
\includegraphics[scale=0.3]{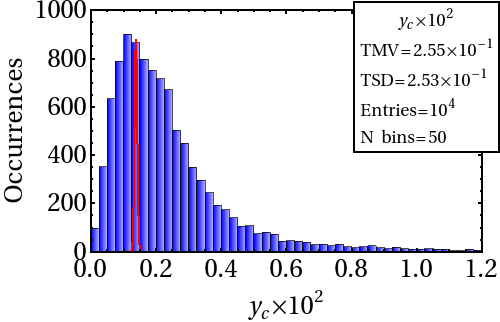}
\includegraphics[scale=0.3]{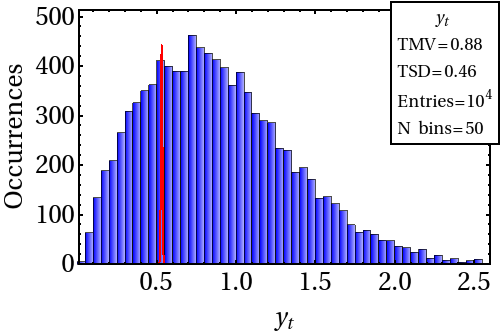}
\includegraphics[scale=0.3]{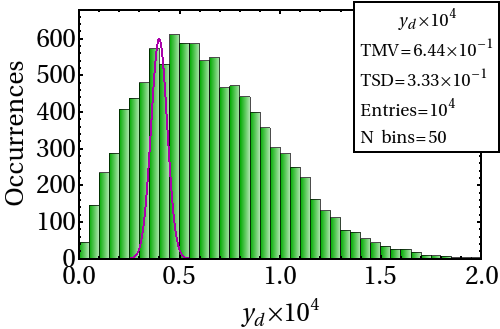}
\includegraphics[scale=0.3]{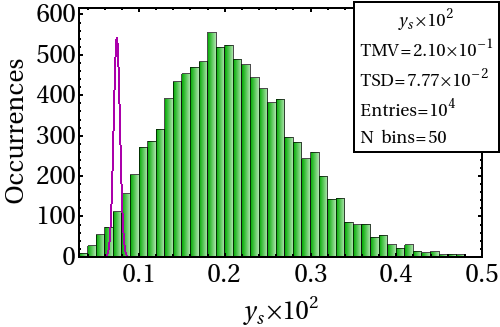}
\includegraphics[scale=0.3]{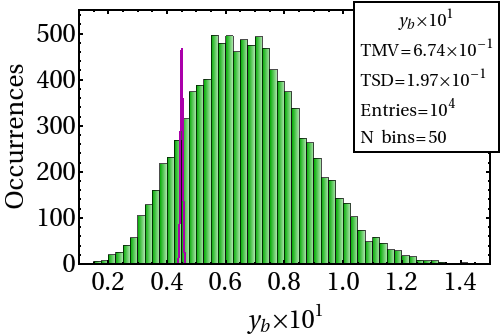}
\includegraphics[scale=0.3]{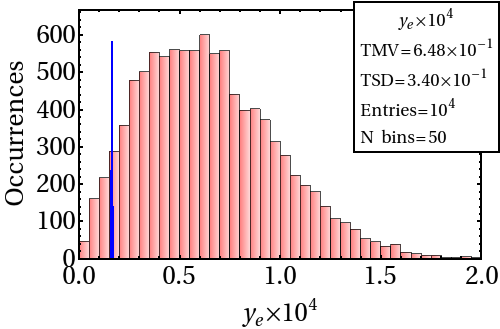}
\includegraphics[scale=0.3]{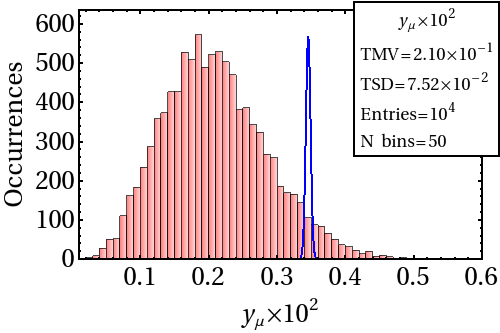}
\includegraphics[scale=0.3]{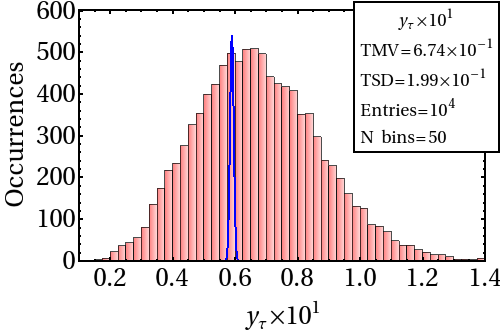}
\includegraphics[scale=0.3]{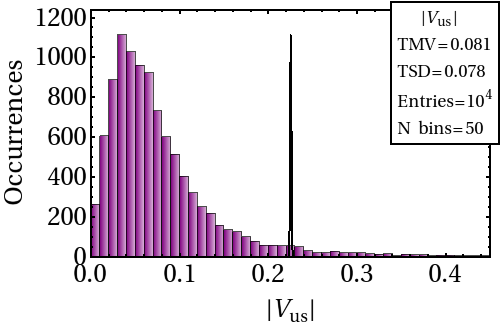}
\includegraphics[scale=0.3]{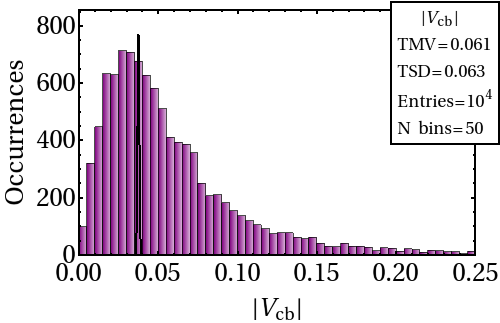}
\includegraphics[scale=0.3]{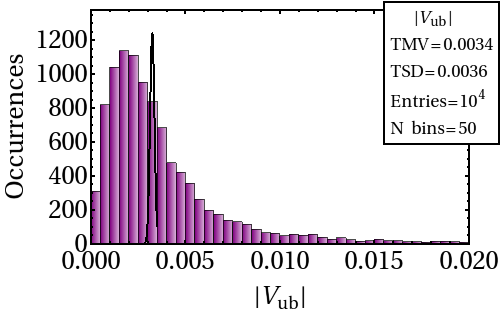}
\includegraphics[scale=0.3]{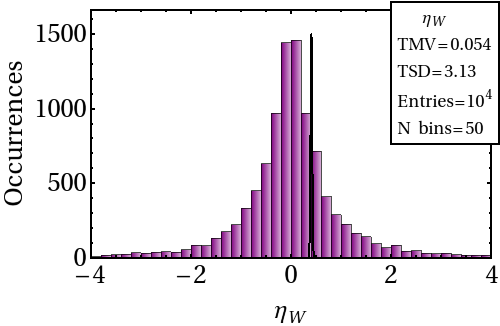}
\caption{ Histogram plots showing the distributions of the observables in the charged fermion sector. Blue (green, pink and purple) plots are the theoretical distributions of the up-type quarks (down-type quarks, charged leptons and CKM mixing parameters) according to the $SU(5)$-based GUTs  with $10^{4}$ occurrences for the case of $\tan\beta=10$ corresponding to the model parameters given in Table \ref{table:10}. Red (magenta, blue and black) curves represent the corresponding experimental 1$\sigma$ uncertainty range. For the charged leptons, a relative uncertainty of $1\%$ is assumed in order to take into account theoretical uncertainties arising from SUSY and GUT scale threshold effects. The number of bins (N bins) is chosen to be 50. }\label{fig:001}
\end{figure}

The blue plots in Fig. \ref{fig:001} show histograms of the theoretical distributions of the up-type quark Yukawa couplings.
Overlaid on these distributions are the experimental values of these couplings. We find very good agreement between theoretical expectations and observations. Among all the charged fermions, the eigenvalue spectrum of the up-type quarks shows the most hierarchical structure which is nicely reproduced. This is not surprising, as the stronger hierarchy is built into the model, see Eqs. (\ref{U4})-(\ref{L4}).

For the down-type quark Yukawa couplings, theoretical distributions are shown in green in Fig. \ref{fig:001}. Overlaid on these distributions are the experimental values of these parameters. These are in good agreement with observations for down-quark and bottom-quark, whereas for the strange-quark, the theoretical mean value tends to be a little higher than the experimentally measured value, but it is still within acceptable range. In the eigenvalue spectrum of charged leptons, which is shown in pink in Fig.  \ref{fig:001},  the theoretical mean value for the muon Yukawa coupling tends to be a little lower than the experimental central value. The reason for these small discrepancies can be understood from the approximate relations
$\frac{y_{s}}{y_{b}} \sim \epsilon_{2} \; \; \rm{and} \; \; \frac{y_{\mu}}{y_{\tau}} \sim \epsilon_{2}$ present in the model.
At the GUT scale one has roughly $y_{b} \sim y_{\tau}$, which implies within the model $y_{s} \sim y_{\mu}$. This is why the histograms of Yukawa couplings for both strange-quark and muon Yukawa couplings are almost identical with approximately the same theoretical mean values, but observation dictates, $y_{s} \sim 4 y_{\mu}$ at the GUT scale. This small discrepancy, inherent to these models, is still not major and is within acceptable range.

The probability distributions of the CKM parameters are shown in purple in Fig. \ref{fig:001}.  Overlaid on these distributions are the experimental values of these observables.  These distributions \footnote{Similar distributions for the CKM parameters are obtained in Ref.  \cite{Donoghue:2005cf} from a completely different statistical approach.}  are also in very good agreement with data.
The theoretical distribution for $V_{us}$ has a mean value that tends to be somewhat smaller than the experimental value. This feature may be understood since the model has $V_{us} \sim \epsilon_{1}/ \epsilon_{2}$.  It also predicts $y_d/y_s \sim 0.05 \sim \epsilon_{1}/ \epsilon_{2}$, which makes $V_{us}$ to peak around $0.05$, rather than the observed value of $\sim 0.2$.  But there is still acceptable agreement.

\begin{figure}[th!]
\centering
\includegraphics[scale=0.35]{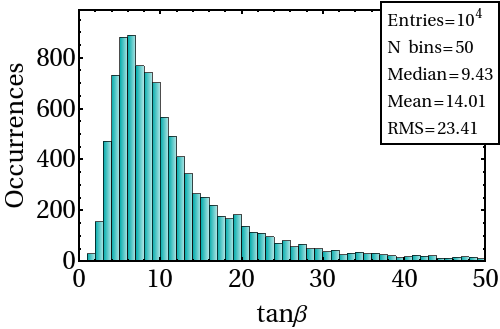}
\includegraphics[scale=0.35]{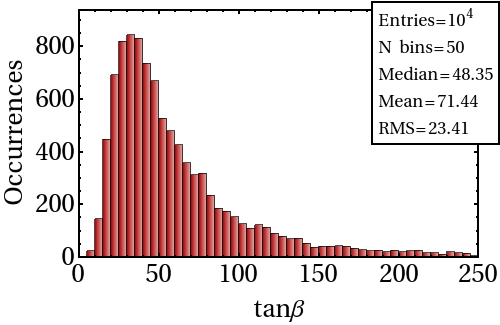}
\caption{Histograms showing theoretical distributions of $\tan\beta$ given by Eq. \eqref{tanbeta} for the $SU(5)$-based GUTs with sample size of $10^{4}$. Left plot corresponds to the case where $\tan\beta=$ 10 and the right plot for $\tan\beta=$ 50. The number of bins (N bins) is chosen to be 50.}\label{fig:003}
\end{figure}

We can do a consistency check for the value of $\tan\beta$ used.
From Eq. \eqref{tanbeta} we have, $\tan \beta \simeq \epsilon_4\; m_t/m_b\;  \frac{|(\overrightarrow{d_{0}})_{3}|}{(Y^0_U)_{33}}$. Since $O(1)$ random variables are present in this equation, $\tan\beta$ in these models follows a distribution shown in Fig. \ref{fig:003}. Both  histograms have a long tail behaviour with the mean values of the distributions being $\tan\beta =$ 14 and 71.4 respectively. For histograms with such behaviour, median may be a better measure, which are $\tan\beta =$ 9.4 and 48.3 respectively.  We see broad consistency with the input values of $\tan\beta$ used in each case.

Since the small parameters $\epsilon_{i}$ do not enter into the neutrino sector, in the optimization process we did not include the neutrino observables. Once the model parameters  are fixed as in Table \ref{table:10}, one can include the neutrino sector in the sampling process  and investigate how well the observed quantities in this sector are reproduced by these models. Since the matrix structure is the same as the ones considered in earlier works assuming anarchical hypothesis only in the neutrino sector~\cite{Hall:1999sn,Haba:2000be},  the histogram distributions of the neutrino observables should be similar, which is what we find. In Figs. \ref{fig:002A} and \ref{fig:002B} we present plots for the theoretical predictions of the neutrino observables. The theoretical average values of these observables resulting from the Monte Carlo analysis are shown in Table \ref{table:5}. The input values for neutrino observables are taken from Ref.~\cite{Fogli:2012ua} corresponding to the case of normal ordering of the neutrino mass spectrum. We restrict our analysis to normal ordering, since the random matrix structure for the neutrinos strongly prefers this over inverted ordering. In our Monte Carlo simulations we found a $95.6 \%$ probability for normal ordering and a $4.4 \%$ probability for  inverted ordering, which is  similar to the results of Ref.~\cite{Lu:2014cla}).
To ensure normal ordering, we assume $m_{1} \leq m_{2} < m_{3}$  and  we put the constraint $r < 1$ ($r \equiv \Delta m^{2}_{\rm{sol}}/\Delta m^{2}_{\rm{atm}}$ with $\Delta m^{2}_{\rm{sol}}=m^{2}_{2}-m^{2}_{1}$ and $\Delta m^{2}_{\rm{atm}}=m^{2}_{3}-m^{2}_{2}$) in the sampling procedure.

\begin{figure}[th!]
\centering
\includegraphics[scale=0.25]{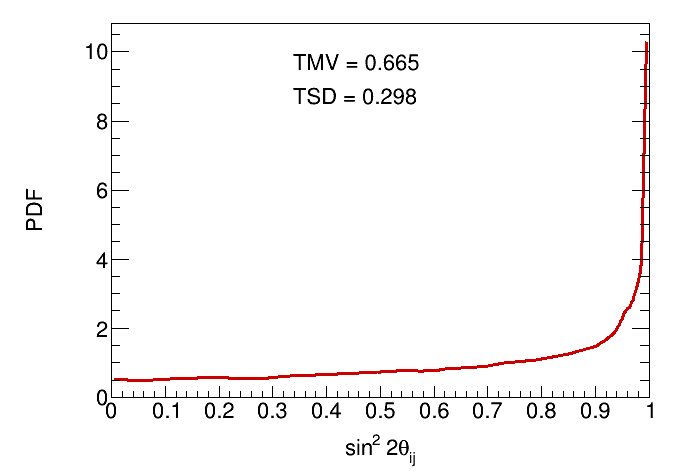}
\includegraphics[scale=0.25]{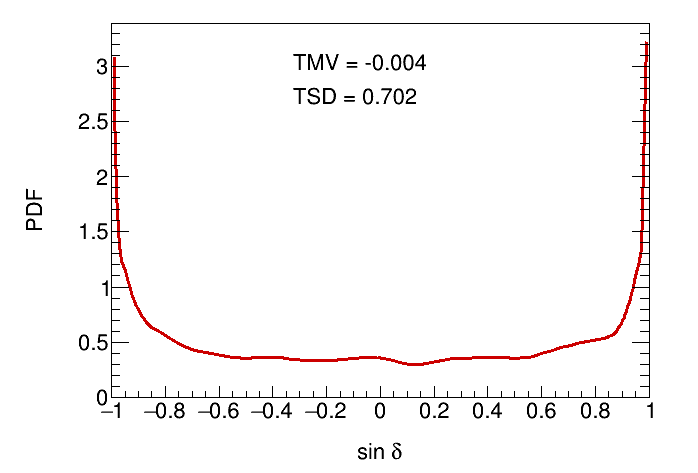}
\caption{Probability density plots for the neutrino mixing parameters for $SU(5)$-based GUTs. The left plot is for the mixing angles, $\sin^{2}2 \theta_{ij}$ for $(ij) = (12), (23)$ and $(13)$, and the right plot is for the  CP-violating parameter $\sin\delta$. In these probability density plots, the area under the curve within a certain range represents the probability of finding the quantity within that particular range. Here TMV$=$theoretical mean value, TSD$=$theoretical standard deviation. }\label{fig:002A}
\end{figure}

\begin{table}[th!]
\centering
\footnotesize
\resizebox{0.6\textwidth}{!}{
\begin{tabular}{|c|c|c||c|c|c|c|}
\hline
Observables & ECV & $1\sigma$ exp & TMV & TSD & $\frac{\rm{TMV}}{\rm{EMV}}$ & pull \\ [1ex] \hline
$\frac{\Delta m^{2}_{sol}}{\Delta m^{2}_{atm}}$ & 0.031 & 0.001 & 0.135 & 0.186 & 4.37 & 0.56 \\ \hline
$\sin^{2}\theta_{12}$ & 0.308 & 0.017 & 0.504 & 0.287 & 1.63 & 0.68 \\ \hline
$\sin^{2}\theta_{23}$ & 0.3875 & 0.0225 & 0.501 & 0.290 & 1.29 & 0.39 \\ \hline
$\sin^{2}\theta_{13}$ & 0.0241 & 0.0025 & 0.334 & 0.235 & 13.8 & 1.31   \\ [1ex]  \hline
\end{tabular}
}
\caption{Theoretical sampling results of the $SU(5)$-based model obtained from Monte Carlo simulation in the neutrino sector. Experimental central values with associated one sigma uncertainties are also quoted taken  from Ref.~\cite{Fogli:2012ua}. Here TMV$=$theoretical mean value, TSD$=$theoretical standard deviation, ECV$=$experimental central value and pull is defined in Eq. \eqref{eq:dis}. The theoretical results presented here are for sample size of $10^{4}$. The best fit values shown in this table correspond to $\chi^{2}/n_{obs}=$ 0.66.}
\label{table:5}
\end{table}

\begin{figure}[h]
\centering
\includegraphics[scale=0.3]{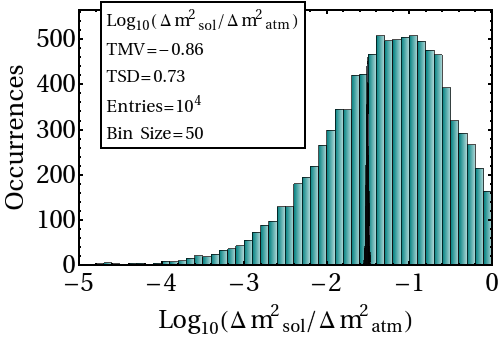}
\includegraphics[scale=0.3]{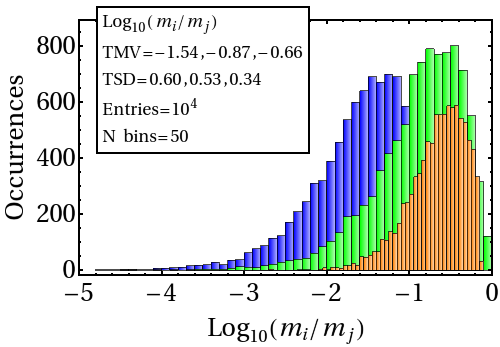} \\
\includegraphics[scale=0.23]{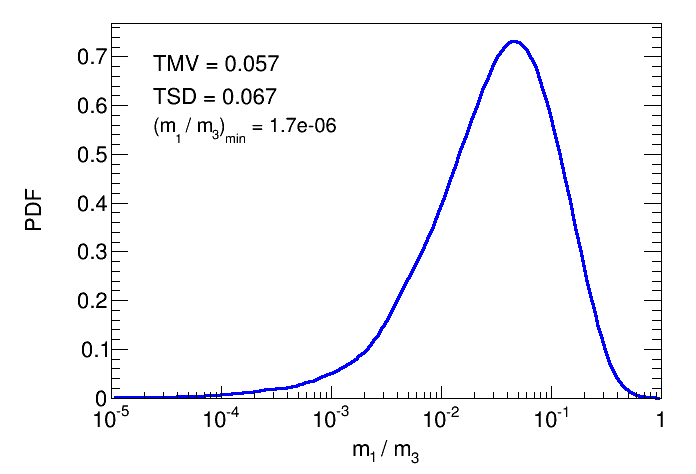}
\includegraphics[scale=0.23]{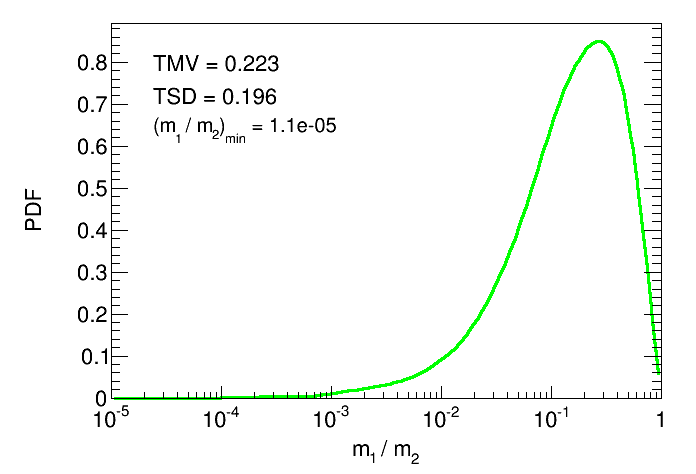}
\caption{ Two histogram plots showing the theoretical distributions of $\log_{10}(\Delta m^{2}_{\rm{sol}}/\Delta m^{2}_{\rm{atm}})$ (upper left) and
$\log_{10}(m_{ij})=\log_{10}(m_{i}/m_{j})$ (upper right; blue, green and orange histograms are for $\log_{10}(m_{1}/m_{3})$, $\log_{10}(m_{1}/m_{2})$ and $\log_{10}(m_{2}/m_{3})$ respectively). The black curve in the upper left plot represents the experimental 1$\sigma$ uncertainty range.
The two bottom plots are the probability density functions for the neutrino mass ratios $m_{i}/m_{j}$ (blue and green  plots are for $m_{1}/m_{3}$ and $m_{1}/m_{2}$). In these probability density plots,  the area under the curve within a certain range represents the probability of finding the quantity within that particular range. These plots are the results from our Monte Carlo analysis for the anarchical neutrino mass models with normal mass ordering. Here TMV$=$theoretical mean value, TSD$=$theoretical standard deviation.  For the two histogram distributions the number of bins is chosen to be 50 and for all the plots the sample size is taken to be $10^{4}$. }\label{fig:002B}
\end{figure}

In Fig. \ref{fig:002A} we plot the probability density for the neutrino mixing parameters. The area under the curve in a probability density plot between any two values of the observable represents the probability of finding the observable within that particular range and the total area is normalized to unity. From these plots it is clear that for this class of models all the mixing  parameters  $\sin^{2}2\theta_{ij}$ in the neutrino sector take preferentially large values. The  CP-violating parameter $\sin\delta$ is peaked at its maximal values  of $\pm1$. Preference of all the mixing parameters to be large is a consequence of the complete anarchical form of the neutrino mass matrix as their distributions are uniquely fixed by the invariant Haar measure.

In Fig. \ref{fig:002B} we plot theoretical distributions of $\log_{10}(\Delta m^{2}_{\rm{sol}}/\Delta m^{2}_{\rm{atm}})$ and $\log_{10}(m_{i}/m_{j})$.
The upper left plot in Fig. \ref{fig:002B} shows that the anarchic structure of the neutrino mass matrix prefers small values of the ratio of the two mass squared differences, $r$  and the theoretical mean value is quite close to the experimental central value. The upper right plot reveals that anarchy predicts mild hierarchy in the neutrino mass spectrum. The lower plots in  Fig. \ref{fig:002B} exhibits the probability densities for the two different neutrino mass ratios, $m_{1}/m_{3}$ and $m_1/m_2$.  As can be seen, the ratio $m_1/m_2$ peaks around 0.3.  Extreme small values of $m_1$ are strongly disfavored in this model.  For example, $m_1/m_2 < 0.01$ will be favored only with a 4\% probability.

\subsection{Monte Carlo analysis of \boldmath{$SU(5)$}-inspired  $U(1)$ flavor models}
\label{SU5FHD}
\subsubsection{Models with two parameters $\{\epsilon, \epsilon_4\}$}
\label{CRG1}
In this subsection, we present our Monte Carlo results for the $SU(5)$-inspired  $U(1)$ flavor models with $U(1)$ charges chosen to be $\{q_1=1, q_2=0, p=0\}$   as explained in Sec. \ref{SU5F}. Models of this type have two  parameters, $\{\epsilon, \epsilon_4\}$}. The only modification needed compared to our general setup is  in the charged fermions sector where the matrix $H$ is given by Eq. \eqref{HH}. This set of models has one less parameter compared to the general model. We have performed a fit as before in this two parameter case and the fitted model parameters are presented in Table~\ref{two_para_eps}. From this Table one finds, $\epsilon \sim \lambda^2$, where $\lambda \sim 0.22$.  With this fixed parameters, the corresponding best fit values of the observables are shown in Table \ref{two_para_fit} and the theoretical distributions of these quantities are presented in Fig. \ref{two_para_fig} in Appendix \ref{App:A01}. By comparing the fit results of Tables \ref{table:4} and \ref{two_para_fit} one sees that a slightly better fit is obtained for the three parameter case compared to the analysis done here with one less parameter. In Table \ref{table:4}, all the observables  are reproduced within 2$\sigma$ error on average, whereas in Table \ref{two_para_fit}, with one less parameter, two of the observables are in the $(2-3) \sigma$ range  for the case of $\tan \beta=10$ and for the case of $\tan \beta=50$, one of the observables is little above 2$\sigma$ error on average.  Since the neutrino sector is exactly the same for all these models belonging to $SU(5)$-based GUTs, the analysis in the previous subsection remains unchanged.

\begin{table}[th]
\centering
\footnotesize
\resizebox{0.45\textwidth}{!}{
\begin{tabular}{|c|c|c|}
\hline
$\tan\beta$ & $10$ & $50$ \\ [1ex] \hline
$\epsilon$ & 0.02855$\pm$0.00150 & 0.03847$\pm$0.00215  \\ \hline
$\epsilon_{4}$ & 0.03909$\pm$0.00220 & 0.14537$\pm$0.00826 \\ [0.5ex] \hline
\end{tabular}
}
\caption{  Model parameters determined by $\chi^{2}$ minimization  for the $SU(5)$-inspired   $U(1)$ flavor symmetry models with two parameters. }
\label{two_para_eps}
\end{table}

\begin{table}[!ht]
\centering
\footnotesize
\resizebox{0.9\textwidth}{!}{
\begin{tabular}{|c|c|c||c|c|c|c|}\hline
Observables & \multicolumn{2}{c|}{TMV$\pm$TSD} & \multicolumn{2}{c|}{$\frac{\rm{TMV}}{\rm{ECV}}$} & \multicolumn{2}{c|}{pull} \\\cline{2-7}
 & $\tan\beta=10$ & $\tan\beta=50$ & $\tan\beta=10$ & $\tan\beta=50$ & $\tan\beta=10$ & $\tan\beta=50$ \\\hline
$y_{u}/10^{-6}$ & 3.49$\pm$3.89 & 4.96$\pm$5.55    & 1.35 &  1.97 & 0.23 &  0.43  \\ \hline
$y_{c}/10^{-3}$ & 2.08$\pm$2.15 &  2.50$\pm$2.57   & 1.51 &  1.82 & 0.32 &  0.43   \\ \hline
$y_{t}$ & 0.88$\pm$0.46 &  0.88$\pm$0.46    & 1.65 &  1.63 & 0.76 &  0.74  \\ \hline
$y_{d}/10^{-4}$ & 0.44$\pm$0.23 &  1.92$\pm$1.00     & 1.10 &  1.32 & 0.17 &  0.46  \\ \hline
$y_{s}/10^{-3}$ & 1.90$\pm$0.69 &  7.45$\pm$2.71     & 2.56 &  2.62 & 1.67 &  1.69  \\ \hline
$y_{b}/10^{-1}$ & 0.67$\pm$0.19 &   2.42$\pm$0.71   & 1.49 &  1.40 & 1.11 &  0.96 \\ \hline
$y_{e}/10^{-4}$ & 0.44$\pm$0.23 &  1.90$\pm$1.00     & 2.69 &  3.21 & 1.41 &  1.31  \\ \hline
$y_{\mu}/10^{-3}$ & 1.90$\pm$0.75 &  7.38$\pm$2.71    & 0.55 &  0.59 & -2.24 &  -1.87   \\ \hline
$y_{\tau}/10^{-1}$ & 0.68$\pm$0.19 &  2.42$\pm$0.70    & 1.15 &  1.09 & 0.42 &   0.28  \\ \hline
$\lvert V_{us} \rvert /10^{-2}$ & 8.17$\pm$7.80 &  6.81$\pm$6.86     & 0.36 &  0.30 & -2.68 &  -2.29  \\ \hline
$\lvert V_{cb} \rvert /10^{-2}$ & 5.75$\pm$5.93 &  6.19$\pm$6.30   & 1.53 &  1.78 & 0.33 &  0.43  \\ \hline
$\lvert V_{ub} \rvert /10^{-3}$ & 2.73$\pm$3.03 &  2.81$\pm$2.96    & 0.84 &  0.93 & -0.16 &  -0.06  \\ \hline
$\eta_{W}$ & 0.006$\pm$2.509 &  0.003$\pm$2.30     & 0.01 &  0.006 & -0.15 &  -1.13 \\ [1ex]  \hline
\end{tabular}
}
\caption{$\chi^{2}$ best fit values of the observables for the $SU(5)$-inspired   $U(1)$ flavor symmetry models with two  parameters.  The fixed model parameters are given in Table \ref{two_para_eps}. The best fit values shown in this table correspond to $\chi^{2}/n_{obs}=1.44$  and 1.41  for $\tan\beta=$ 10 and 50 respectively.}\label{two_para_fit}
\end{table}

\subsubsection{Monte Carlo analysis of \boldmath{$U(1)$} model with one parameter $\{\epsilon \}$}
\label{CRG2}
In this subsection we apply a Monte Carlo analysis to the $SU(5)$-inspired flavor symmetry model with the $U(1)$-flavor charge assignment of $\{q_1=2, q_2=1, p=0,1,2 \}$ as discussed in Sec. \ref{SU5F}. As explained there, the matrix elements in Eqs. \eqref{eq:100A}-\eqref{eq:100B} have order one complex coefficients $c^f_{ij}$.  We assume that the coefficients are random complex variables with Gaussian distribution of variance 0.5 for both real and imaginary parts. For the off-diagonal terms of the complex symmetric matrix $Y_{R}$ the coefficients have variance of 0.25. We generate this unbiased set of random variables following Gaussian distribution in a manner similar to the one described earlier. By taking the sample size to be $10^{4}$, we study the theoretical probability distributions of the observables in the fermion sector. We carry out the Monte Carlo analysis for  three cases  with $p=0,1,2$ (corresponding to $\tan\beta=55, 25, 5$ respectively) and present the values of the parameter $\epsilon$ that minimizes the $\chi^{2}$ for each case. For these values of $\tan\beta$ the RGE running factors are not given in Ref. \cite{Xing:2007fb} and hence we run the two loop MSSM RGEs \cite{Barger:1992ac,Barger:1992pk} from low scale to the GUT scale \footnote{We also performed the running for the cases with $\tan\beta=$ 10 and 50 and found consistency with Ref. \cite{Xing:2007fb} and hence the values presented in Table \ref{table:2}.}.  We take the low scale central values of the observables from  Table 2 of Ref.~\cite{Antusch:2013jca} at $\mu=1$ TeV where the observables are converted to the $\rm{\overline{DR}}$ scheme, use the SUSY matching formula (without taking into account the threshold corrections) for the Yukawa couplings and evolve them upto the GUT scale and use these values as inputs (shown in Table \ref{table:22}) during the optimization. Like before, for the charged leptons, we assume a relative $1\%$ uncertainty in order to take into account the theoretical uncertainties such as  SUSY and GUT scale threshold effects.

\begin{table}[th]
\centering
\footnotesize
\resizebox{0.75\textwidth}{!}{
\begin{tabular}{|c|c|c|c|}
\hline
\pbox{20cm}{Yukawa Couplings and \\  CKM mixing parameters} & \pbox{20cm}{$\tan\beta =5$ \\ (at $\mu= \rm{M_{GUT}}$)}  & \pbox{20cm}{$\tan\beta =25$ \\ (at $\mu= \rm{M_{GUT}}$)} & \pbox{20cm}{$\tan\beta =55$ \\ (at $\mu= \rm{M_{GUT}}$)} \\ [1ex] \hline
$y_{u}/10^{-6}$ & $2.98 \pm 1.00 $ & $2.88 \pm 0.96 $ & $2.96 \pm 0.99 $  \\ \hline
$y_{c}/10^{-3}$ & $1.45 \pm 0.04$ & $1.4 \pm 0.04$  & $1.44 \pm 0.04 $   \\ \hline
$y_{t}/10^{-1}$ & $5.43 \pm 0.04$ & $5.23 \pm 0.04$  & $5.85 \pm 0.05 $ \\ \hline
$y_{d}/10^{-4}$ & $0.24 \pm 0.02$ & $1.24 \pm 0.12$  & $3.55 \pm 0.36 $  \\ \hline
$y_{s}/10^{-3}$ & $0.48 \pm 0.024$ & $2.47 \pm 0.12$  & $7.04 \pm 0.35 $ \\ \hline
$y_{b}/10^{-2}$ & $2.73 \pm 0.02$ & $14.33 \pm 0.12$  & $49.61 \pm 0.44 $ \\ \hline
$y_{e}/10^{-4}$ & $0.10 \pm 0.001$ & $0.51 \pm 0.005 $ & $1.45 \pm 0.01 $  \\ \hline
$y_{\mu}/10^{-2}$ & $0.21 \pm 0.002$ & $1.08 \pm 0.01  $ & $3.07 \pm  0.03 $  \\ \hline
$y_{\tau}/10^{-1}$ & $0.36 \pm 0.003$ & $1.89 \pm 0.01  $ & $6.53 \pm  0.06 $  \\ \hline
$ \lvert V_{us} \rvert /10^{-2}$ & $22.53 \pm 0.07$ & $22.53 \pm 0.07$  & $22.53 \pm 0.07 $  \\ \hline
$\lvert V_{cb} \rvert /10^{-2}$ & $3.72 \pm 0.05$ & $3.70 \pm 0.05$  & $3.37 \pm 0.05 $  \\ \hline
$\lvert V_{ub} \rvert /10^{-3}$ & $3.22 \pm 0.11$  & $3.21 \pm 0.11$  & $2.92 \pm 0.10 $  \\ \hline
$\eta_{W}$ & $0.35 \pm 0.01$ & $0.35 \pm 0.01$  & $0.35 \pm 0.01 $  \\ [1ex] \hline
\end{tabular}
}
\caption{Experimental central values with associated 1$\sigma$ uncertainties at  $\rm{M_{GUT}}$ scale used in our fits. The low scale central values of the observables are taken from the Table 2 of Ref.~\cite{Antusch:2013jca} at $\mu=1$ TeV.  For the charged leptons, a relative uncertainty of $1\%$ is assumed in order to take into account the theoretical uncertainties as for example SUSY threshold and GUT scale effects. }
\label{table:22}
\end{table}

\begin{table}[th]
\centering
\footnotesize
\resizebox{0.45\textwidth}{!}{
\begin{tabular}{|c|c|c|c|}
\hline
$p$ & 2 & 1 & 0 \\ [1ex] \hline
$\tan\beta$ & 5 & 25 & 55 \\ \hline
$\epsilon$ & 0.1956$\pm$0.0097 & 0.1985$\pm$0.0105 & 0.1755$\pm$0.0098 \\ [0.5ex] \hline
\end{tabular}
}
\caption{  Model parameters fixed by minimization for the flavor symmetry based models defined in Eqs. \eqref{eq:100A}-\eqref{eq:100B} by employing Monte Carlo analysis with different values of $p$.}
\label{table:20}
\end{table}

\begin{figure}[th!]
\centering
\includegraphics[scale=0.21]{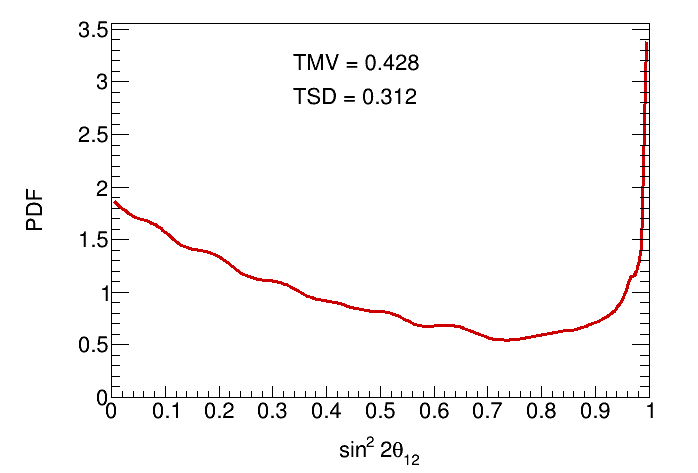}
\includegraphics[scale=0.21]{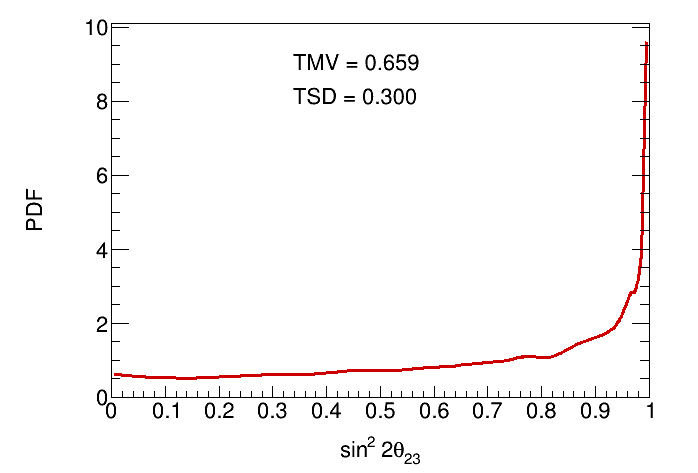}
\includegraphics[scale=0.21]{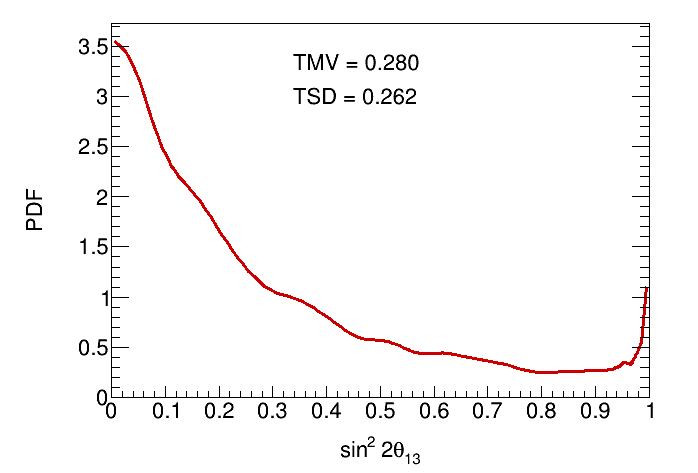}
\includegraphics[scale=0.21]{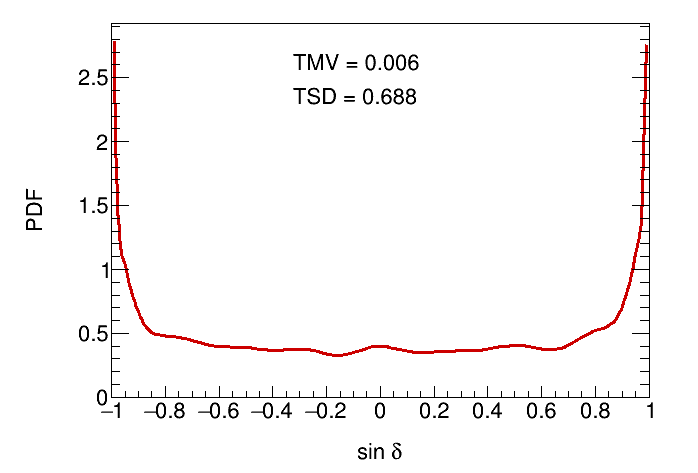}
\caption{Probability density plots for the neutrino mixing parameters for the $SU(5)$-inspired flavor symmetry based models defined in Eqs. \eqref{eq:100A}-\eqref{eq:100B}. The upper plots are for the mixing angles, $\sin^{2}2 \theta_{ij}$ and the lower plot is for CP-violating parameter $\sin\delta$.  }\label{fig:007A}
\end{figure}

\begin{figure}[th!]
\centering
\includegraphics[scale=0.3]{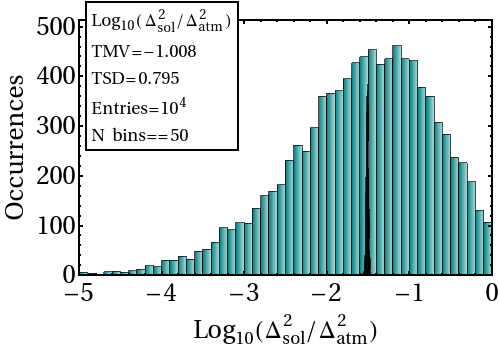}
\includegraphics[scale=0.3]{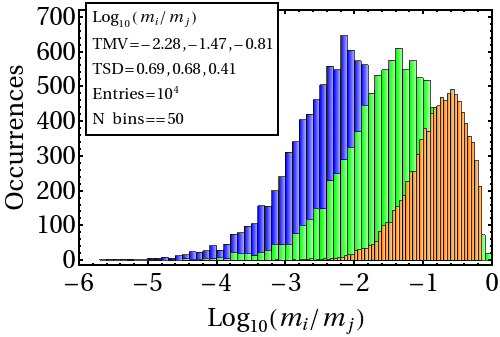} \\
\includegraphics[scale=0.23]{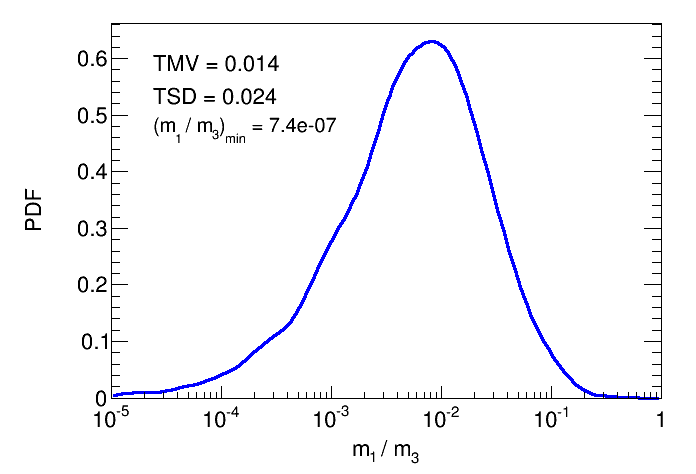}
\includegraphics[scale=0.23]{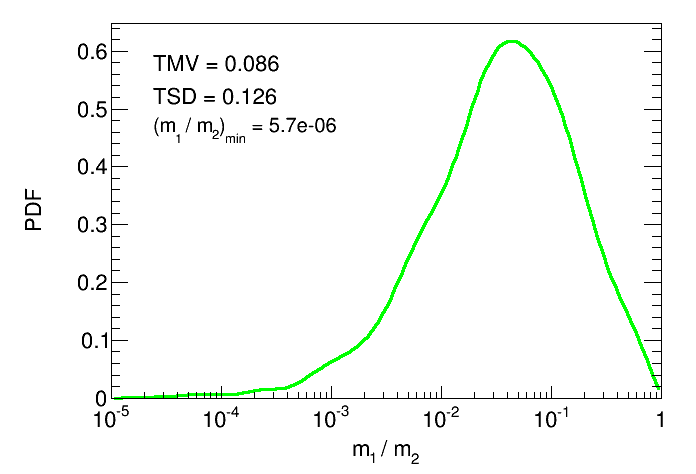}
\caption{The  theoretical distributions and the probability density plots of the observables in the neutrino sector for the $SU(5)$-inspired flavor symmetry based models defined in Eqs. \eqref{eq:100A}-\eqref{eq:100B}. The notation is the same as in Fig. \ref{fig:002B}. }\label{fig:007B}
\end{figure}

The numerical values of the model parameter determined by $\chi^2$-minimization are presented in Table \ref{table:20}. These values are similar to the ones computed in Table 2 of Ref.~\cite{Babu:2003zz}. The best fit values resulting from the $\chi^{2}$ minimization for the three cases with $p=0,1,2$ are presented in Table \ref{table:21}. From this Table one sees that, for this class of models with a single parameter, the fit to the charged fermion observables is not very different from that of the models with 3 parameters. For  $V_{us}$, the pull is greater than $2\sigma$, but the rest of the observables are in good agreement.  The main difference of this model compared to the previous two models is in the neutrino mixing parameters. In the $SU(5)$-based GUTs, the set of models where the left-handed light neutrino Yukawa coupling matrix elements are all $\sim O(1)$, large values of mixing angles are preferred for all three mixing parameters $\sin^{2}2\theta_{ij}$ (see Fig. \ref{fig:002A}). On the other hand, the present model  which is described by the Yukawa matrices given in  Eqs. \eqref{eq:100A}-\eqref{eq:100B},  $O(1)$ entries exist only in the 2-3 sector that give rise to large  $\sin^{2}2\theta_{23}$. But due to a suppression factor $\epsilon$ in the 1-3 sector, $\sin^{2}2\theta_{13}$ naturally comes out to be smaller than unity. The probability density plots of $\sin^{2}2\theta_{ij}$ are shown in Fig. \ref{fig:007A}, the patterns remain the same for different values of $p$  for this set of models (Fig. \ref{fig:007B}) compared to the previous set analyzed before (Fig. \ref{fig:002B}). Except for the three mixing parameters, the theoretical distributions of the observables in the fermion sector remain similar in pattern and are shown in Figs. \ref{fig:006} in Appendix \ref{App:A02} for the case of $p=2$ (histograms for other values of $p$'s are similar, and are not shown).

\begin{table}[!ht]
\centering
\footnotesize
\resizebox{1.05\textwidth}{!}{
\begin{tabular}{|c|c|c|c||c|c|c|c|c|c|}\hline
Observables & \multicolumn{3}{c|}{TMV$\pm$TSD}  & \multicolumn{3}{c|}{$\frac{\rm{TMV}}{\rm{ECV}}$} & \multicolumn{3}{c|}{pull} \\\cline{2-10}
  & $\tan\beta=5$ & $\tan\beta=25$ & $\tan\beta=55$ & $\tan\beta=5$ & $\tan\beta=25$  & $\tan\beta=55$ & $\tan\beta=5$ & $\tan\beta=25$  & $\tan\beta=55$ \\\hline
$y_{u}/10^{-6}$ & 4.88$\pm$5.61 & 5.42$\pm$6.06 & 2.00$\pm$2.26 &   1.63 & 1.88 & 0.67 & 0.33 & 0.41 & -0.38 \\ \hline
$y_{c}/10^{-3}$ & 2.42$\pm$2.47 & 2.59$\pm$ 2.66 & 1.62$\pm$1.76 &  1.66 & 1.84 & 1.12 & 0.39 & 0.44 & 0.10 \\ \hline
$y_{t}$ & 0.89$\pm$0.46 & 0.89$\pm$0.46 & 0.88$\pm$0.46 &  1.64 & 1.70 & 1.51 & 0.76 & 0.79 & 0.64 \\ \hline
$y_{d}/10^{-5}$  & 1.97$\pm$1.39 & 11.0$\pm$7.78 & 30.8$\pm$22.6 & 0.80 & 0.88 & 0.86 & -0.33 & -0.18 & -0.20   \\ \hline
$y_{s}/10^{-3}$  & 1.37$\pm$0.65 & 7.31$\pm$3.49 & 28.4$\pm$13.6 &  2.83 & 2.95 & 4.04 & 1.36 & 1.38 & 1.57 \\ \hline
$y_{b}/10^{-1}$  & 0.51$\pm$0.18 & 2.65$\pm$0.94 & 13.4$\pm$4.77 &   1.86 & 1.85 & 2.71 & 1.30 & 1.29 & 1.77  \\ \hline
$y_{e}/10^{-5}$  & 1.96$\pm$1.14 & 11.10$\pm$7.88 & 31.06$\pm$22.69   & 1.95 & 2.16 & 2.13 & 0.67 & 0.75 & 0.72 \\ \hline
$y_{\mu}/10^{-3}$ & 1.36$\pm$0.64 & 7.24$\pm$3.45 & 28.42$\pm$13.85 & 0.64 & 0.67 & 0.92 & -1.16 & -1.02 & -0.16  \\ \hline
$y_{\tau}/10^{-1}$  & 0.51$\pm$0.18 & 2.66$\pm$0.93 & 13.40$\pm$4.75 &  1.43 & 1.40 & 2.05 & 0.85 & 0.82 & 1.44 \\ \hline
$\lvert V_{us} \rvert /10^{-1}$  & 0.75$\pm$0.72 & 0.77$\pm$0.69 & 0.61$\pm$0.59 &   0.33 & 0.34 & 0.27 & -2.05 & -2.11 & -2.75 \\ \hline
$\lvert V_{cb} \rvert /10^{-1}$  & 0.65$\pm$0.62 & 0.66$\pm$ 0.65 & 0.53$\pm$0.54 &   1.74 & 1.79 & 1.57 & 0.44 & 0.45 & 0.35 \\ \hline
$\lvert V_{ub} \rvert /10^{-2}$  & 0.31$\pm$0.36 & 0.32$\pm$0.36 & 0.20$\pm$0.24 &   0.98 & 1.01 & 0.69 & -0.01 & 0.01 &-0.36  \\ \hline
$\eta_{W}$  & 0.04$\pm$5.56 & 0.01$\pm$2.49 & 0.04$\pm$2.72 &   0.11 & 0.02 & 0.11 & -0.05 & -0.13 & -0.11 \\ \hline
$\frac{\Delta m^{2}_{sol}}{\Delta m^{2}_{atm}}$ & 0.09$\pm$0.16 & 0.10$\pm$ 0.16 & 0.09$\pm$0.16 &  3.17 & 3.27 & 3.21 & 0.42 & 0.43 & 0.41 \\ \hline
$sin^{2}\theta^{PMNS}_{12}$ & 0.17$\pm$0.19 & 0.17$\pm$0.19 & 0.15$\pm$0.18 &  0.56 & 0.57 & 0.50 & -0.70 & -0.66 & -0.84 \\ \hline
$sin^{2}\theta^{PMNS}_{23}$ & 0.47$\pm$0.29 & 0.47$\pm$0.29 & 0.48$\pm$0.29 &   1.22 & 1.24 & 1.22 & 0.31 & 0.30 & 0.30 \\ \hline
$sin^{2}\theta^{PMNS}_{13}$ & 0.09$\pm$0.12 & 0.10$\pm$0.12 & 0.08$\pm$0.11 &  3.97 & 4.14 & 3.44 & 0.57 & 0.58 & 0.51 \\ [1ex]  \hline
\end{tabular}
}
\caption{$\chi^{2}$ best fit values of the observables for the $SU(5)$-inspired flavor symmetry based models defined in Eqs. \eqref{eq:100A}-\eqref{eq:100B} with fixed values of the model parameters given in Table \ref{table:20}. The best fit values shown in this table correspond to $\chi^{2}/n_{obs}=$ 0.73, 0.74 and 1.05  for $p =$ 2, 1 and 0 respectively. Here TMV$=$theoretical mean value, TSD$=$theoretical standard deviation, ECV$=$experimental central value and pull is defined in Eq. \eqref{eq:dis}. }\label{table:21}
\end{table}

\section{A Variant Monte Carlo Analysis of the \boldmath{$SU(5)$}-based Models}\label{SU5P}

The Monte Carlo analysis of Sec. \ref{SU5A} treats the random variables
as unbiased set with Gaussian distribution and investigates the likelihood of
these models to procreate the experimental values. The results presented in the
previous section show that, on average, the agreement of the theoretical
mean values with the experimental central values is very good, except for few
observables for which the theoretical mean values do not coincide with the
experimental central values but still the  experimental central values lie
within the range of values predicted by the theory. Since we have no control
over the random variables, the theoretical standard deviations of each
observables are quite large (as can be seen from  columns 4 and 5 of Table
\ref{table:4}) and of the same order as the theoretical mean values. In this
section, we present a modified version of the Monte Carlo analysis, where the
model  parameters, $\epsilon_{i}$ are not fixed but rather treated as
constrained random parameters. As before, we start with the set of uncorrelated
random variables having Gaussian distribution and analyze the class of models with Yuwaka coupling matrices given by Eqs. \eqref{U4}-\eqref{L4}. We consider a projection of
these distributions onto a subspace of the original space of random parameters
defined by the experimental constraints. These constraints create correlations
between the random parameters, and therefore their distributions in the
constrained subspace are in general different from the original (unconstrained)
distributions. We optimize the model parameters by minimizing the difference
between the complete set $\{\bm{r}\}$ of random parameters describing a given
class of models, and the subset $\{\bm{r^*}\}$ of random parameters describing
the models that satisfy the experimental constraints
$O_{i\; \rm{th}}=E_{i\;\rm{exp}}$, which we call the distortion and denote by
$D\left(\{\bm{r}^*\},\{\bm{r}\}\right)$.
The condition of optimization is then
\vspace*{-5pt}
\begin{equation}
\bm{\epsilon}_{\rm best} = \argmin_{\bm{\epsilon}}
D\left(\{\bm{r}^*\},\{\bm{r}\}\right).
\label{eq:besteps}
\end{equation}

\begin{table}[t]
\centering
\footnotesize
\resizebox{0.4\textwidth}{!}{
\begin{tabular}{|c|c|c|c|}\hline
Observables &TMV$\pm$TSD & $\frac{\rm{TMV}}{\rm{ECV}}$ & pull  \\\hline
$y_{u}/10^{-6}$ & 2.57$\pm$0.09 &1.00&0.00   \\ \hline
$y_{c}/10^{-3}$ & $1.40\pm$0.03 &1.02&0.39  \\ \hline
$y_{t}$ & 0.545$\pm$0.053 &1.02&0.25  \\ \hline
$y_{d}/10^{-4}$ & 0.39$\pm$0.04 &0.99&-0.05  \\ \hline
$y_{s}/10^{-3}$ & 0.75$\pm$0.03 &1.02&0.28  \\ \hline
$y_{b}/10^{-2}$ & 4.49$\pm$0.22 &0.99&-0.02 \\ \hline
$y_{e}/10^{-5}$ & 1.64$\pm$0.001 &1.00&0.18  \\ \hline
$y_{\mu}/10^{-3}$ & 3.46$\pm$0.002 &1.00&0.11   \\ \hline
$y_{\tau}/10^{-1}$ & 0.589$\pm$0.001 &0.99&-0.09  \\ \hline
$\lvert V_{us} \rvert$ & 0.225$\pm$0.0009 &0.99&-0.29  \\ \hline
$\lvert V_{cb} \rvert /10^{-2}$ & 3.75$\pm$0.017 &1.00&0.04 \\ \hline
$\lvert V_{ub} \rvert /10^{-3}$ & 3.24$\pm$0.03 &0.99&-0.01 \\ \hline
$\eta_{W}$ & 0.35$\pm$0.004 &1.00&0.00 \\ [1ex]  \hline
\end{tabular}
}
\caption{Best fit values of the observables for the $SU(5)$-based GUTs defined in Eqs. \eqref{U4}-\eqref{L4}  by employing the modified  Monte Carlo analysis. Here we have  considered the case with $\tan\beta =10$ as input.  As explained in the text, this results correspond to minimization of the function  $D=D(O,E)+D\left(\{\bm{r}^*\},\{\bm{r}\}\right)$. This fit corresponds to $D(O,E)/n_{obs}=0.03$. Here TMV$=$theoretical mean value, TSD$=$theoretical standard deviation, ECV$=$experimental central value and pull is defined in Eq. \eqref{eq:dis}.}\label{tab:n10rnd}
\end{table}

To implement the optimization procedure, we modify the $\chi^2$ minimization
approach described in the previous sections by introducing an additional step
which, starting from initial set of random parameters $\{\bm{r}_0\}$,
tries to update the current set of random parameters $\{\bm{r}\}$
by minimizing $D=D(O,E)+D\left(\{\bm{r}\},\{\bm{r}_0\}\right)$, where
\vspace*{-5pt}
\begin{equation}\label{DOE}
D(O,E)=\sum\left(
\frac{O_{i\; \rm{th}}-E_{i\; \rm{exp}}}{\sigma_{i\; \rm{exp}}}
\right)^2
\end{equation}
accounts for discrepancy between the model prediction and experiment, and
the measure of distortion is chosen to be
\vspace*{-5pt}
\begin{equation}
D\left(\{\bm{r}\},\{\bm{r}_0\}\right) =
\sum\frac{\left(C_{jk}-{\bf E}\left[C_{jk}\right]\right)^2}{{\bf E}\left[C_{jk}\right]},
\end{equation}

\noindent where $C_{jk}$ is the number of occurrences of the binned value of the expected
cumulative distribution function (cdf) of random variable $r_j$, and the sum is
taken over all cdf bins $k$ and all elements of all random matrices $j$ in the
model. The method we use is an iterative procedure that alternates the
$\chi^2$ minimization and $\{\bm{r}\}$ optimization steps. The best fit results of this
procedure obtained for the $SU(5)$-based GUTs  defined in Eqs. \eqref{U4}-\eqref{L4} is presented in Table \ref{tab:n10rnd}. Here we have  considered the case with $\tan\beta =10$ as input. The models parameters that are extracted from this procedure are given in Eq. \eqref{eps}.
\vspace*{-10pt}
\begin{align}\label{eps}
&\varepsilon_1 = 0.00106\pm 0.00001, \nonumber \\
&\varepsilon_2 = 0.08023\pm 0.00044,\\
&\varepsilon_4 = 0.03294\pm 0.00024. \nonumber
\end{align}

\begin{table}[th]
\centering
\footnotesize
\resizebox{0.45\textwidth}{!}{
\begin{tabular}{|c|c|c|c|}
\hline
Observables & TMV$\pm$TSD & $\frac{\rm{TMV}}{\rm{ECV}}$ & pull \\ [1ex]
\hline
$\frac{\Delta m^{2}_{sol}}{\Delta m^{2}_{atm}}$ &  0.031 $\pm$  0.0002 &  1.0 &  0.01 \\ \hline
$\sin^{2}\theta_{12}$  &  0.31 $\pm$  0.02 &  0.99 &  0.17 \\ \hline
$\sin^{2}\theta_{23}$       &  0.39 $\pm$  0.03 &  0.99 &  0.23 \\ \hline
$\sin^{2}\theta_{13}$  &  0.024 $\pm$  0.001 &  1.0 & 0.12 \\ \hline
\end{tabular}
}
\caption{
Best fit values of observables
using the modified approach of Monte Carlo analysis in the neutrino sector for $SU(5)$-based GUTs defined in Eqs. \eqref{U4}-\eqref{L4}. The best fit values shown in this table correspond to $\chi^{2}/n_{obs}=$ 0.1. Here TMV$=$theoretical mean value, TSD$=$theoretical standard deviation, ECV$=$experimental central value and pull is defined in Eq. \eqref{eq:dis}.
\label{tab:n10rndN}
}
\end{table}

\FloatBarrier
\begin{figure}[t!]
\centering
\includegraphics[scale=0.25]{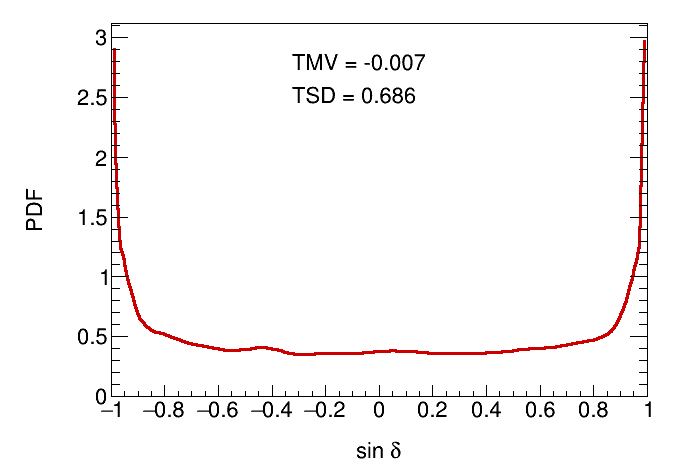}\\
\includegraphics[scale=0.25]{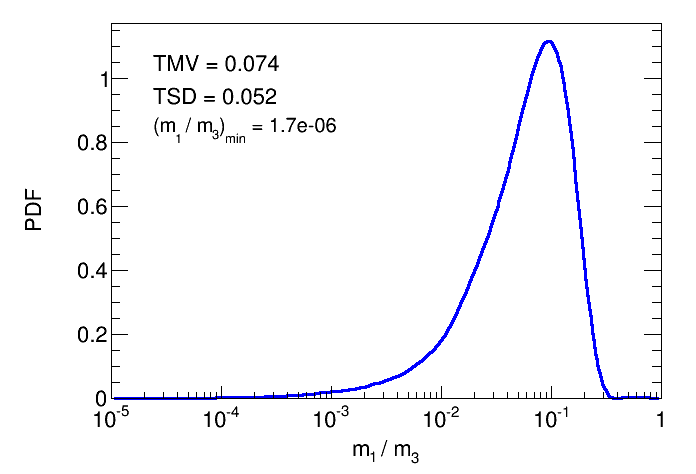}
\includegraphics[scale=0.25]{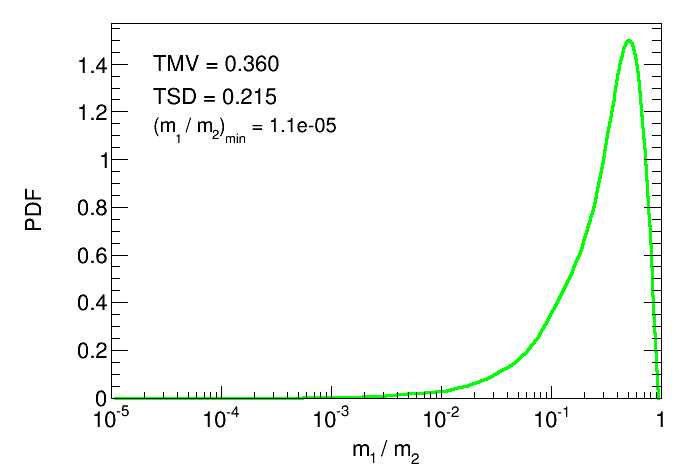}
\caption{Probability density plots of the  experimentally unmeasured quantities in the neutrino sector, the sine of the Dirac type phase (upper) and neutrino mass ratios $m_1/m_3$ (lower left) and $m_1/m_2$ (lower right) by employing the modified Monte Carlo analysis  for $SU(5)$-based GUTs defined in Eqs. \eqref{U4}-\eqref{L4}. }\label{fig:006NN}
\end{figure}

The best fit values presented in Table \ref{tab:n10rnd} corresponds to $D(O,E)=0.43$. In this modified approach, all the theoretically predicted values of the observables almost coincide with the experimental measured values. Compared to the approach explained in the previous sections, theoretical errors are greatly reduced and comparable to the experimental uncertainties. Histogram distributions of the observables  in the charged fermion sector corresponding to this result are presented in Fig. \ref{fig:004c} in Appendix \ref{App:B01} and the distributions of the restricted set $\{\bm{r^*}\}$ are shown in Figs. \ref{fig:005A3}, \ref{fig:005B3} and \ref{fig:005C3} in Appendix \ref{App:B02} for the matrices $Y^{0}_U$, $Y^{0}_D$ and $Y^{0}_L$ respectively.  We also employ this approach in the neutrino sector Eq. \eqref{nu} separately, where the model parameters $\epsilon_{i}$ are absent. The results are presents in Table \ref{tab:n10rndN} that correspond to $D(O,E)=0.1$. The histogram distributions of the theoretical predictions of these quantities in the neutrino sector are shown in Fig. \ref{fig:004N} in Appendix \ref{App:B03} and the modified set $\{\bm{r^*}\}$ in Figs. \ref{fig:005D3} and \ref{fig:005E3} in Appendix \ref{App:B04}. The $\sin \delta$ and the two neutrino mass ratios $m_1/m_3$ and $m_1/m_2$ are shown in Fig. \ref{fig:006NN}. This variant of the Monte Carlo analysis   shows that with the subspace $\{\bm{r^*}\}$ which does not have much deviation from the original landscape $\bm{r}$,  excellent agreement of the observables  to the experimental measured values can be achieved. One can in principle apply this modified approach to the special cases of the $SU(5)$-based GUTs explained in Sec. \ref{SU5F} but we do not include those analysis here.

\section{Conclusion}
\label{section:CC}
In this paper we have extended the idea of anarchy from the neutrino sector to the quark and charged lepton sectors. This is made possible in the context of $SU(5)$ unified theories where the $10_i$ fermions mix with vector-like $10_\alpha+\overline{10}_\alpha$ fermions having GUT scale masses.  While all the Yukawa couplings in these models are of order one, these mixings provide three hierarchical parameters which explain all the hierarchies in the charged fermion masses and quark mixing angles.  The neutrino sector is immune to such mixings, and remain anarchical.
We have also studied special cases of this general $SU(5)$ setup with smaller number of input parameters -- either 2 or 1 -- by introducing a flavor $U(1)$ symmetry that distinguishes the three families of $10_i$ fermions.

We have presented detailed quantitative analysis of these models following a probabilistic approach. The Yukawa couplings of the model are assumed to be uncorrelated random variables obeying Gaussian distributions.  Our Monte Carlo analysis shows that the combined anarchy-hierarchy scenario gives very good fit to all the fermion masses and mixings.   We have also presented a variant Monte Carlo method where the model parameters are not kept fixed but have certain distributions constrained by the phenomenological considerations.  This approach is proposed to systematically explore the subspace of the original  Gaussian landscape that becomes consistent with all experimental constraints with greater accuracy. A figure of merit in this approach is the distortion of the distributions compared to the original Gaussian distributions. The framework is found to provide a good quality fit.

The theoretical distributions of the observables in the charged fermion sector remain roughly the same for the various  models studied here.
There is one important difference in  the neutrino mixing parameters in the flavor $U(1)$ model that distinguishes the $\overline{5}_1$ from
$\overline{5}_{2,3}$ fields:  The mixing parameter $\sin\theta_{13}$ comes out to be somewhat smaller than $\sin\theta_{23}$.
Anarchy prefers normal ordering of neutrino mass spectrum with a mild hierarchy in the masses.
A comparison of the two experimentally unmeasured quantities in the neutrino sector, the mass ratio $m_{1}/m_{2}$ and the CP-violating parameter $\sin \delta$ predicted by our statistical analysis for the two different sets of models studied here is presented in Table \ref{table:NN}.

\begin{table}
\centering
\begin{tabular}{ |c|c|c|c|} \hline
\multicolumn{2}{|c|}{Quantity} & Structureless Neutrino Matrix & Hierarchical Neutrino Matrix \\ \cline{1-4} \hline \hline
\multirow{3}{*}{$m_{1}/m_{2}$}  & $\leq$0.01 & 4.24$\%$ & 20.38$\%$ \\\cline{2-4} & $\leq$0.1 & 33.77$\%$ & 74.57$\%$ \\\cline{2-4} & $\leq$0.2 & 56.23$\%$ & 88.33$\%$ \\\cline{2-4}  \hline   \hline
\multirow{5}{*}{$\sin\delta$} & [0,0.25] & 8.15$\%$ & 8.9$\%$  \\\cline{2-4}
 & (0.25,0.5] & 8.79$\%$ & 9.82$\%$  \\\cline{2-4}
 & (0.5,0.75] & 9.68$\%$ & 10.16$\%$  \\\cline{2-4}
 & (0.75,1.0] & 23.87$\%$ & 21.18$\%$  \\\cline{2-4} \hline
\end{tabular}
\caption{Comparison of probabilities of the two unmeasured quantities  in the neutrino sector for the $SU(5)$-based GUTs with different neutrino mass matrix structures. For the quantity $\sin\delta$, these probabilities in the negative side remain roughly the same in the separate domains as for the positive side. Square bracket represents the end points are included in the set whereas for the round bracket the end points are not included.  }\label{table:NN}
\end{table}

\section*{Acknowledgments}
K.S.B. and S.S. would like to thank the organizers of CETUP* 2015 for hospitality and partial support during the 2015 Summer Program at Lead, South Dakota where part of the work was done. They would also like to thank the  participants of CETUP* 2015 for helpful discussions and comments. This work has been supported in part by the U.S. Department of Energy Grant No. No. de-sc0016013. Part of the numerical calculations was performed using the High Performance Computing Center at Oklahoma State University (NSF grant no. OCI-1126330).

\begin{appendices}
\appendixpageoff

\section{\large Distributions of the observables in the charged fermion sector for the \boldmath{$SU(5)$}-inspired \boldmath{$U(1)$} flavor models}
\label{App:A}

\subsection{\normalsize Models with two parameters} \label{App:A01}
\begin{figure}[th!]
\centering
\includegraphics[scale=0.3]{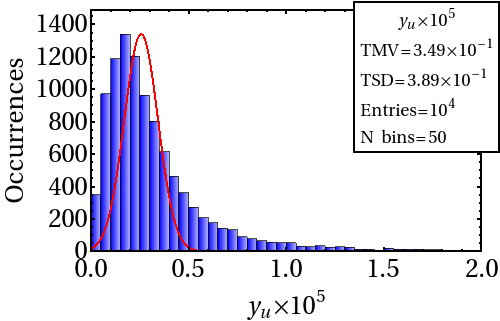}
\includegraphics[scale=0.3]{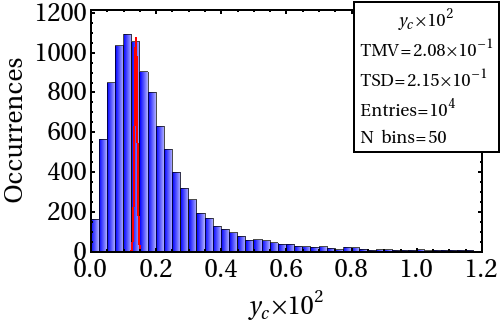}
\includegraphics[scale=0.3]{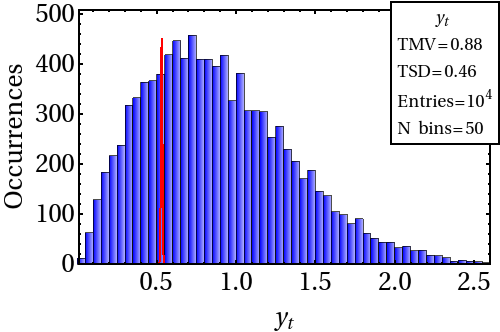}
\includegraphics[scale=0.3]{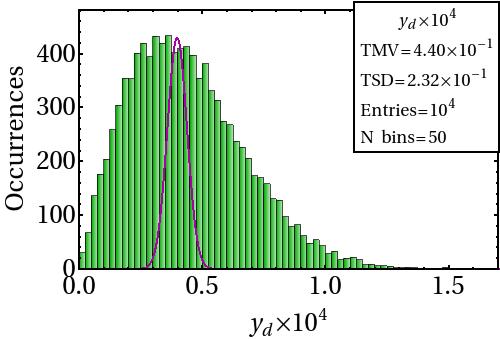}
\includegraphics[scale=0.3]{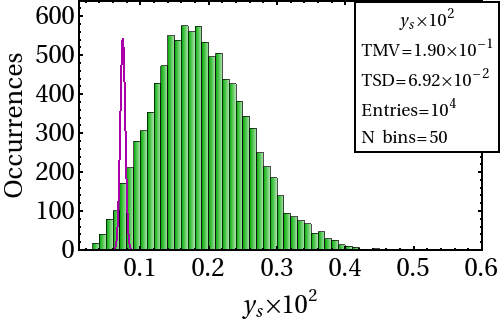}
\includegraphics[scale=0.3]{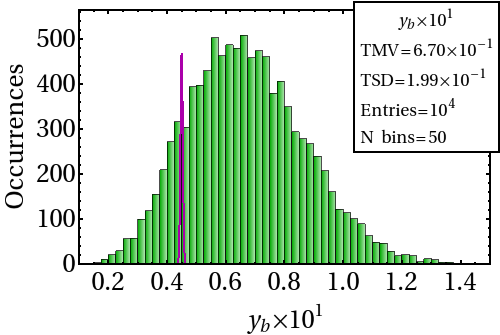}
\includegraphics[scale=0.3]{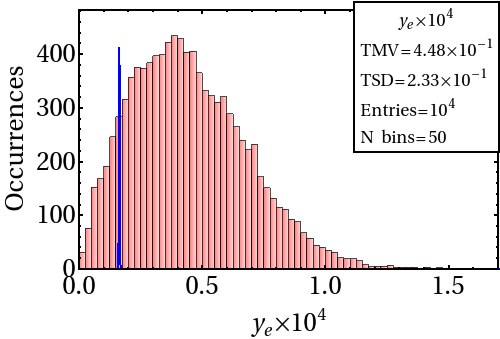}
\includegraphics[scale=0.3]{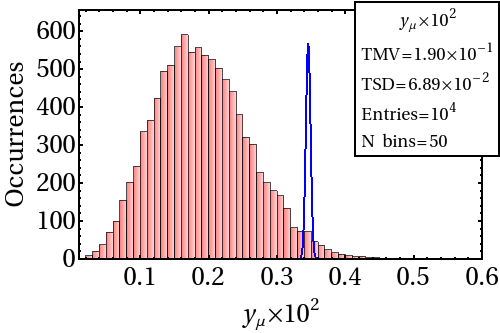}
\includegraphics[scale=0.3]{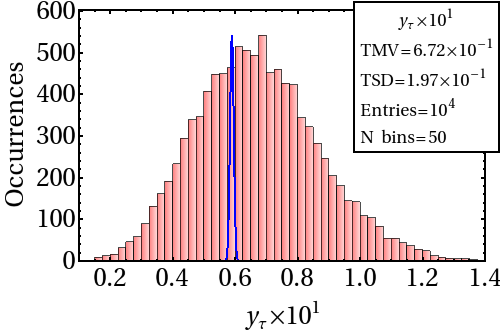}
\includegraphics[scale=0.3]{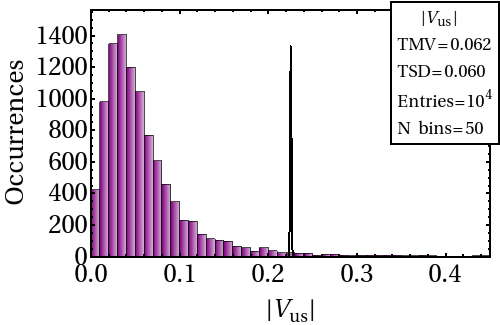}
\includegraphics[scale=0.3]{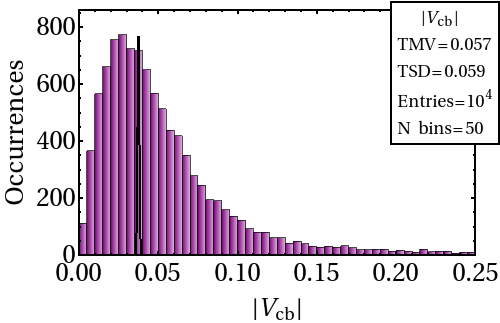}
\includegraphics[scale=0.3]{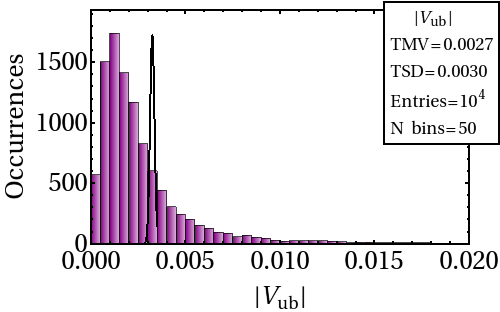}
\includegraphics[scale=0.3]{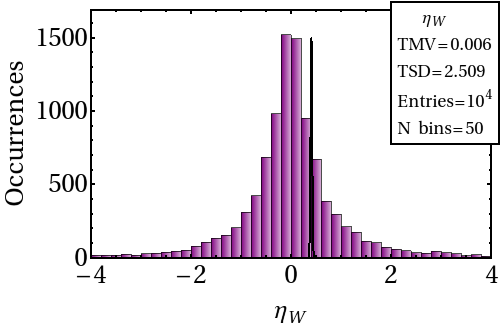}
\caption{ Histograms showing the theoretical distributions of the observables in the charged fermion sector in the $SU(5)$-inspired  $U(1)$ flavor symmetric models with the charge assignment $\{q_1=1,q_2=0,p=0\}$ defined by Eqs. \eqref{U4}-\eqref{L4} and \eqref{HH}  ($\tan\beta=10$).  The color code is the same as in Fig. \ref{fig:001}. }\label{two_para_fig}
\end{figure}

Here we present the theoretical distributions of the observables in the charged fermion sector  for the $SU(5)$-inspired $U(1)$ flavor symmetry models with the charge assignment $\{q_1=1,q_2=0,p=0\}$ defined by Eqs. \eqref{U4}-\eqref{L4} and \eqref{HH} and with two parameters $\{\epsilon,\,\epsilon_4\}$. These are shown in in Fig. \ref{two_para_fig}.

\FloatBarrier

\newpage
\subsection{\normalsize Models with single parameter} \label{App:A02}
\begin{figure}[t!]
\centering
\includegraphics[scale=0.3]{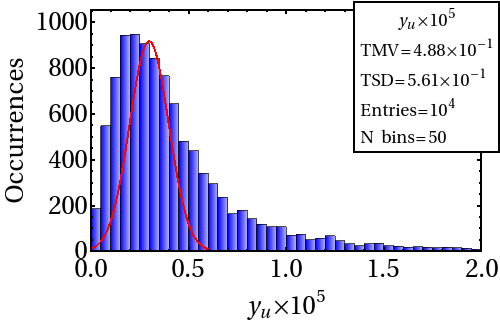}
\includegraphics[scale=0.3]{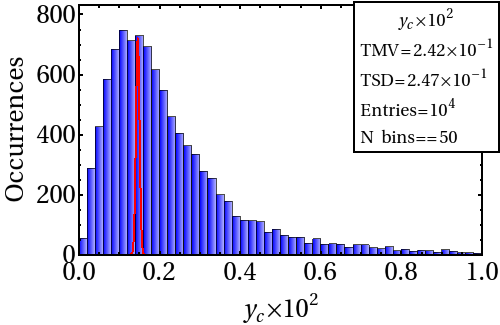}
\includegraphics[scale=0.3]{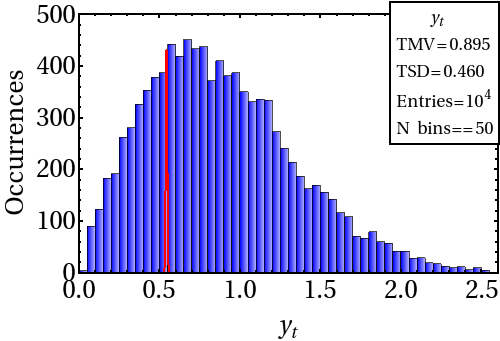}
\includegraphics[scale=0.3]{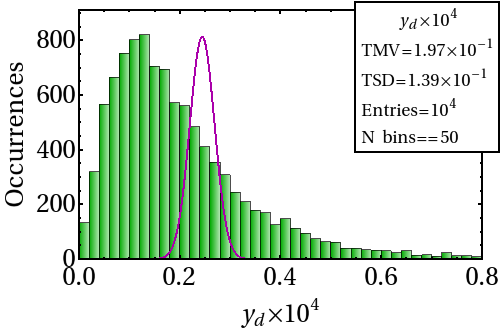}
\includegraphics[scale=0.3]{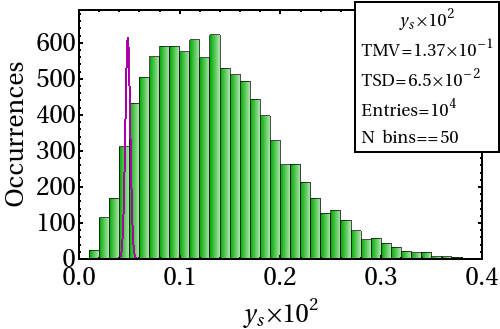}
\includegraphics[scale=0.3]{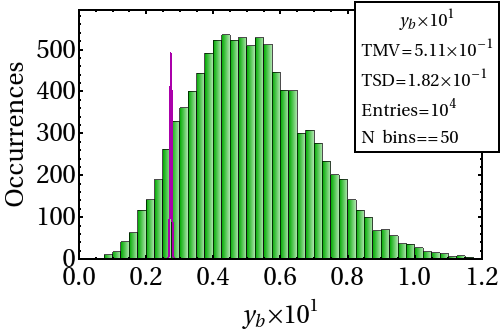}
\includegraphics[scale=0.3]{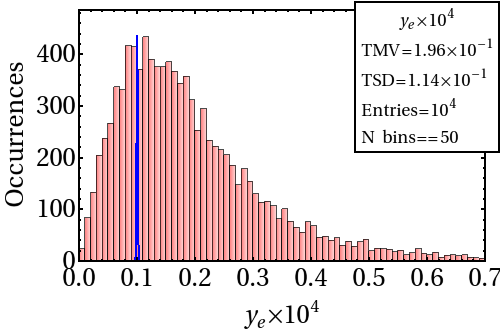}
\includegraphics[scale=0.3]{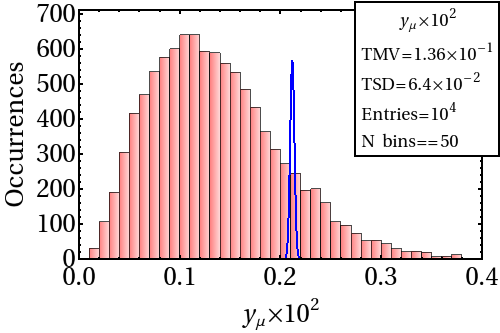}
\includegraphics[scale=0.3]{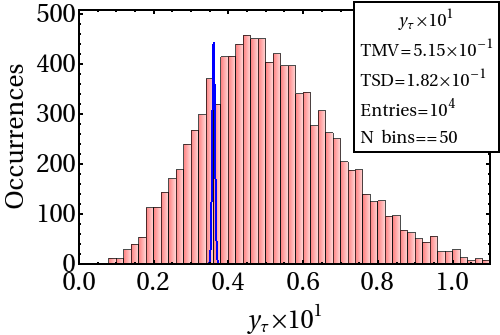}
\includegraphics[scale=0.3]{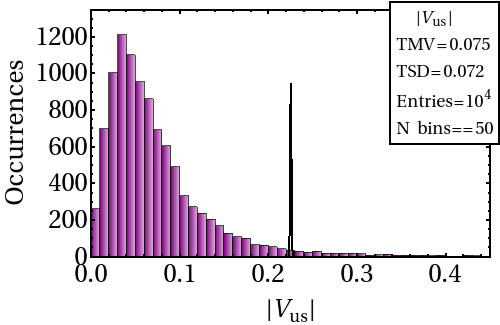}
\includegraphics[scale=0.3]{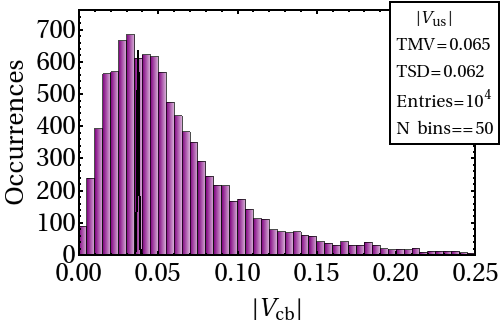}
\includegraphics[scale=0.3]{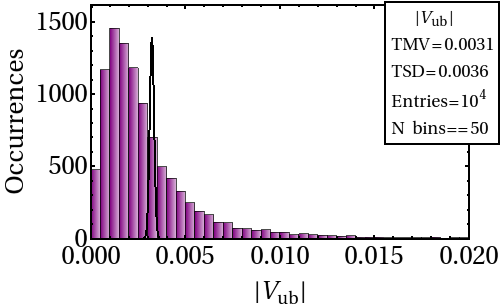}
\includegraphics[scale=0.3]{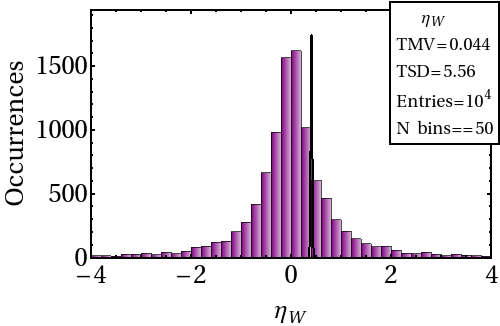}
\caption{Histograms showing the theoretical distributions of the observables in the charged fermion sector according to the $SU(5)$-inspired $U(1)$ flavor symmetry based  models with the charge assignment $\{q_1=2,q_2=1,p=2\}$ defined by Eqs. \eqref{eq:100A}-\eqref{eq:100B}  ($\tan\beta=5$).  The color code is the same as in Fig. \ref{fig:001}. }\label{fig:006}
\end{figure}

Here we present the theoretical distributions of the observables in the charged fermion sector for the $SU(5)$-inspired $U(1)$ flavor symmetry models with the charge assignment $\{q_1=2,q_2=1,p=2\}$ defined by Eqs. \eqref{eq:100A}-\eqref{eq:100B} and with a single parameter $\{\epsilon\}$.  The results are shown in Fig. \ref{fig:006}.

\FloatBarrier

\newpage
\section{\large Distributions of the observables and random entries resulting from the modified Monte Carlo analysis} \label{App:B}

\subsection{\normalsize Distributions of the observables resulting from the subset obtained by the modified Monte Carlo analysis in the charged fermion sector} \label{App:B01}

\begin{figure}[th!]
\centering
\includegraphics[width=0.32\linewidth]{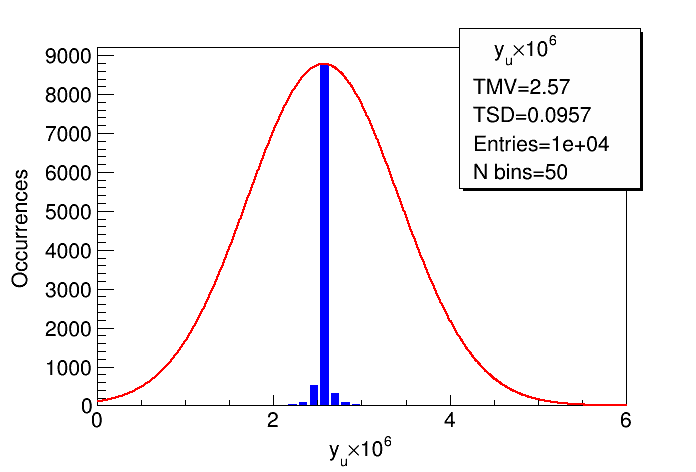}
\includegraphics[width=0.32\linewidth]{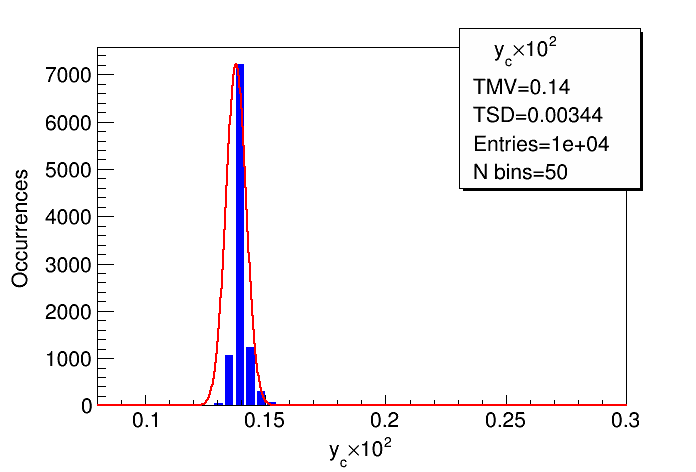}
\includegraphics[width=0.32\linewidth]{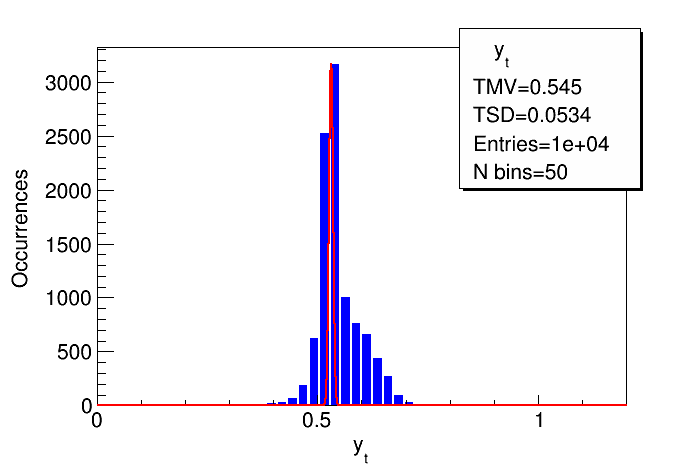}
\includegraphics[width=0.32\linewidth]{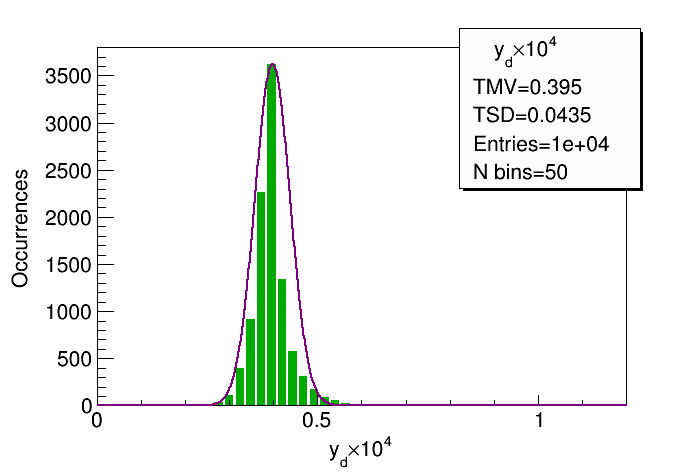}
\includegraphics[width=0.32\linewidth]{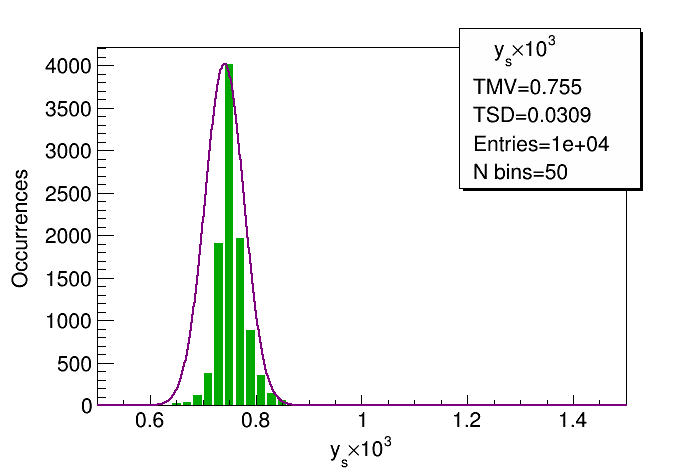}
\includegraphics[width=0.32\linewidth]{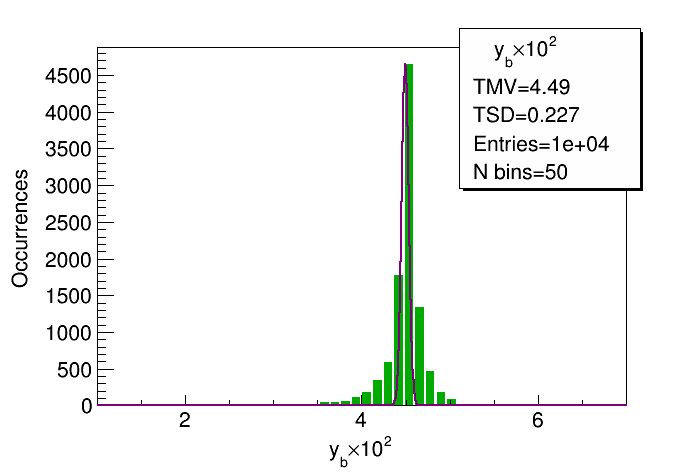}
\includegraphics[width=0.32\linewidth]{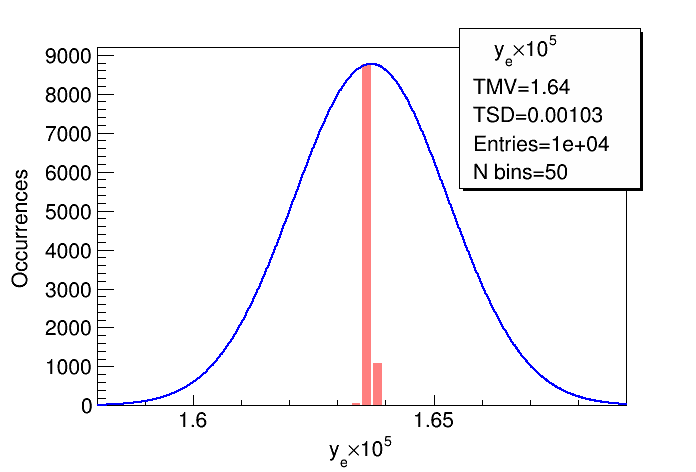}
\includegraphics[width=0.32\linewidth]{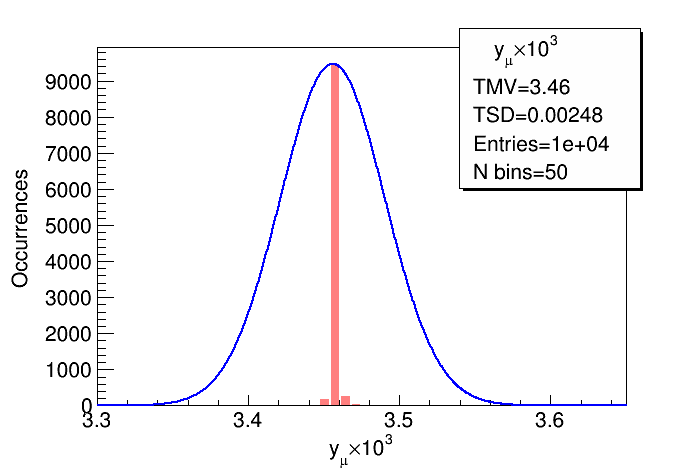}
\includegraphics[width=0.32\linewidth]{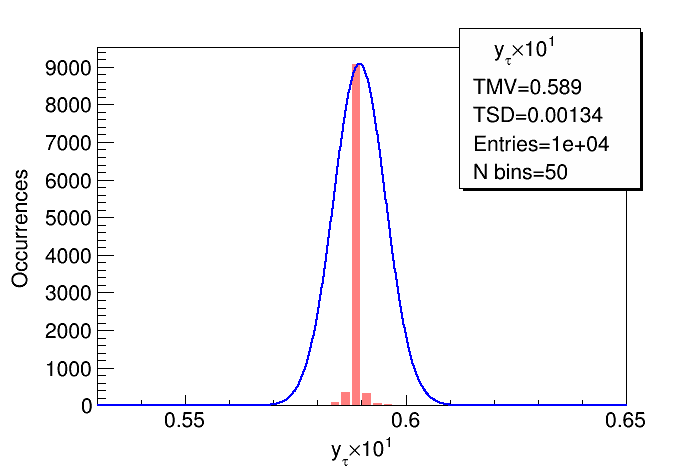}
\includegraphics[width=0.32\linewidth]{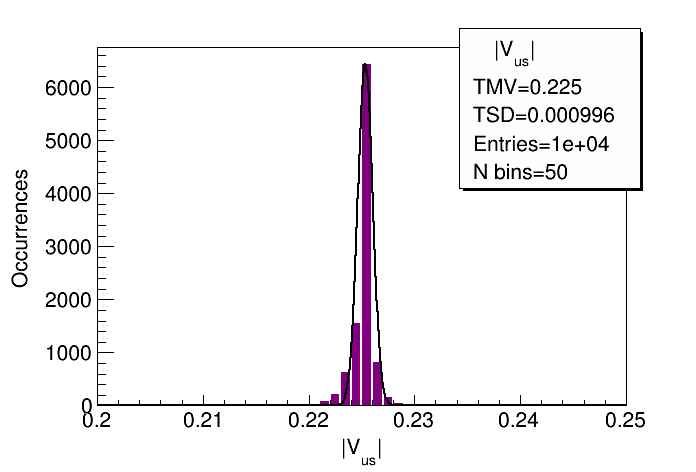}
\includegraphics[width=0.32\linewidth]{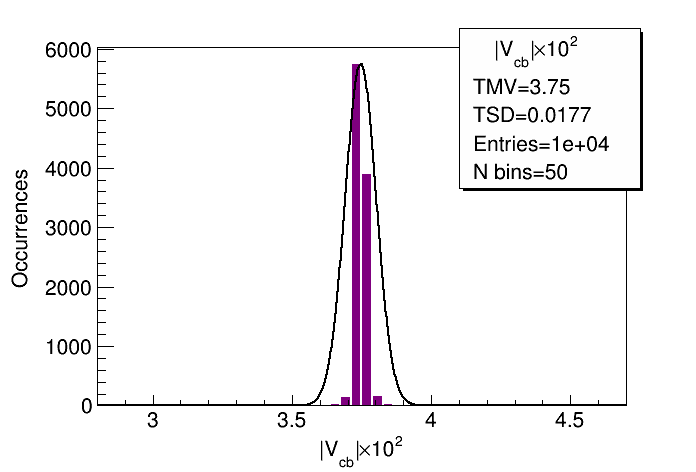}
\includegraphics[width=0.32\linewidth]{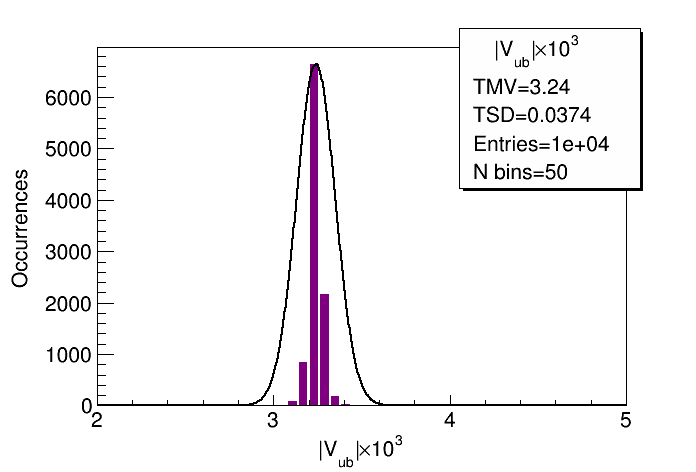}
\includegraphics[width=0.32\linewidth]{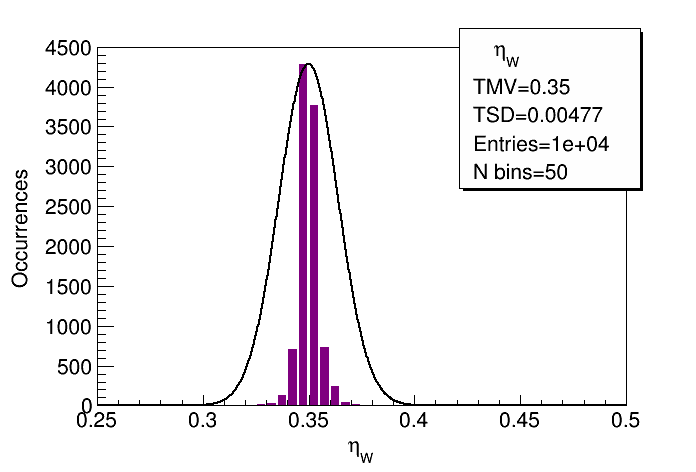}
\caption{Histogram distributions of the observables in the charged fermion sector according to the modified Monte Carlo  method for $SU(5)$-based GUTs defined in Eqs. \eqref{U4}-\eqref{L4} with $\tan\beta=10$. Color code is the same as Fig. \ref{fig:001}. Note the change of scales compared to Fig. \ref{fig:001} for few of the plots ($y_{u}\times 10^{5} \rightarrow y_{u}\times 10^{6}$,  $y_{s}\times 10^{2} \rightarrow y_{s}\times 10^{3}$,  $y_{e}\times 10^{4} \rightarrow y_{e}\times 10^{5}$, $y_{\mu}\times 10^{2} \rightarrow y_{\mu}\times 10^{3}$, $|V_{cb}| \rightarrow |V_{cb}|\times 10^{2}$, $|V_{ub}| \rightarrow |V_{ub}|\times 10^{3}$). }\label{fig:004c}
\end{figure}

Here we present the distributions  of the observables in Fig. \ref{fig:004c} in the  charged fermion sector   that resulted from $D=D(O,E)+D\left(\{\bm{r}^*\},\{\bm{r}\}\right)$  minimization procedure following the modified  Monte Carlo analysis as explained in Sec. \ref{SU5P} for the $SU(5)$-based GUTs defined in Eqs. \eqref{U4}-\eqref{L4}. The histogram plots of the observables in Fig. \ref{fig:004c} show excellent agreement with the observation. All these quantities are reproduced roughly within their 1$\sigma$ range even though the random matrices remain mostly random with only slight distortions. The modified random entries that predict these distributions of observables are shown in Figs. \ref{fig:005A3}, \ref{fig:005B3} and \ref{fig:005C3}.

\FloatBarrier

\newpage
\subsection{\normalsize Distributions of the projected random entries resulting from the modified Monte Carlo analysis in the charged fermion sector} \label{App:B02}

\begin{figure}[th!]
\centering
\includegraphics[width=0.3\linewidth]{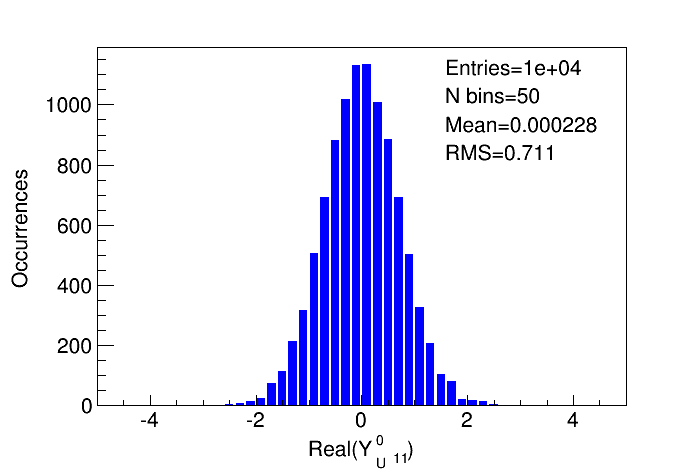}
\includegraphics[width=0.3\linewidth]{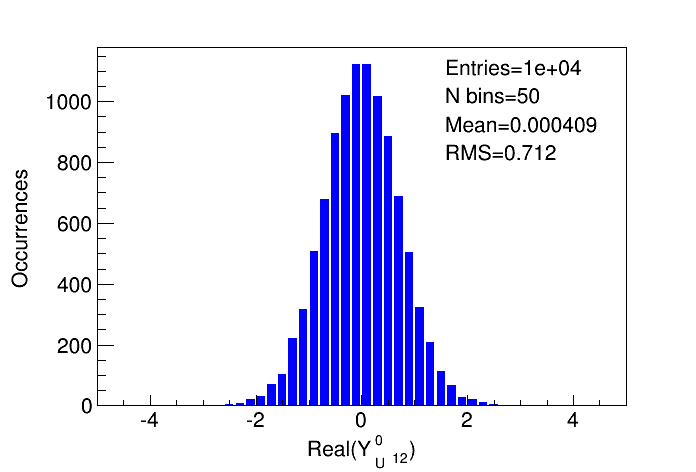}
\includegraphics[width=0.3\linewidth]{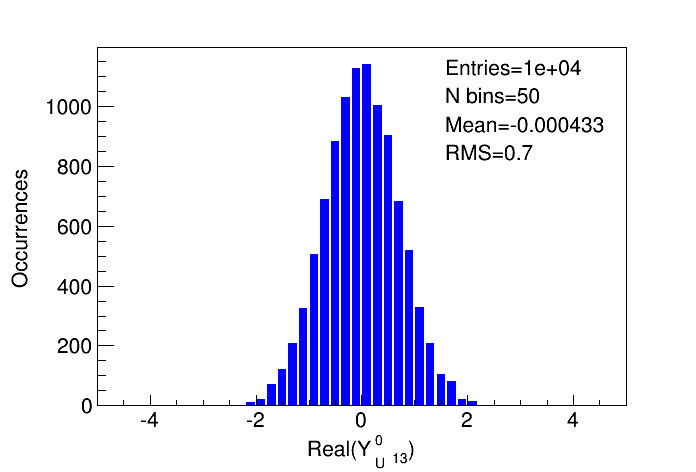}
\includegraphics[width=0.3\linewidth]{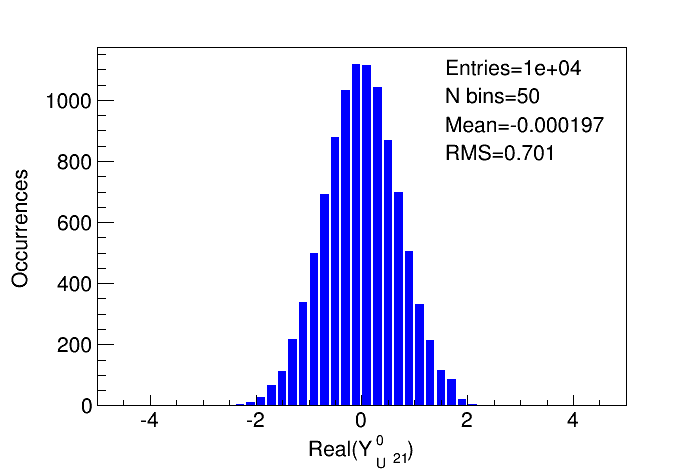}
\includegraphics[width=0.3\linewidth]{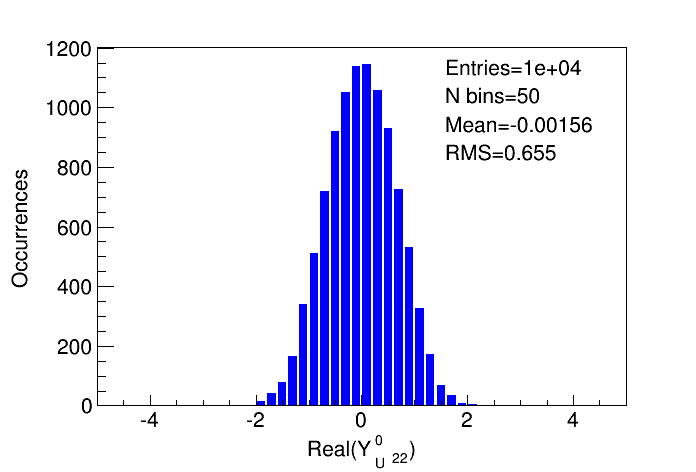}
\includegraphics[width=0.3\linewidth]{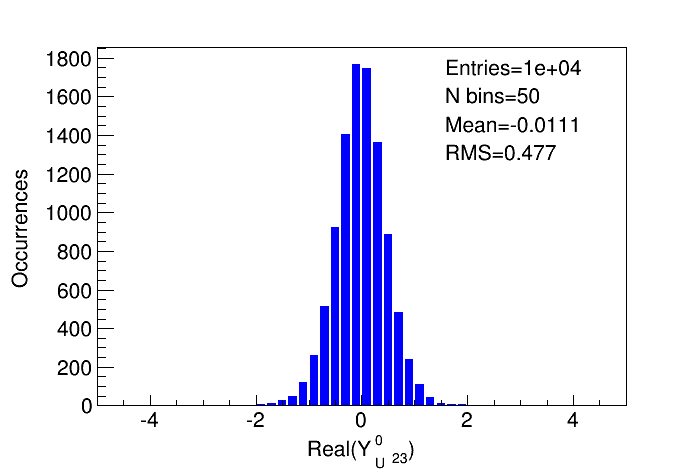}
\includegraphics[width=0.3\linewidth]{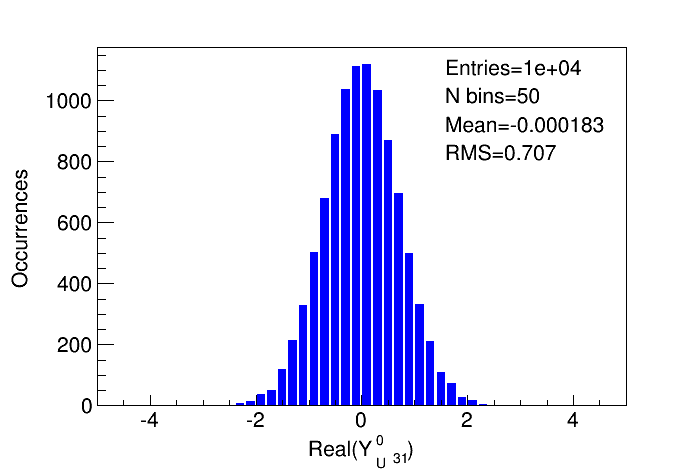}
\includegraphics[width=0.3\linewidth]{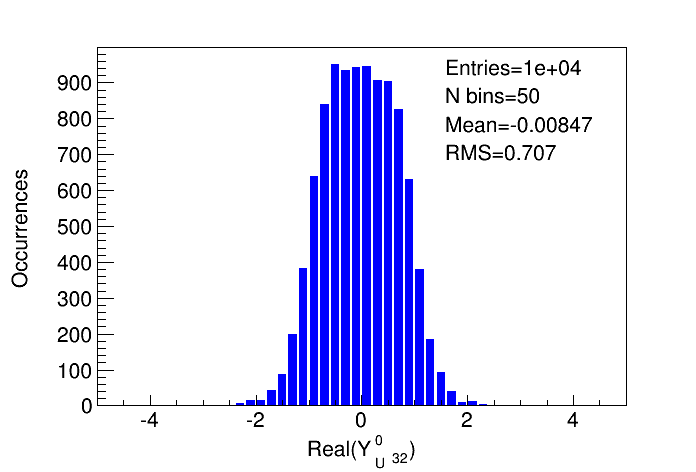}
\includegraphics[width=0.3\linewidth]{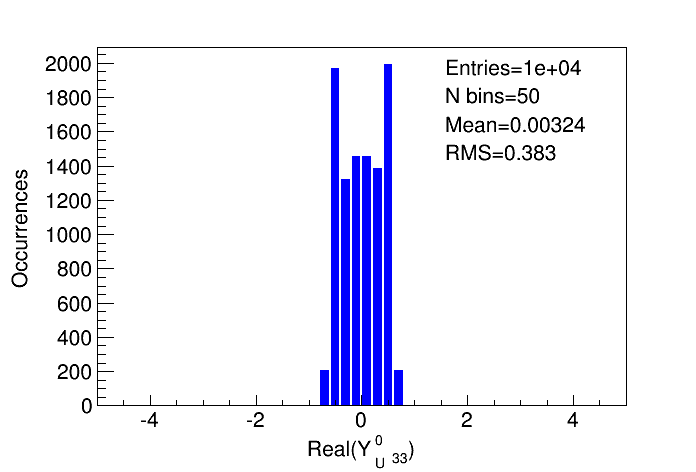}
\includegraphics[width=0.3\linewidth]{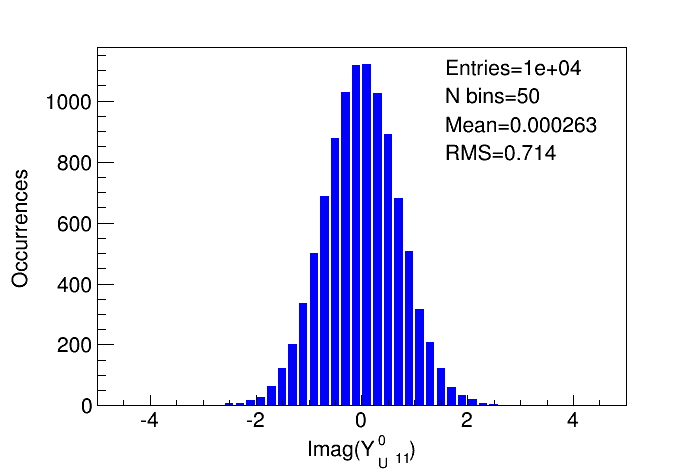}
\includegraphics[width=0.3\linewidth]{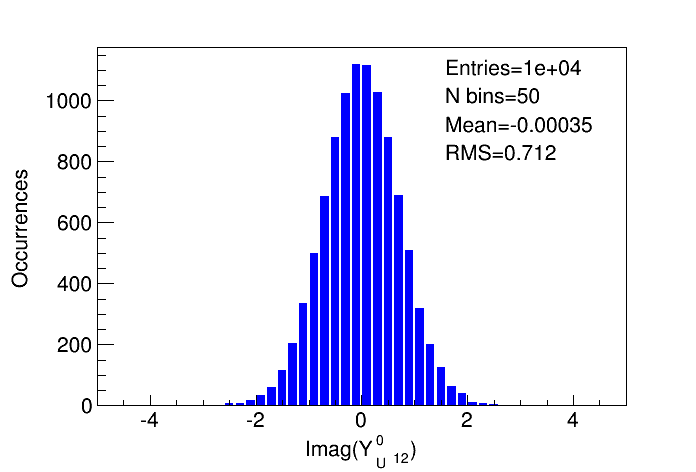}
\includegraphics[width=0.3\linewidth]{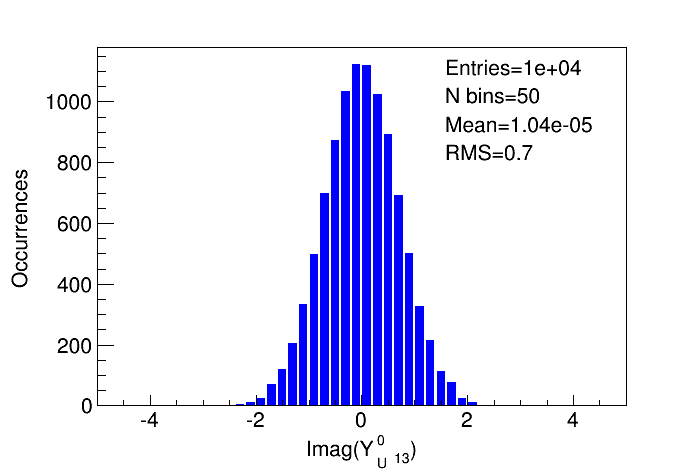}
\includegraphics[width=0.3\linewidth]{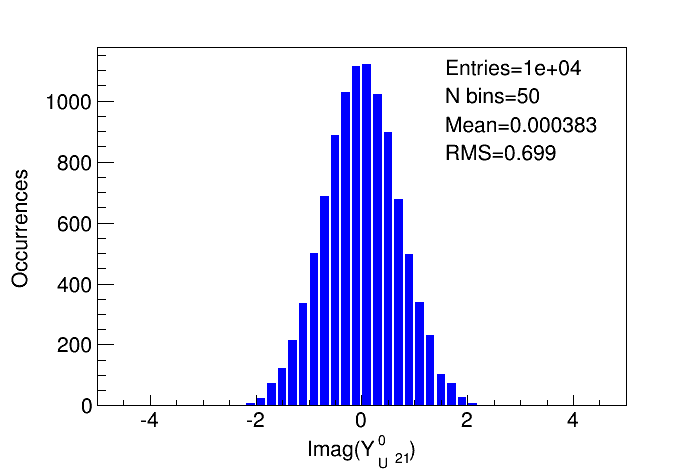}
\includegraphics[width=0.3\linewidth]{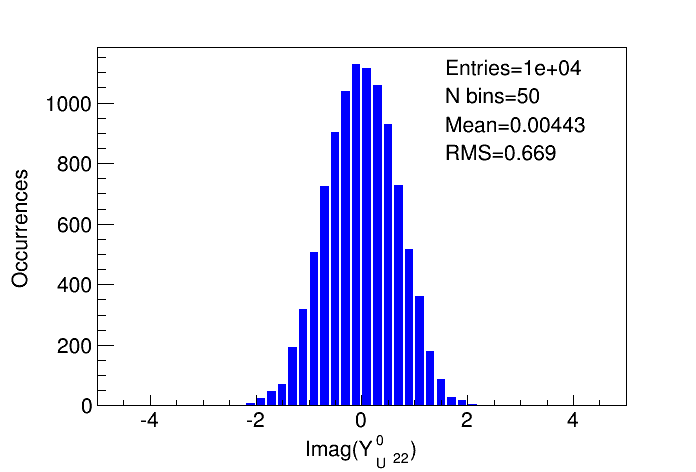}
\includegraphics[width=0.3\linewidth]{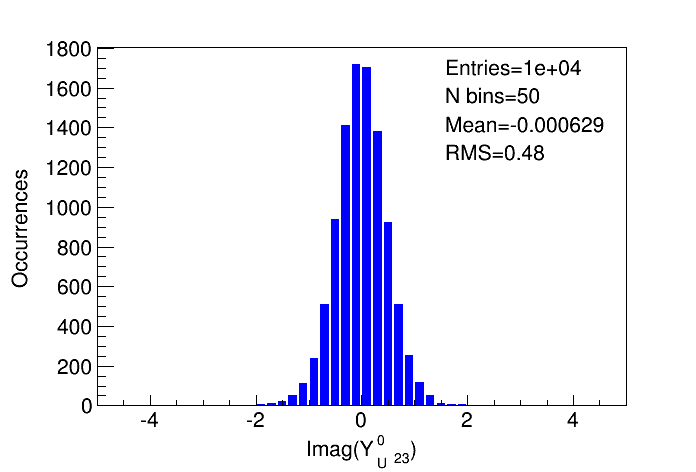}
\includegraphics[width=0.3\linewidth]{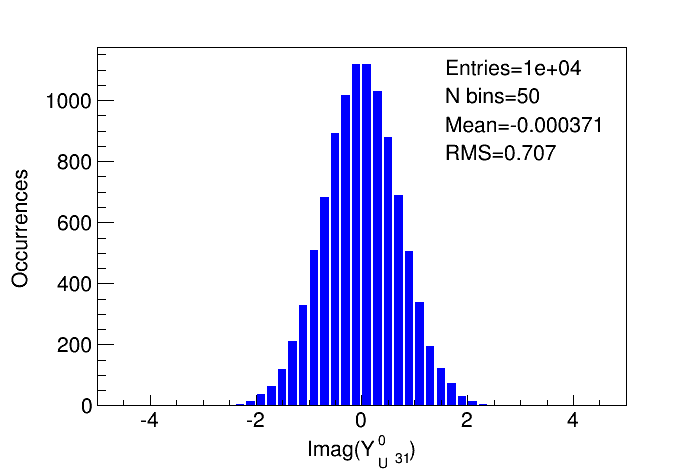}
\includegraphics[width=0.3\linewidth]{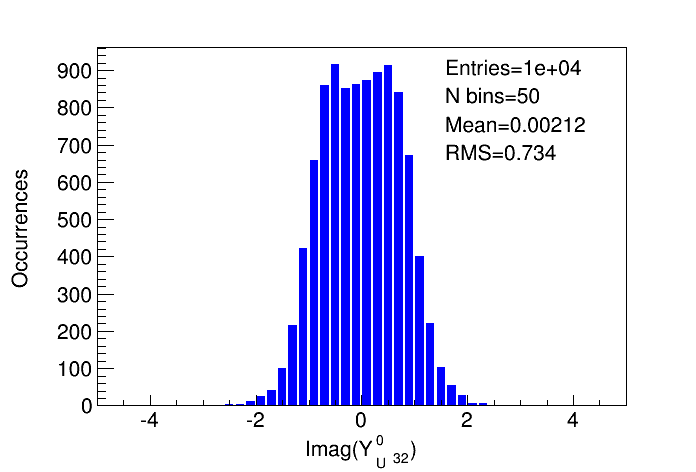}
\includegraphics[width=0.3\linewidth]{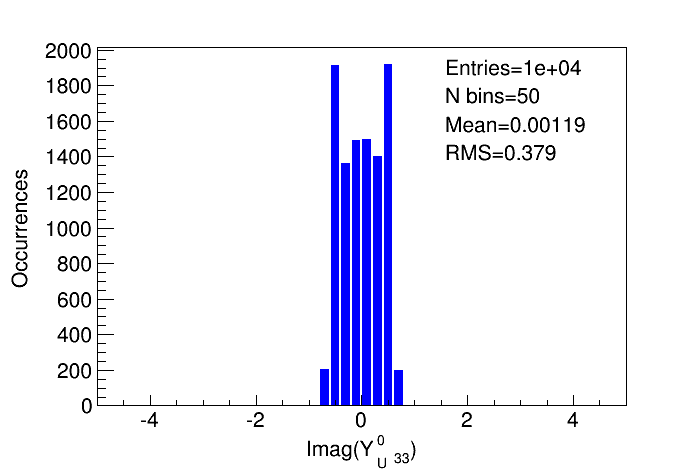}
\caption{Distributions of the $O(1)$ random entries in the matrix $Y^{0}_U$ from the modified Monte Carlo analysis that produce the observables in Fig.\ref{fig:004c} for $\tan\beta=10$. The first nine of the plots are for the real parts and the next nine for imaginary parts of the matrix, $Y^{0}_U$. For all these plots sample size and number of bins are taken to be $10^{4}$ and 50 respectively.  }\label{fig:005A3}
\end{figure}

\begin{figure}[th!]
\centering
\includegraphics[width=0.3\linewidth]{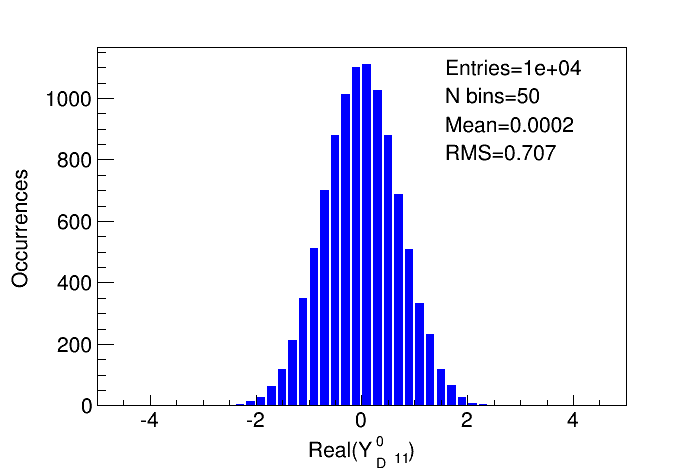}
\includegraphics[width=0.3\linewidth]{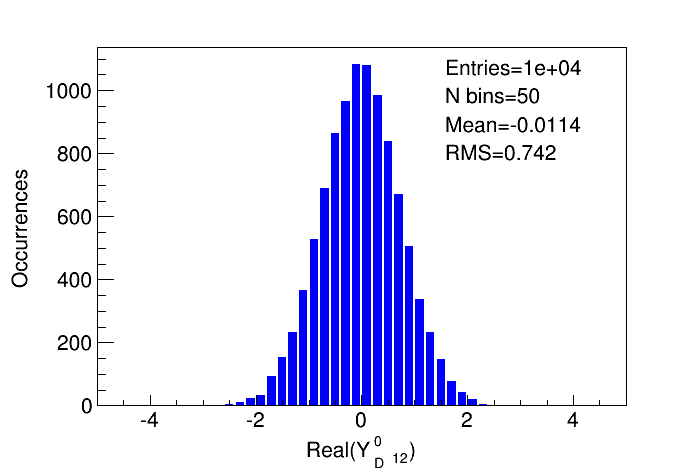}
\includegraphics[width=0.3\linewidth]{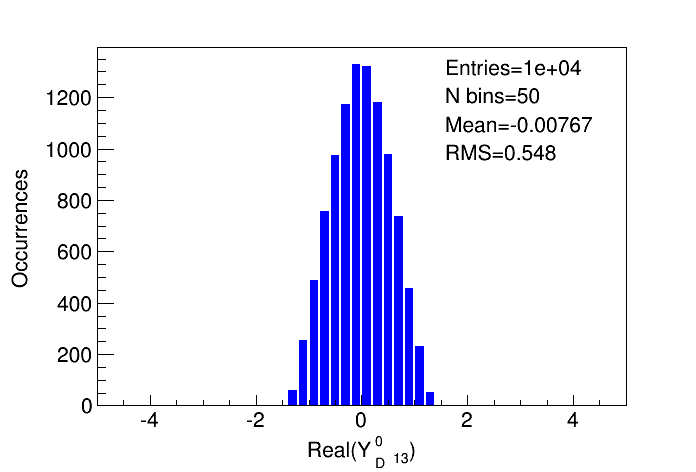}
\includegraphics[width=0.3\linewidth]{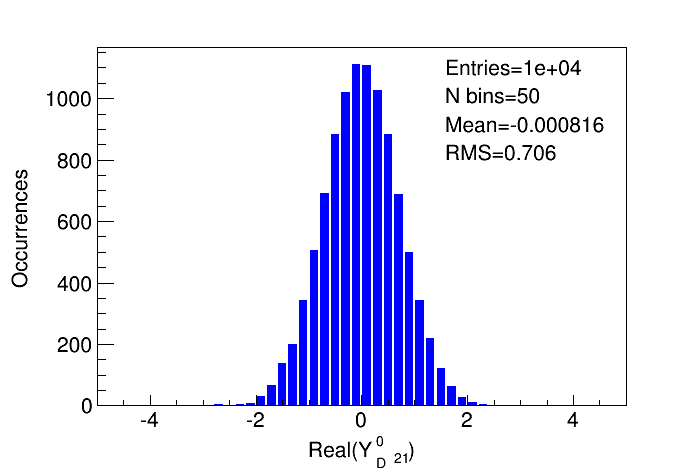}
\includegraphics[width=0.3\linewidth]{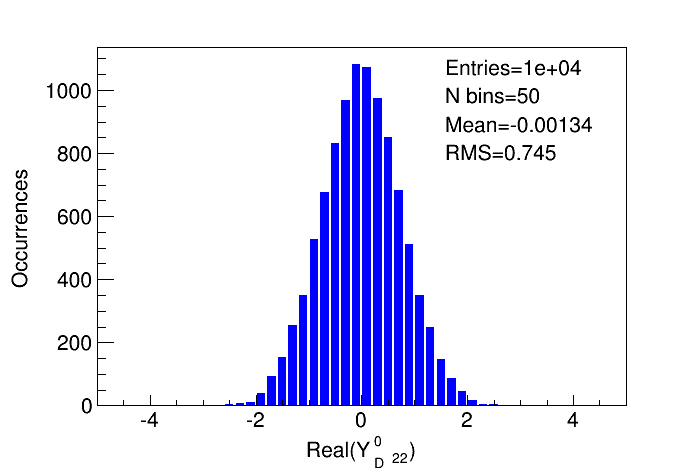}
\includegraphics[width=0.3\linewidth]{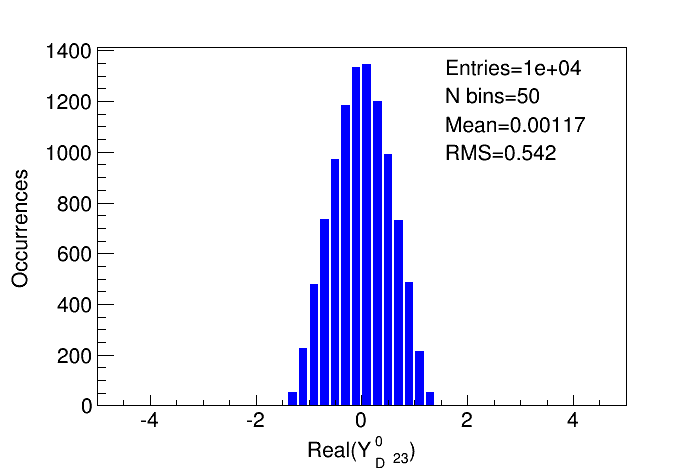}
\includegraphics[width=0.3\linewidth]{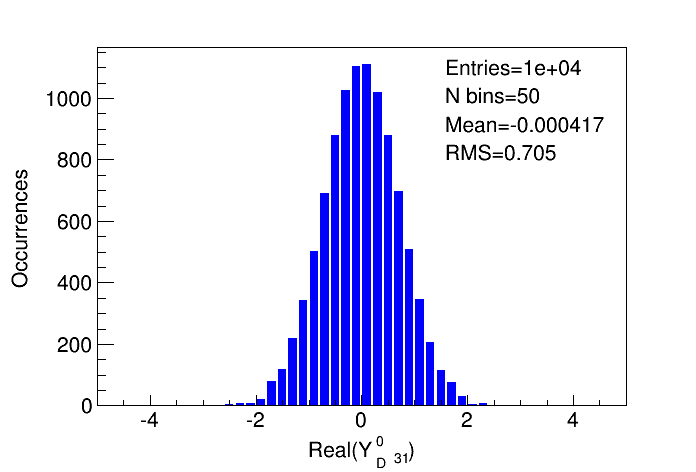}
\includegraphics[width=0.3\linewidth]{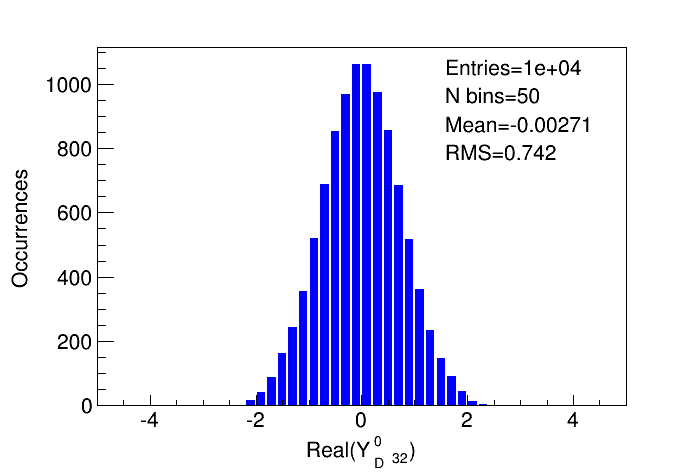}
\includegraphics[width=0.3\linewidth]{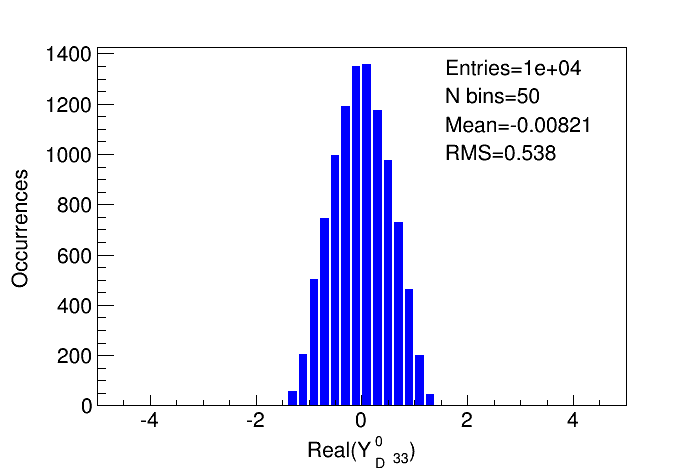}
\includegraphics[width=0.3\linewidth]{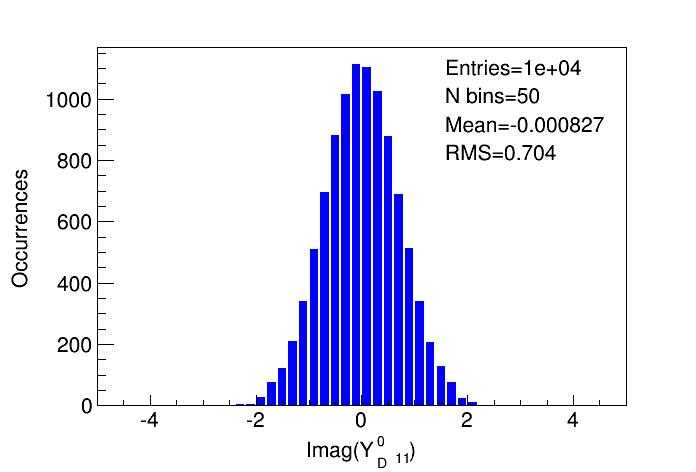}
\includegraphics[width=0.3\linewidth]{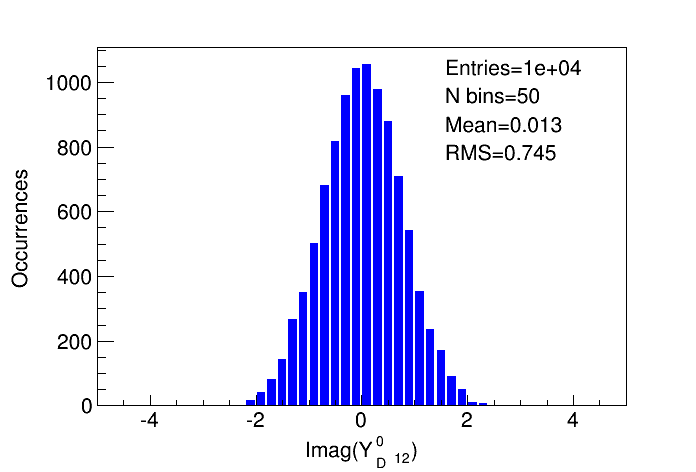}
\includegraphics[width=0.3\linewidth]{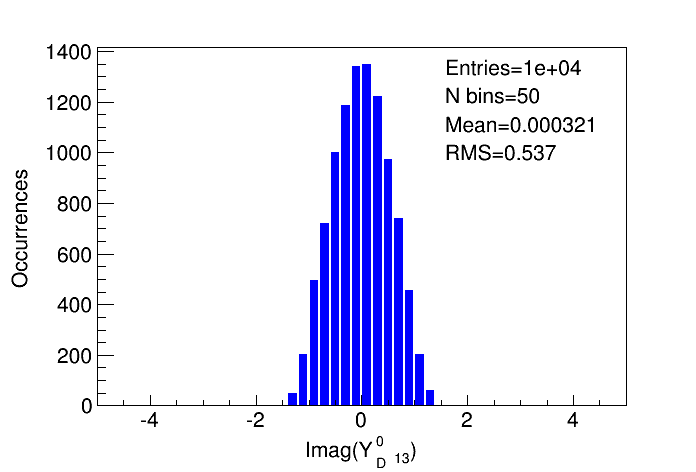}
\includegraphics[width=0.3\linewidth]{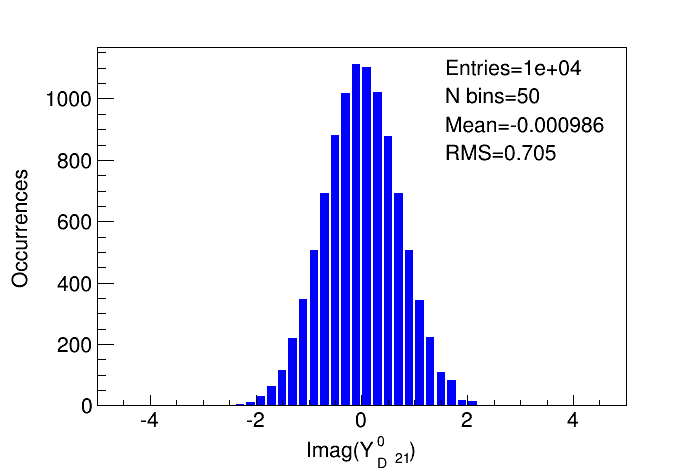}
\includegraphics[width=0.3\linewidth]{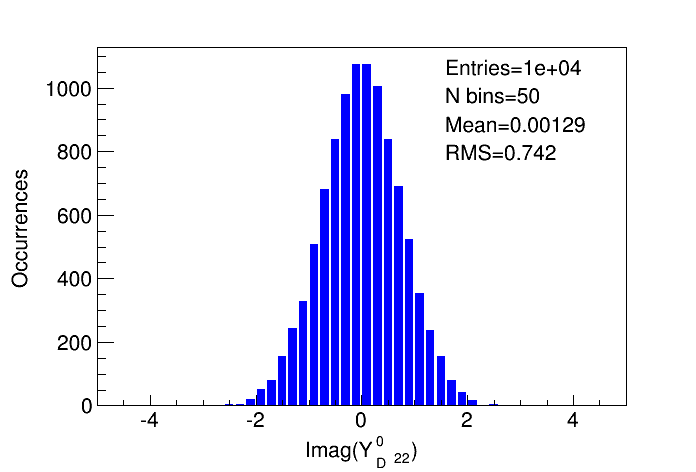}
\includegraphics[width=0.3\linewidth]{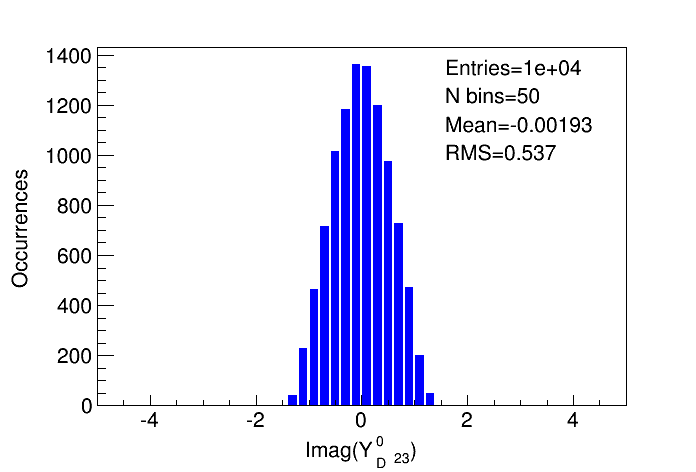}
\includegraphics[width=0.3\linewidth]{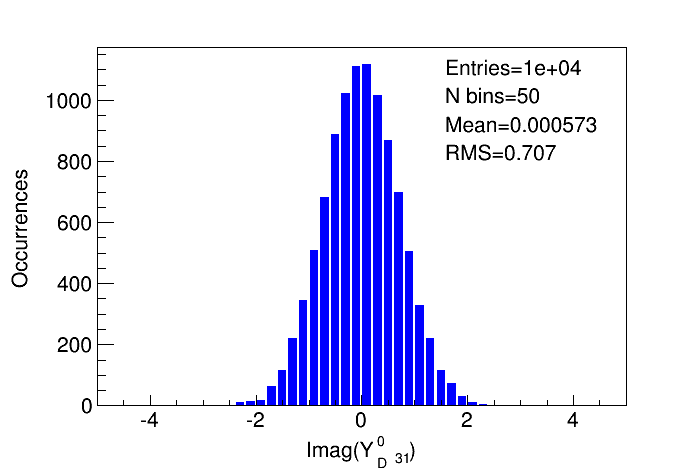}
\includegraphics[width=0.3\linewidth]{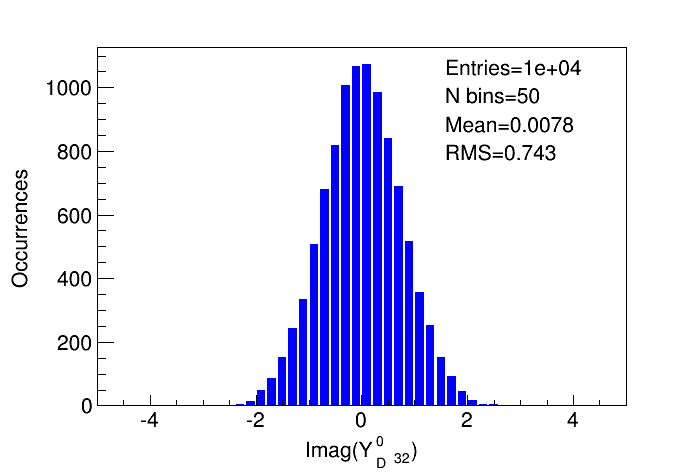}
\includegraphics[width=0.3\linewidth]{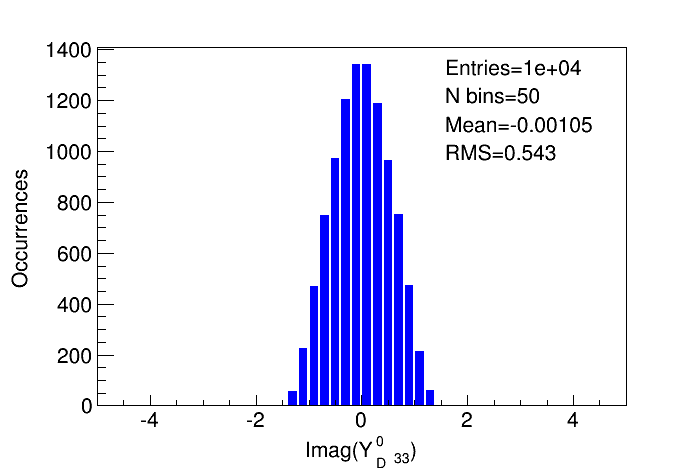}
\caption{Distributions of the $O(1)$ random entries in the matrix $Y^{0}_D$ from the modified Monte Carlo analysis  that produce the observables in Fig.\ref{fig:004c} for $\tan\beta=10$. The first nine of the plots are for the real parts and the next nine for imaginary parts of the matrix, $Y^{0}_D$. For all these plots sample size and number of bins  are taken to be $10^{4}$ and 50 respectively.  }\label{fig:005B3}
\end{figure}

\begin{figure}[th!]
\centering
\includegraphics[width=0.3\linewidth]{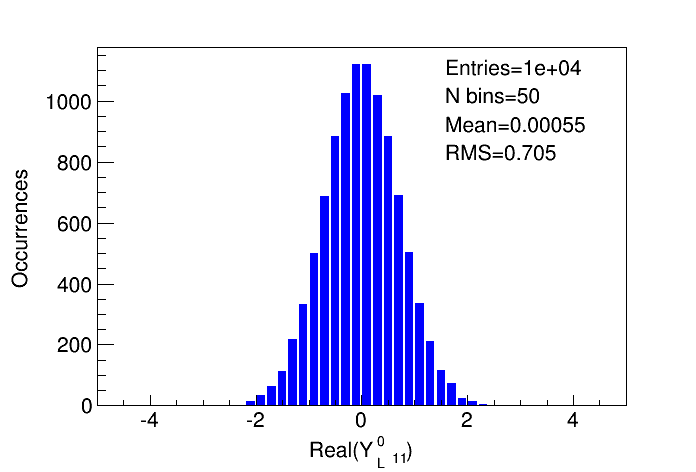}
\includegraphics[width=0.3\linewidth]{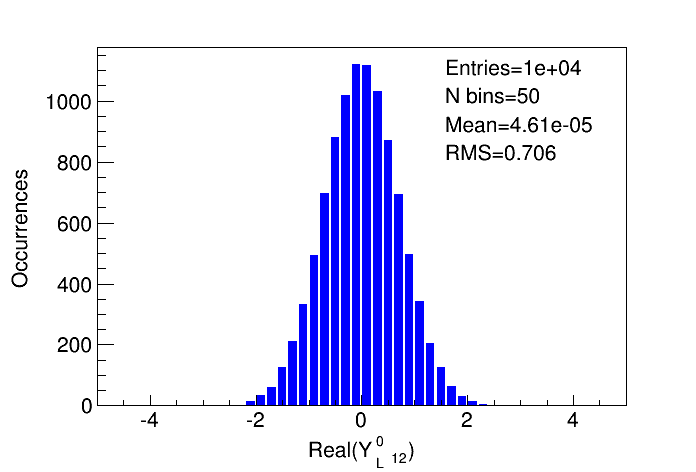}
\includegraphics[width=0.3\linewidth]{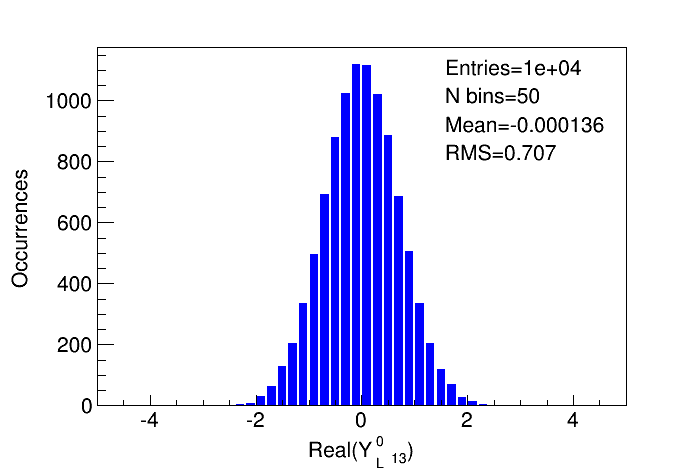}
\includegraphics[width=0.3\linewidth]{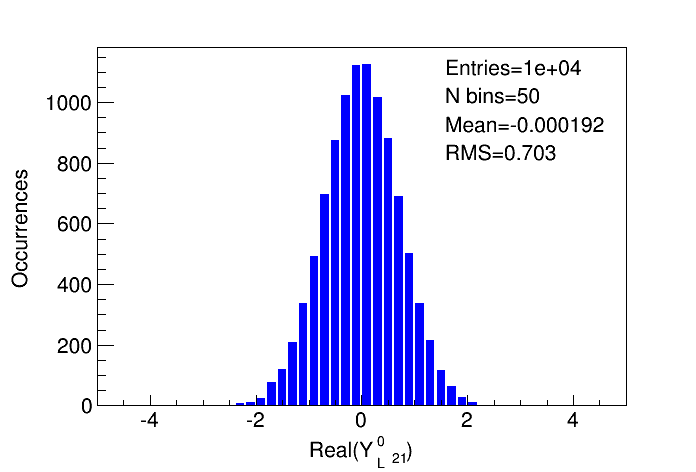}
\includegraphics[width=0.3\linewidth]{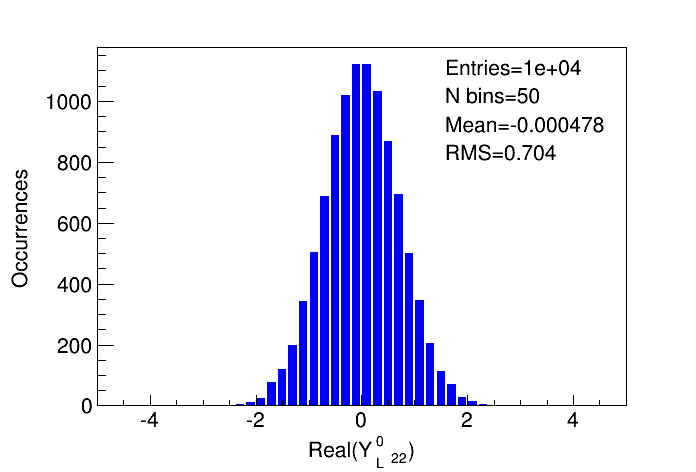}
\includegraphics[width=0.3\linewidth]{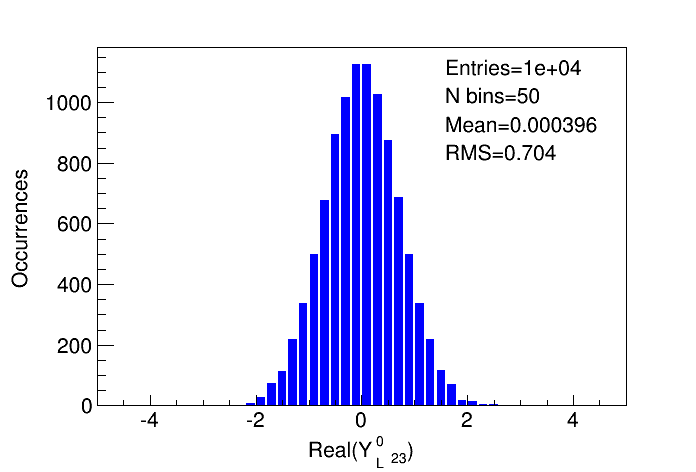}
\includegraphics[width=0.3\linewidth]{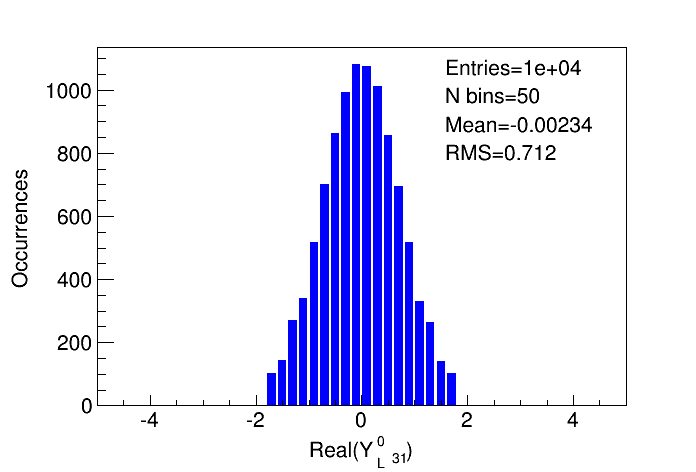}
\includegraphics[width=0.3\linewidth]{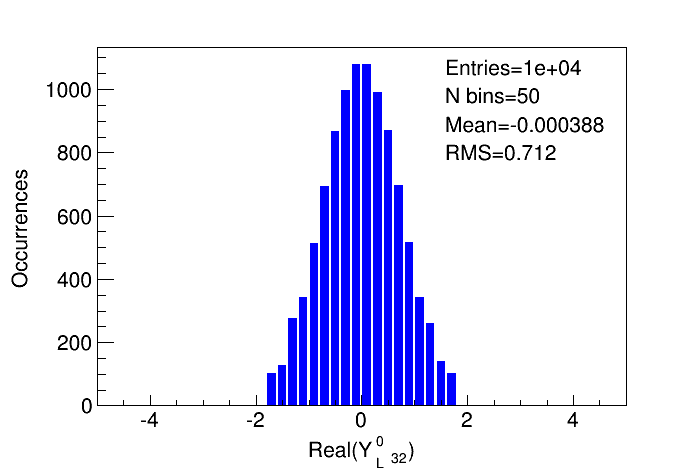}
\includegraphics[width=0.3\linewidth]{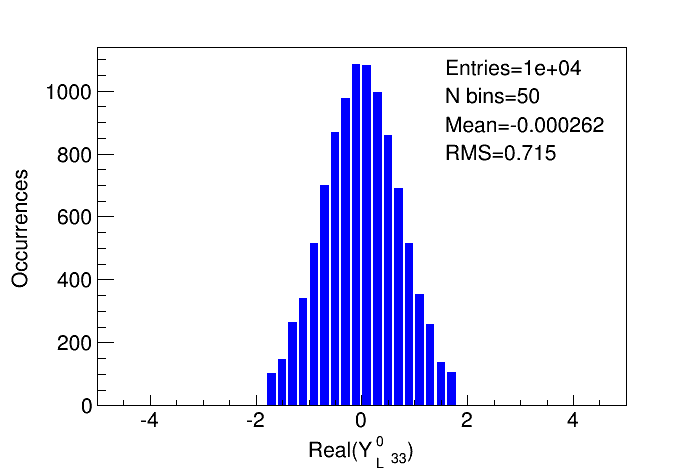}
\includegraphics[width=0.3\linewidth]{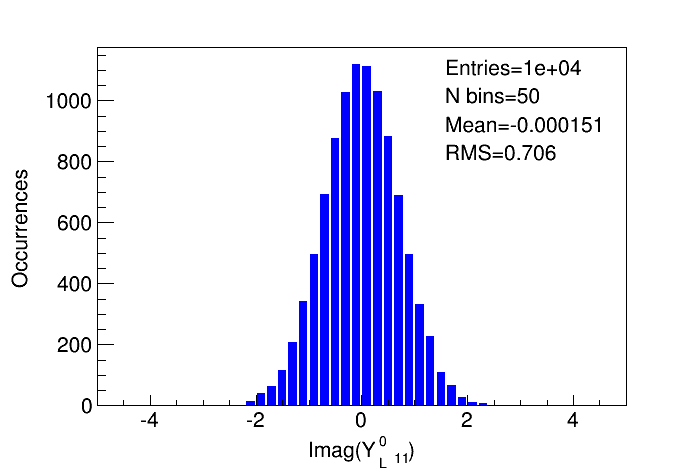}
\includegraphics[width=0.3\linewidth]{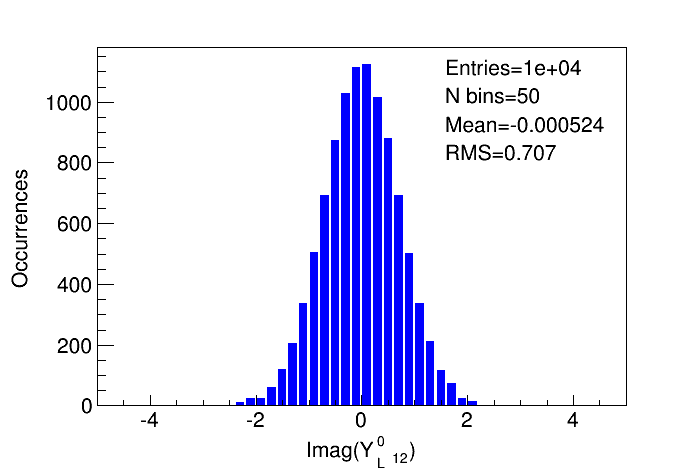}
\includegraphics[width=0.3\linewidth]{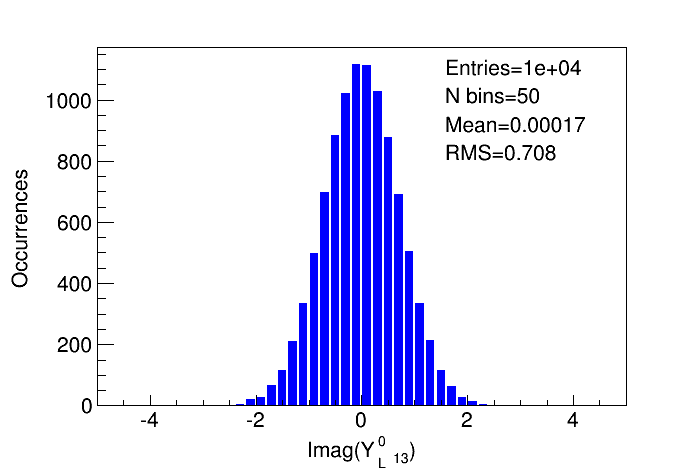}
\includegraphics[width=0.3\linewidth]{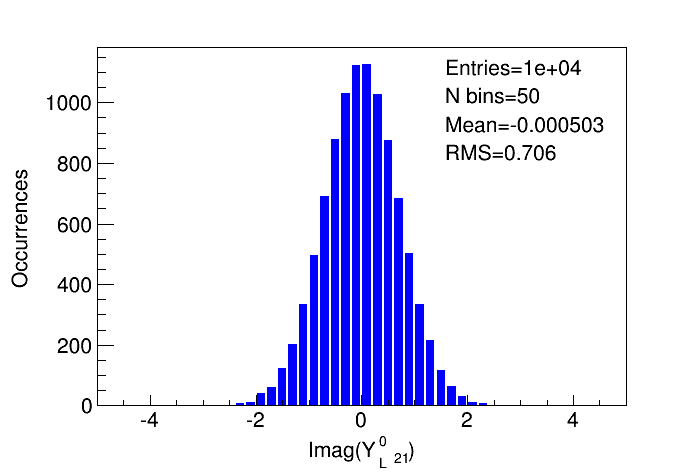}
\includegraphics[width=0.3\linewidth]{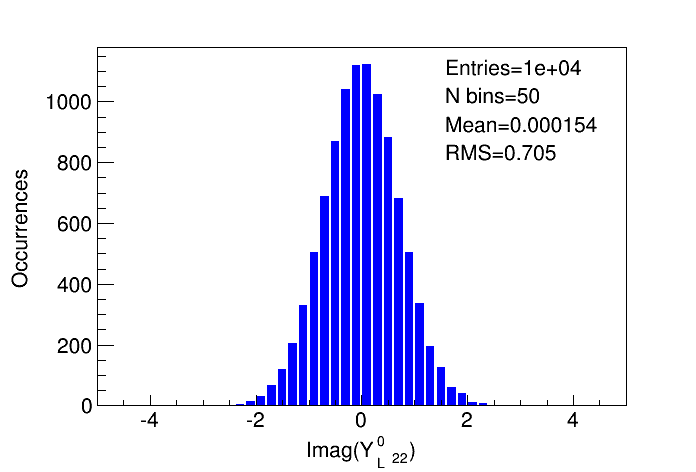}
\includegraphics[width=0.3\linewidth]{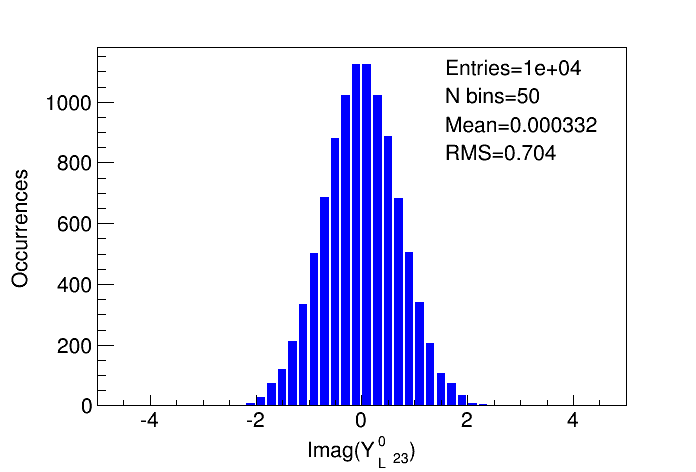}
\includegraphics[width=0.3\linewidth]{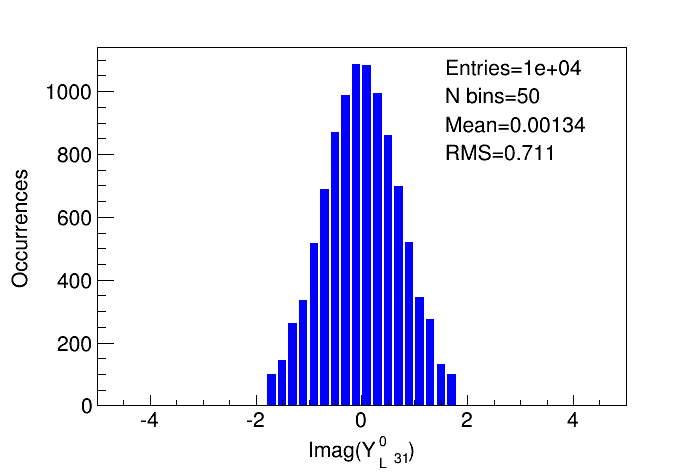}
\includegraphics[width=0.3\linewidth]{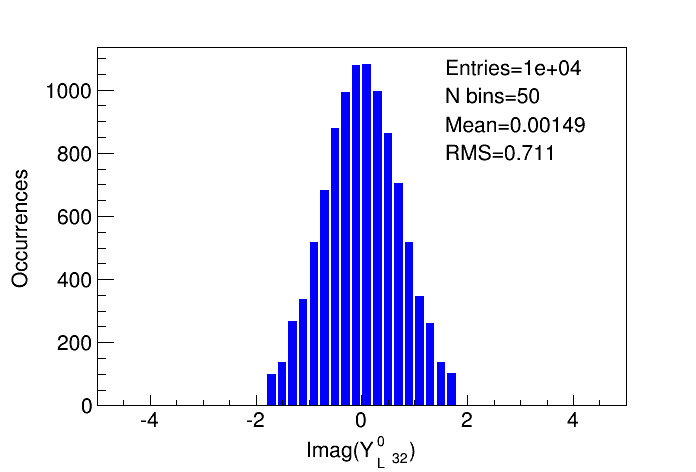}
\includegraphics[width=0.3\linewidth]{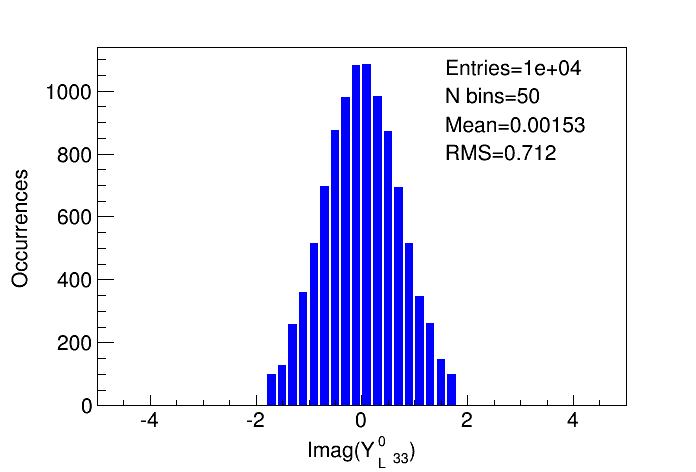}
\caption{Distributions of the $O(1)$ random entries in the matrix $Y^{0}_L$ from the modified Monte Carlo analysis that produce the observables in Fig.\ref{fig:004c} for $\tan\beta=10$. The first nine of the plots are for the real parts and the next nine for imaginary parts of the matrix, $Y^{0}_L$. For all these plots sample size and number of bins  are taken to be $10^{4}$ and 50 respectively.  }\label{fig:005C3}
\end{figure}

Here we present the distributions  of the modified random entries in Fig. \ref{fig:005A3} for up-quark,  \ref{fig:005B3} for down-quark \ref{fig:005C3} and for charged lepton matrices. Theoretical distributions associated with these modified random entries  are shown in Fig. \ref{fig:004c}. These are the result of  $D=D(O,E)+D\left(\{\bm{r}^*\},\{\bm{r}\}\right)$  minimization procedure following the modified Monte Carlo analysis as explained in Ssec. \ref{SU5P} for the $SU(5)$-based GUTs defined in Eqs. \eqref{U4}-\eqref{L4}.  From Figs. \ref{fig:005A3}, \ref{fig:005B3} and \ref{fig:005C3} one can see that majority of the random entries of the matrices, even after the minimization process exhibit Gaussianity and remain similar in distribution as the unbiased set. The (3,3) element in the  up-type quark Yukawa matrix is the only entry that shows somewhat distorted distribution. This analysis shows that the subspace of the random variables that has excellent agreement with experimental data is quite broad.

\FloatBarrier

\newpage
\subsection{\normalsize Distributions of the neutrino observables by applying the modified Monte Carlo analysis} \label{App:B03}

Here we present the theoretical distributions of the neutrino observables by employing the modified  Monte Carlo analysis for the $SU(5)$-based GUTs where the neutrino matrix if given by Eq. \eqref{nu}.

\begin{figure}[th!]
\centering
\includegraphics[width=0.32\linewidth]{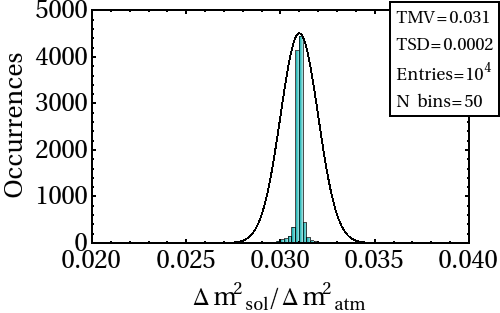} \\
\includegraphics[width=0.32\linewidth]{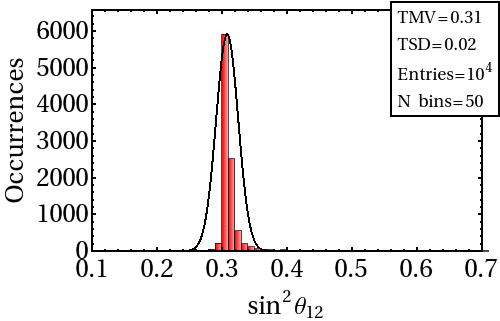}
\includegraphics[width=0.32\linewidth]{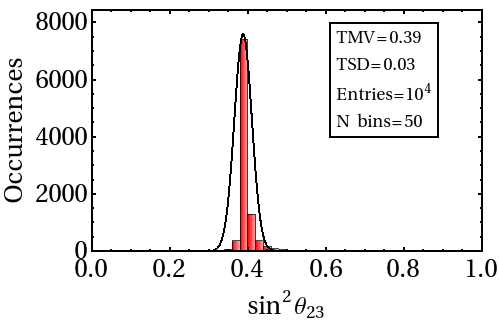}
\includegraphics[width=0.32\linewidth]{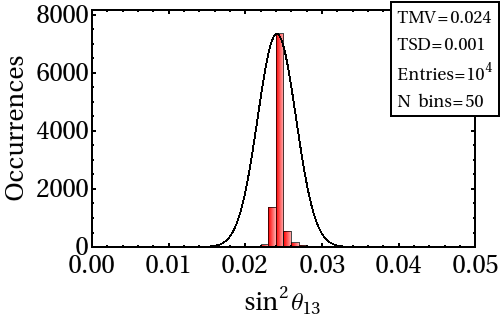}
\caption{Histogram distributions of the observables in the neutrino sector according to the modified Monte Carlo  approach for $SU(5)$-based GUTs with structure-less neutrino mass matrix. The top histogram plot (dark cyan) shows the theoretical distribution of the quantity $\Delta m^{2}_{\rm{sol}}/\Delta m^{2}_{\rm{atm}}$  and the bottom three plots (red) are for the mixing parameters $\sin^{2}\theta_{ij}$. The black curves represent the experimental 1$\sigma$ ranges. The sample size is taken to be $10^{4}$ and number of bins  is taken to be 50. }\label{fig:004N}
\end{figure}

\newpage
\subsection{\normalsize Distributions of the modified random entries in the  neutrino sector by applying the modified  Monte Carlo analysis} \label{App:B04}

\begin{figure}[th!]
\centering
\includegraphics[width=0.3\linewidth]{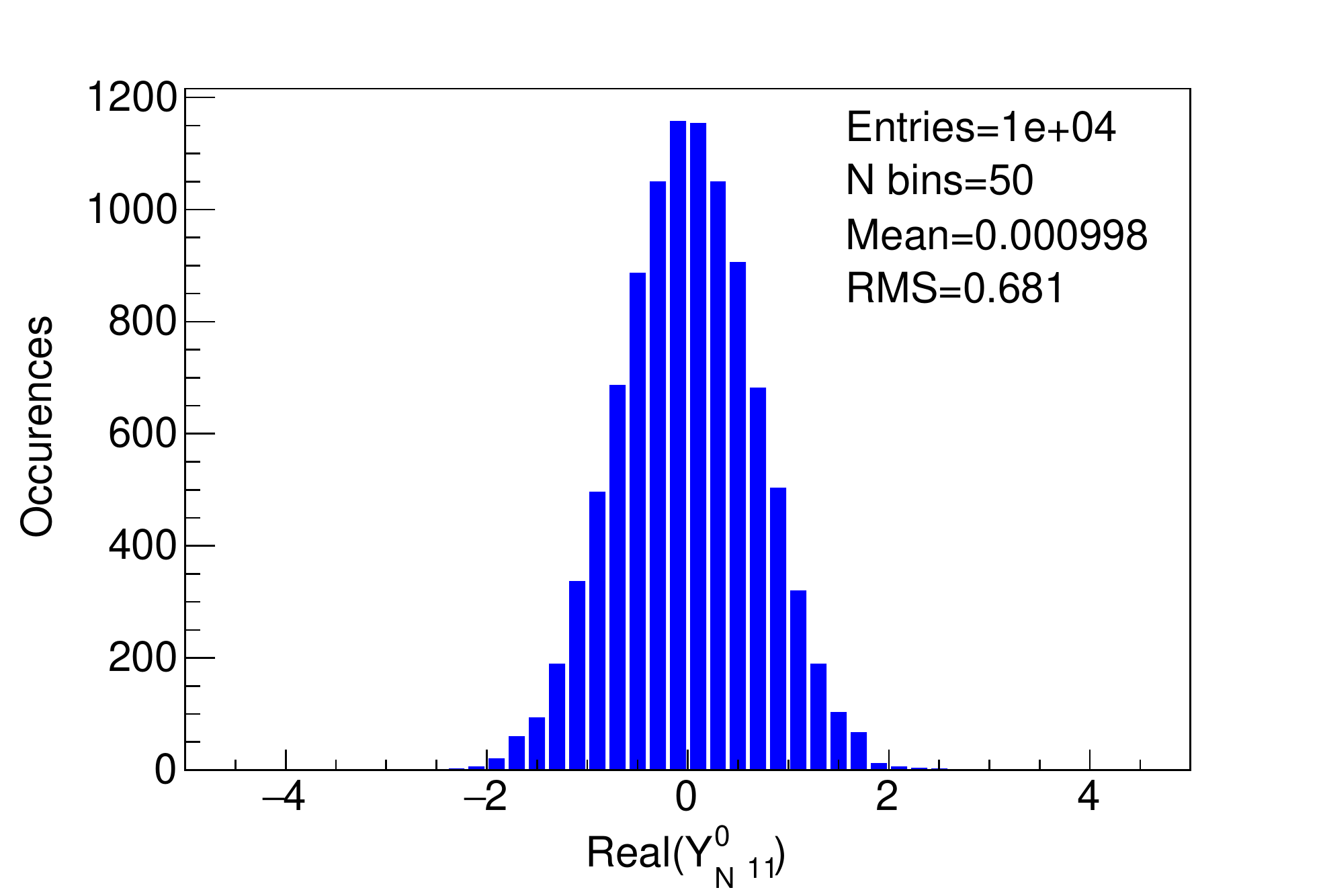}
\includegraphics[width=0.3\linewidth]{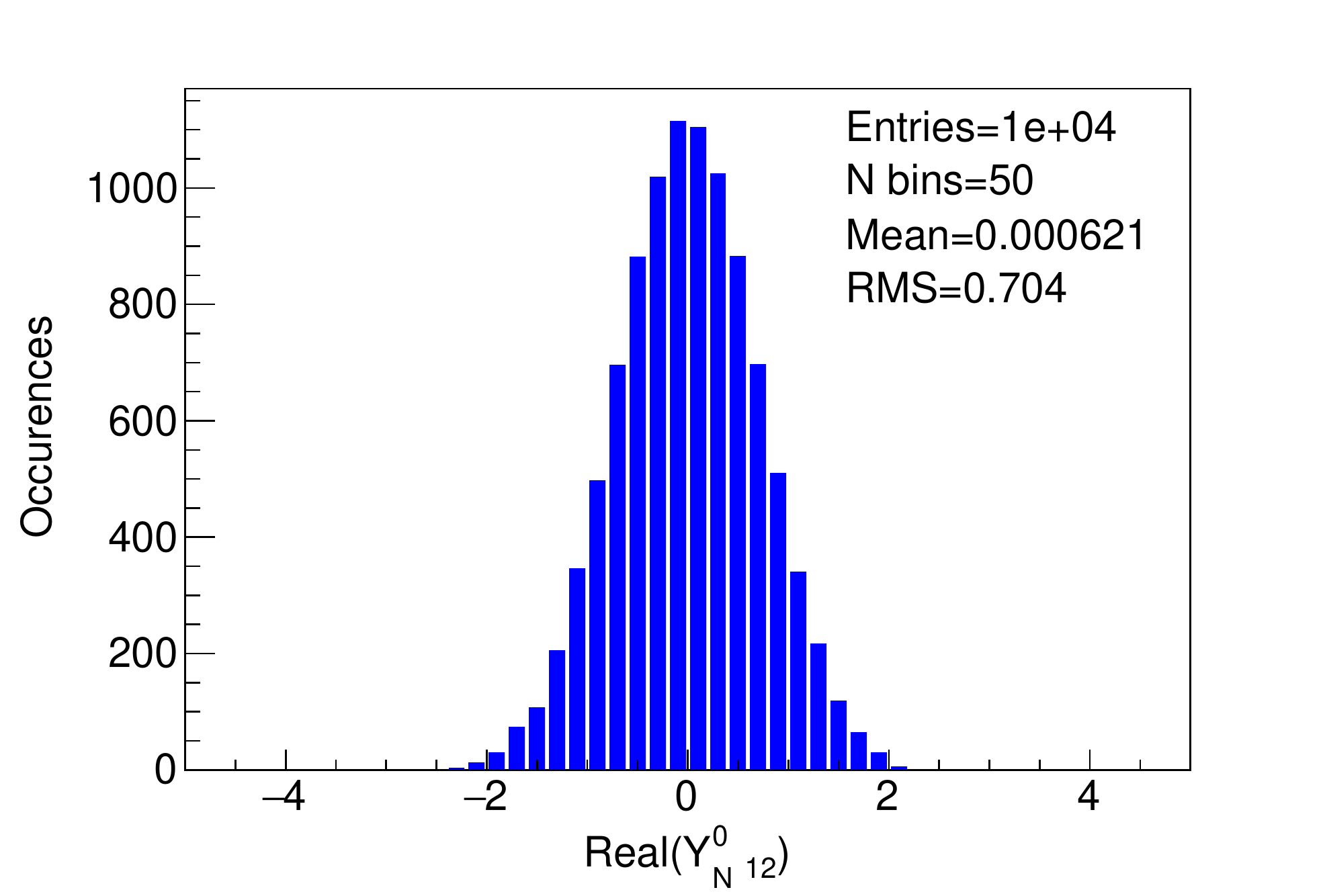}
\includegraphics[width=0.3\linewidth]{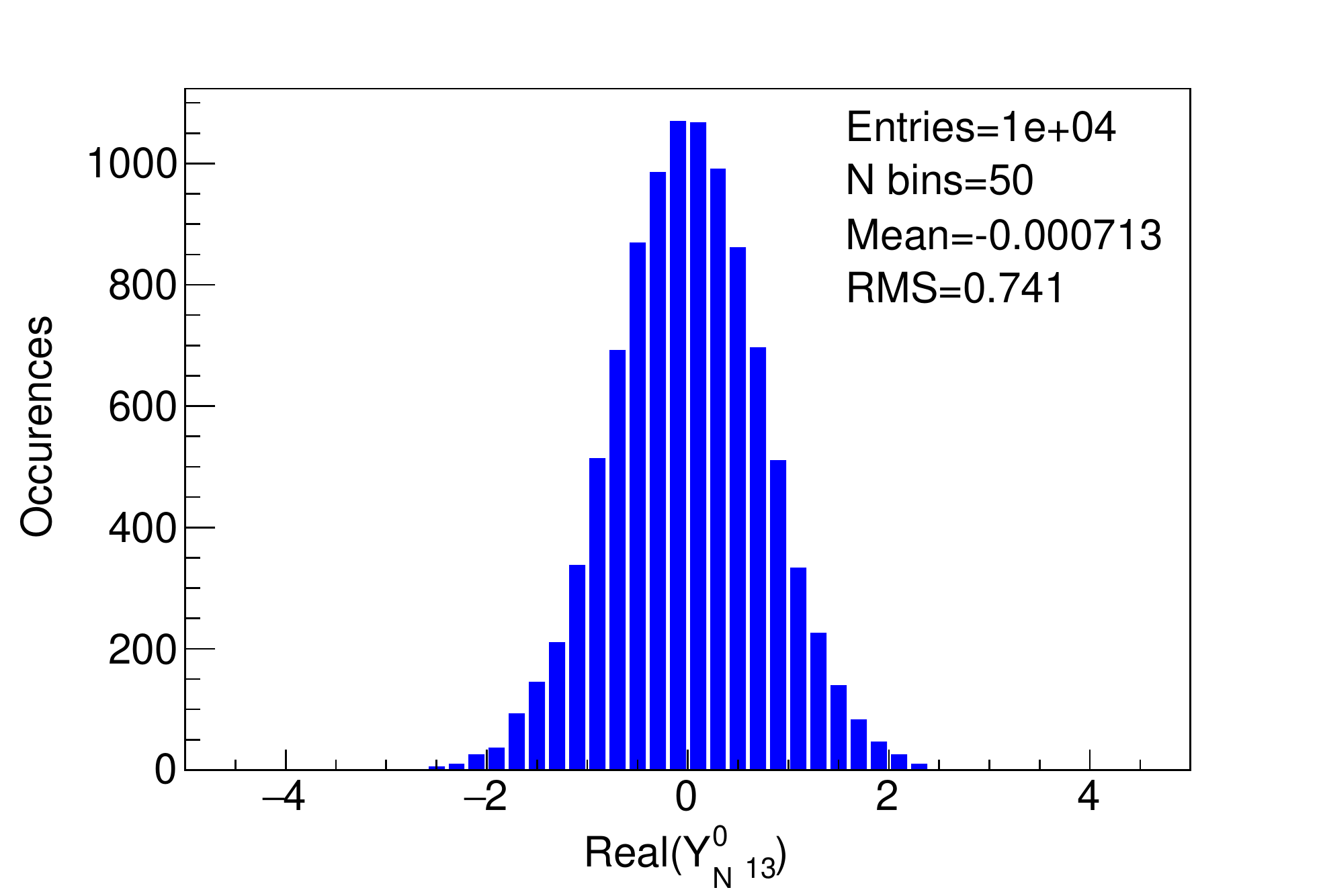}
\includegraphics[width=0.3\linewidth]{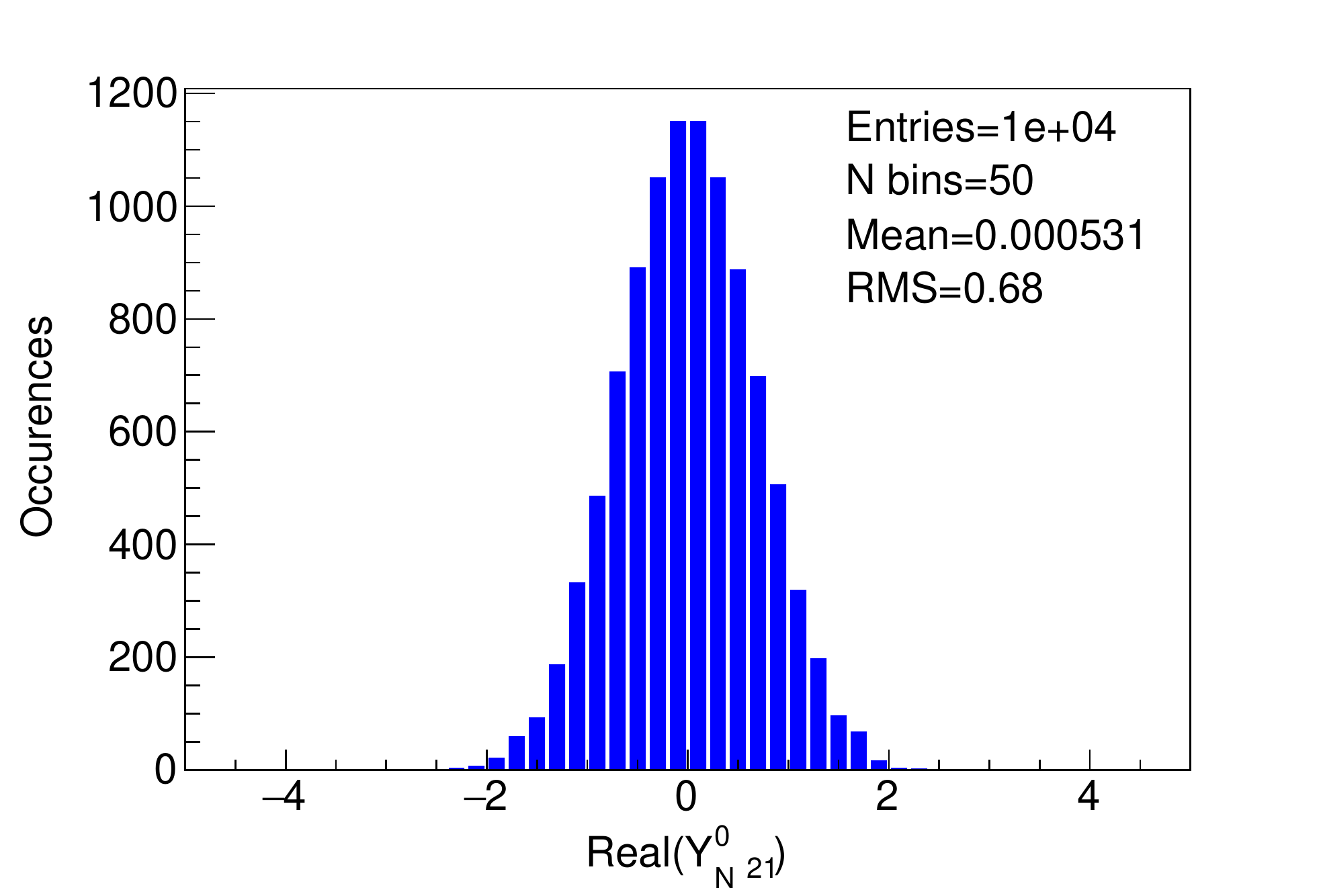}
\includegraphics[width=0.3\linewidth]{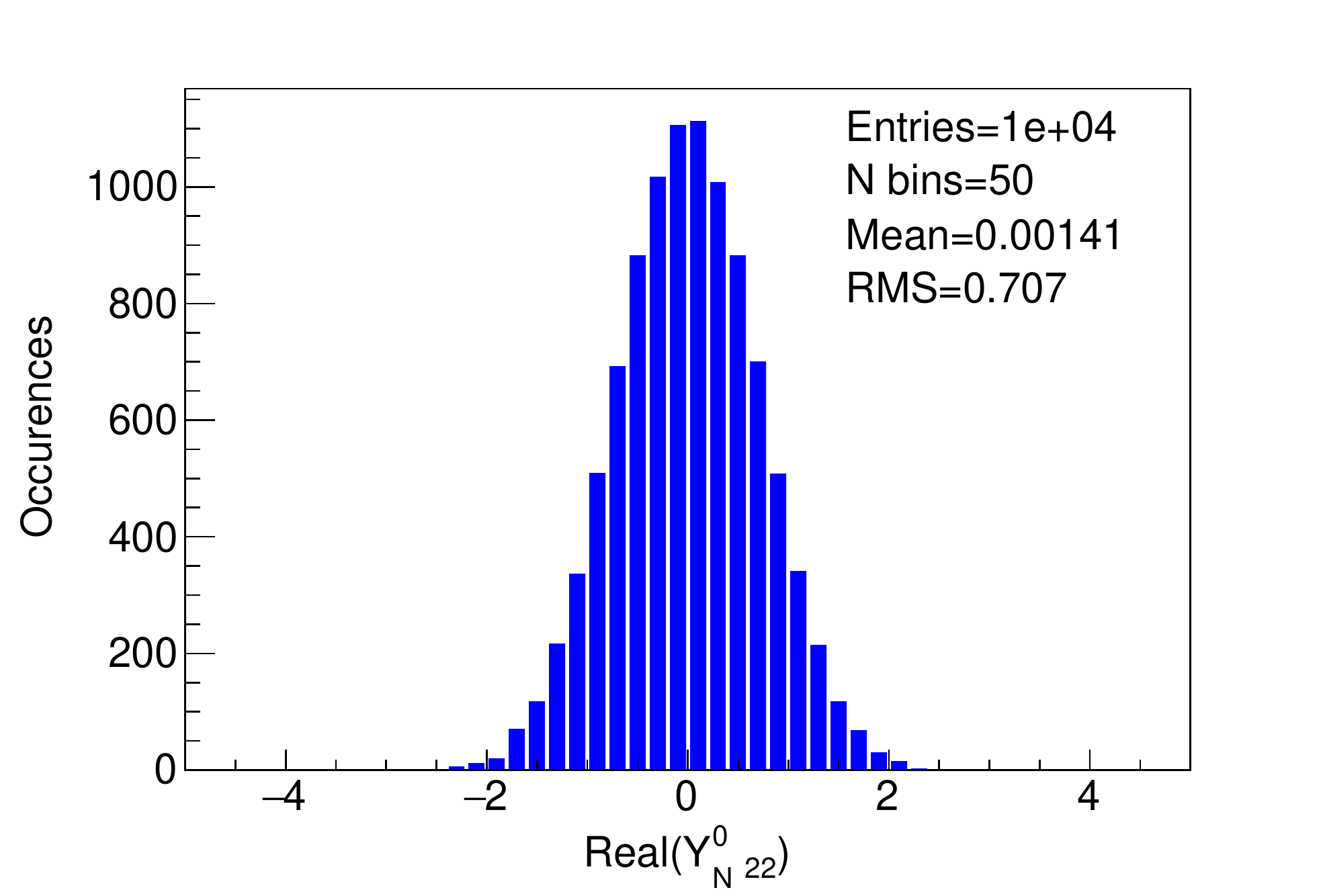}
\includegraphics[width=0.3\linewidth]{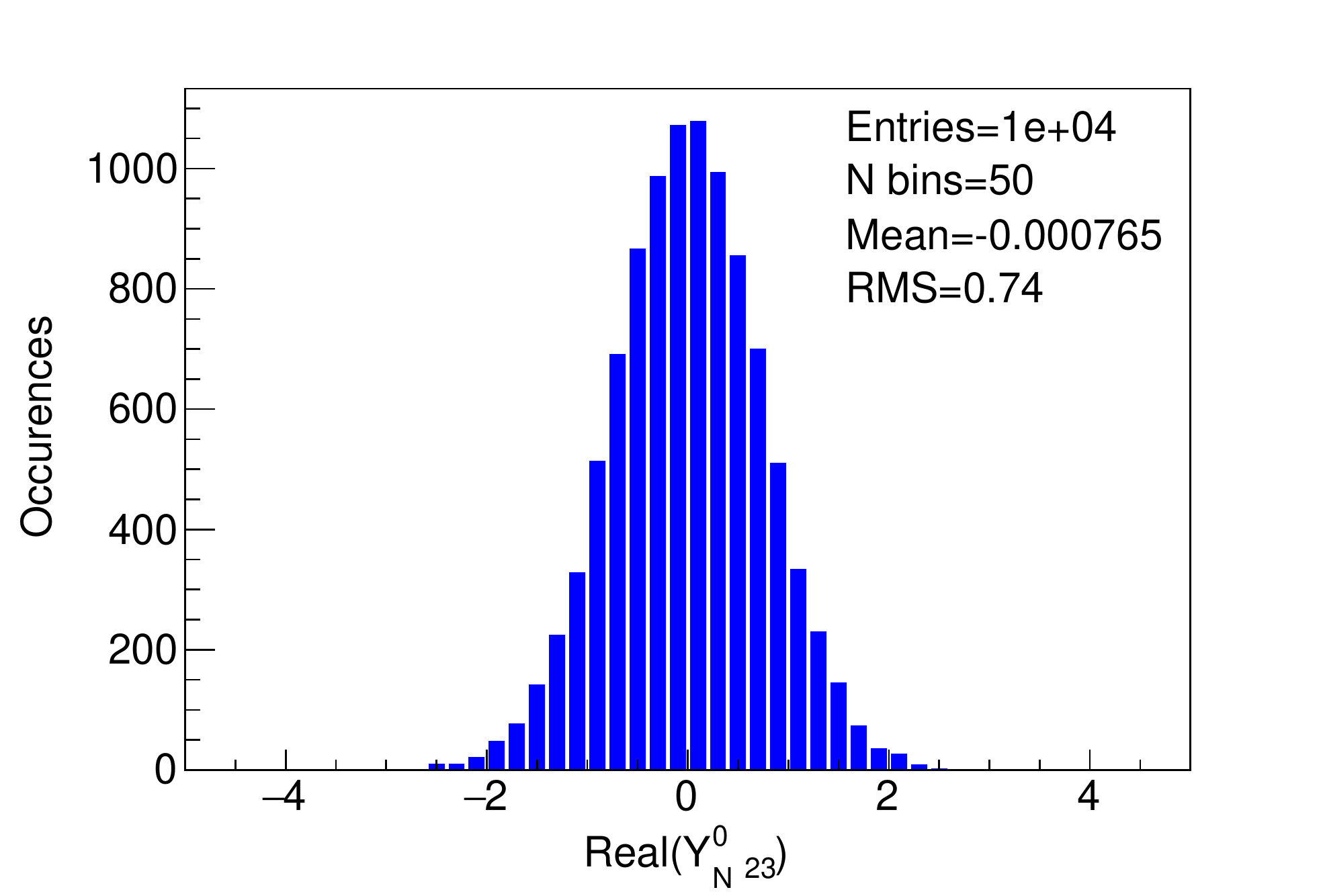}
\includegraphics[width=0.3\linewidth]{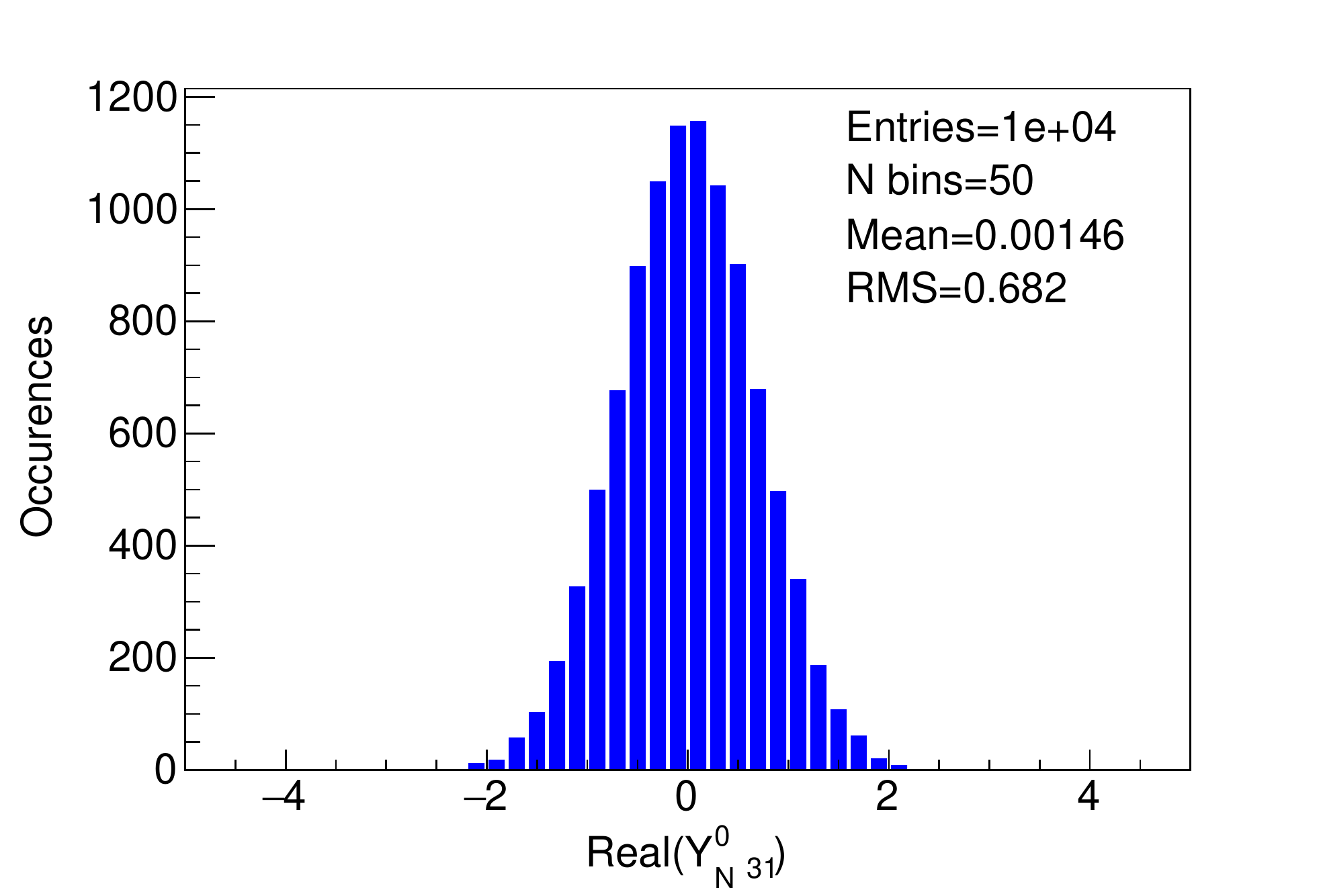}
\includegraphics[width=0.3\linewidth]{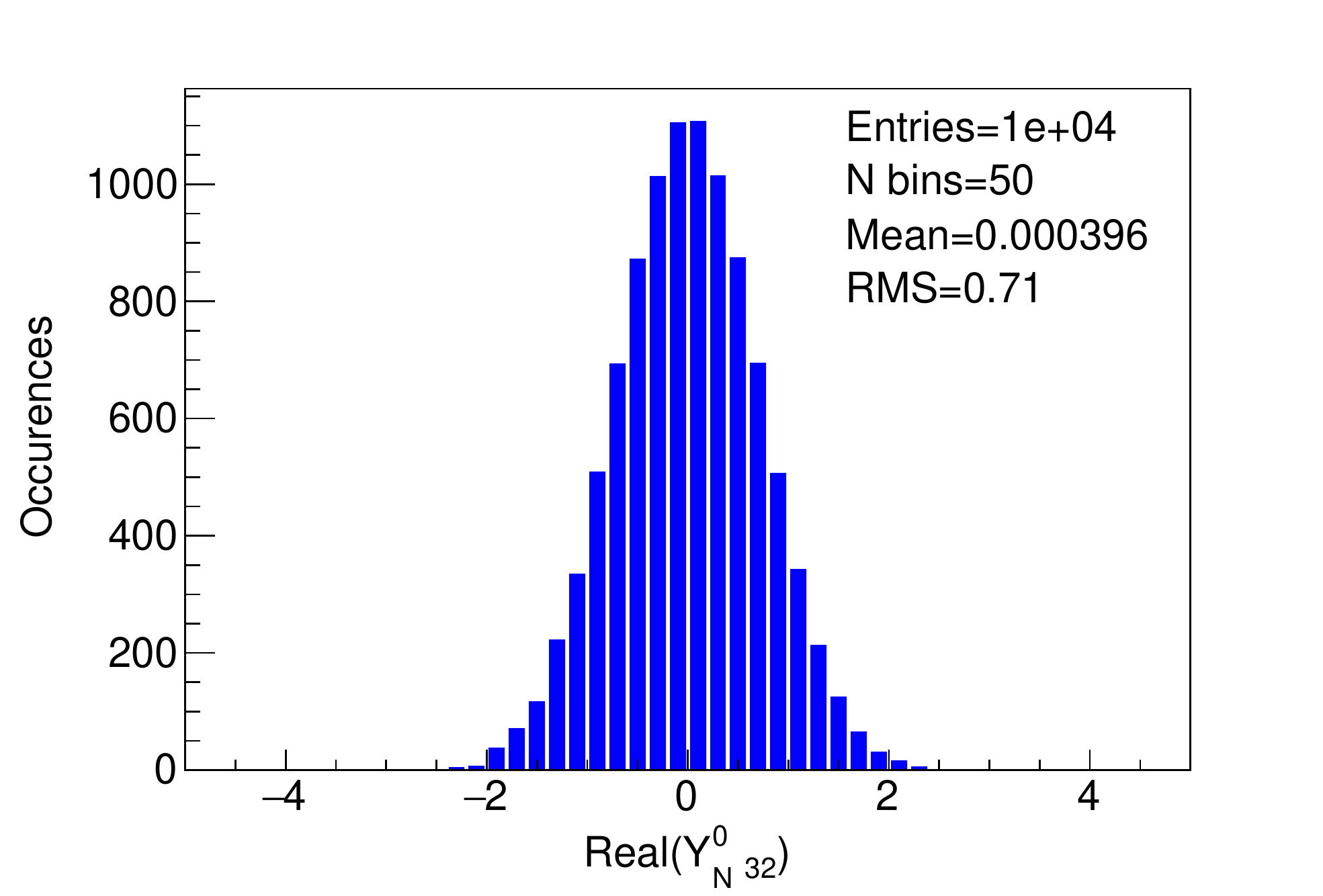}
\includegraphics[width=0.3\linewidth]{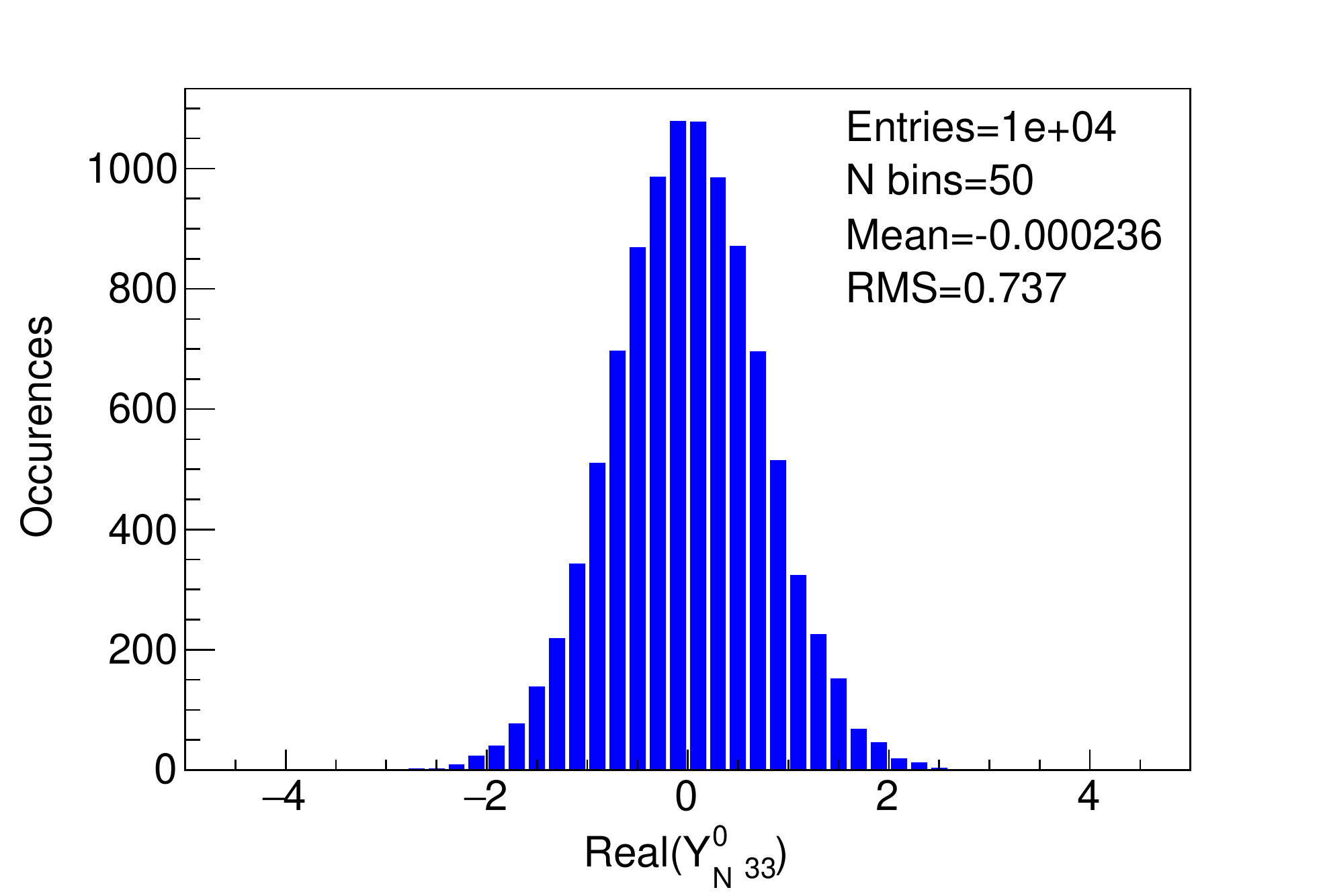}
\includegraphics[width=0.3\linewidth]{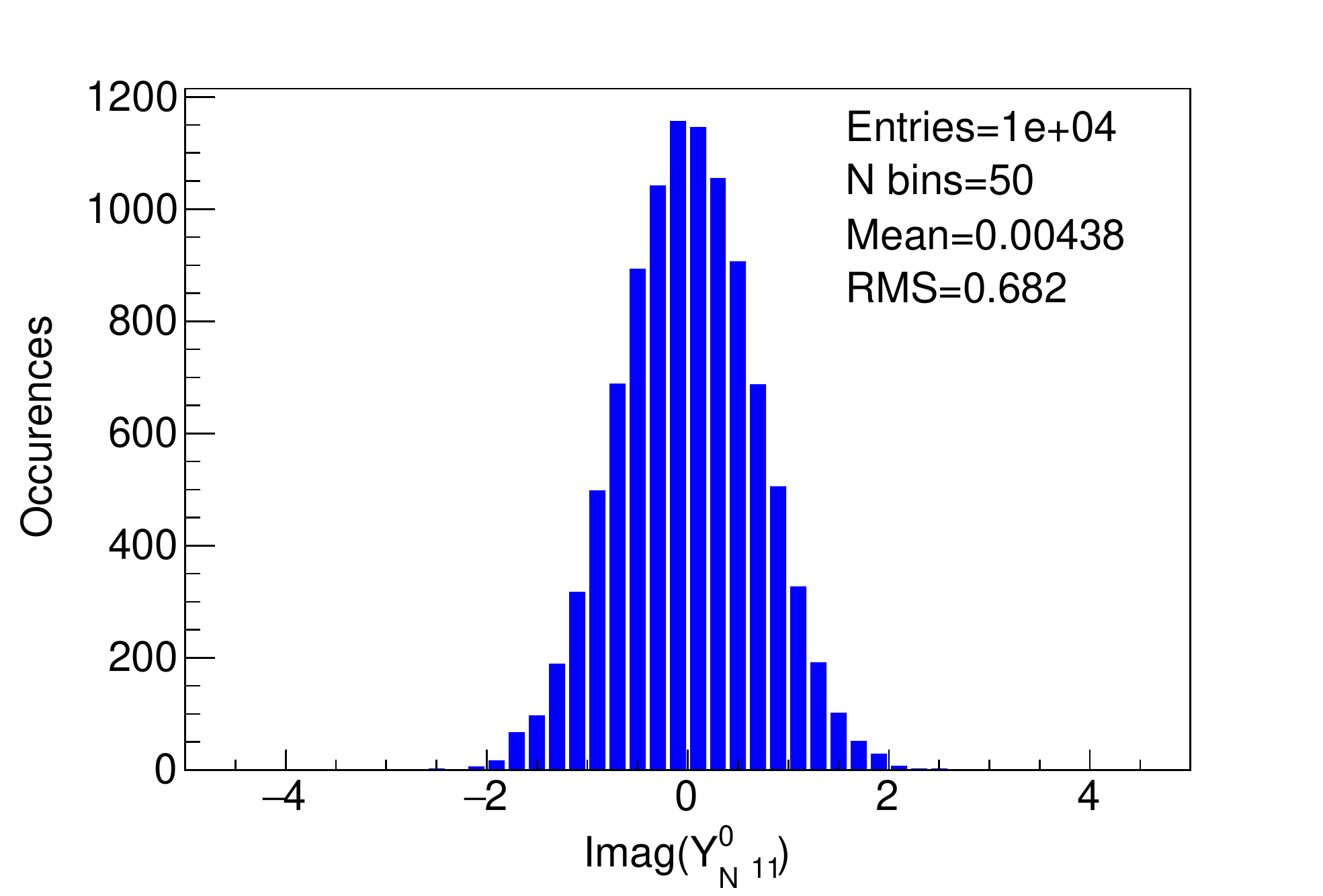}
\includegraphics[width=0.3\linewidth]{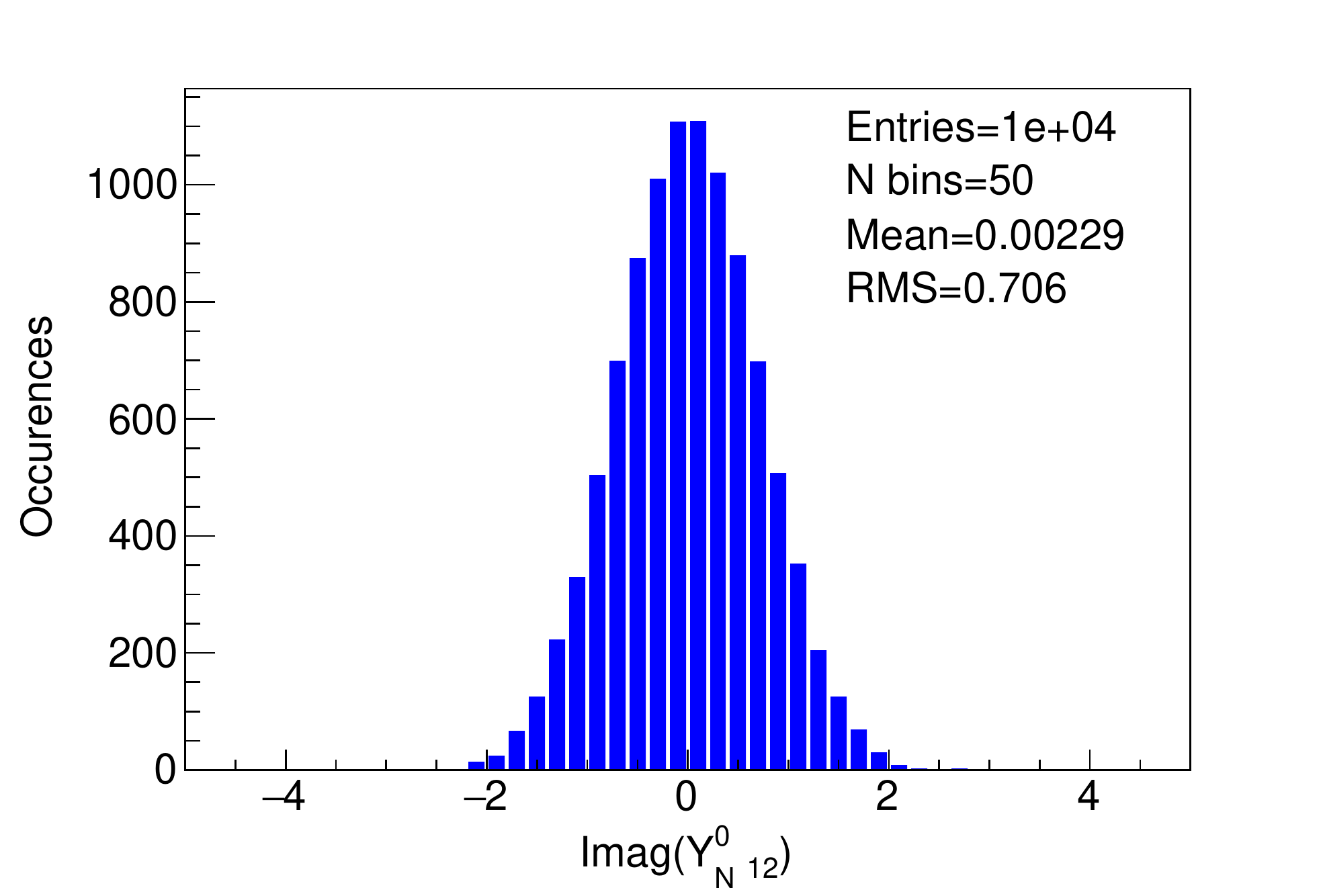}
\includegraphics[width=0.3\linewidth]{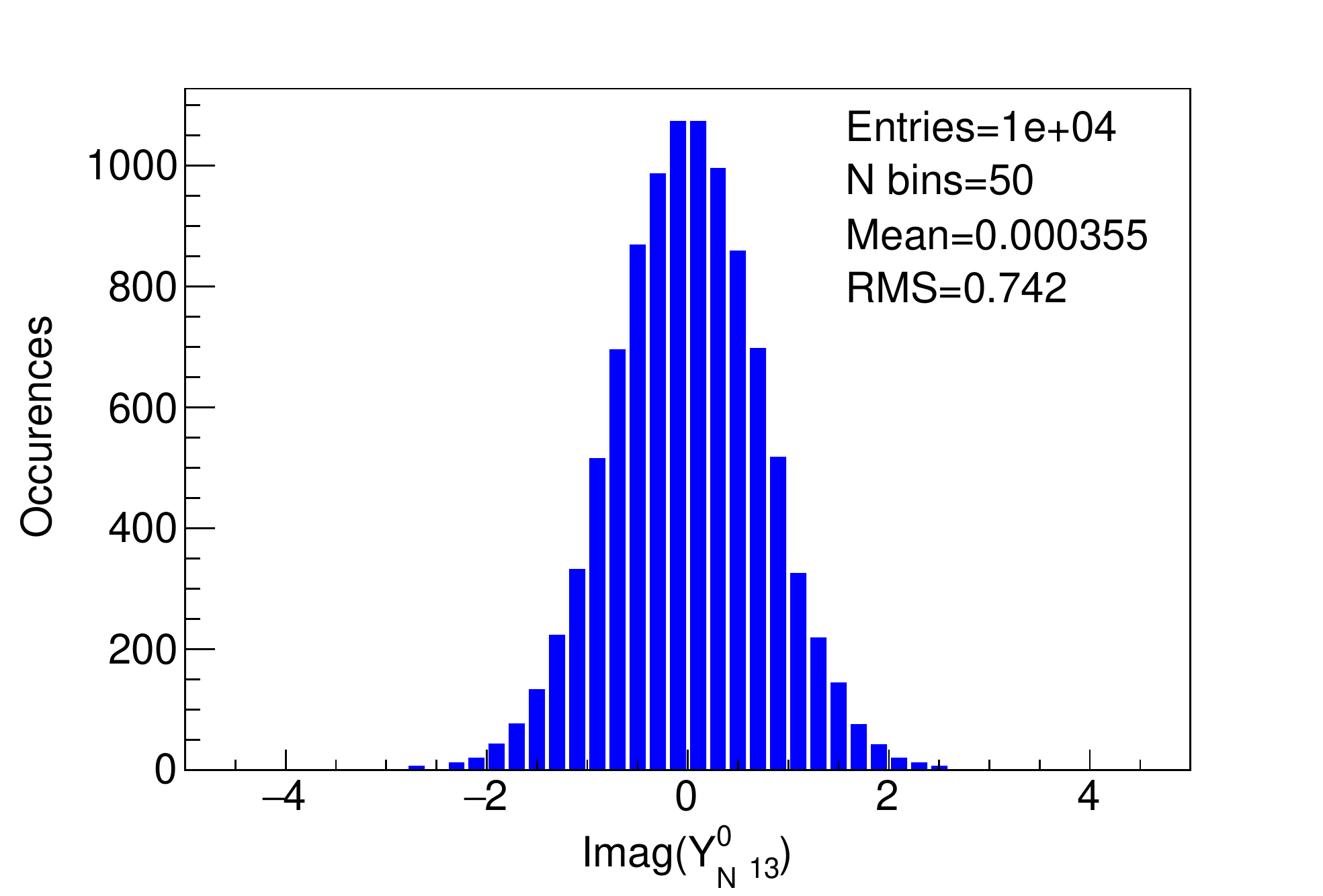}
\includegraphics[width=0.3\linewidth]{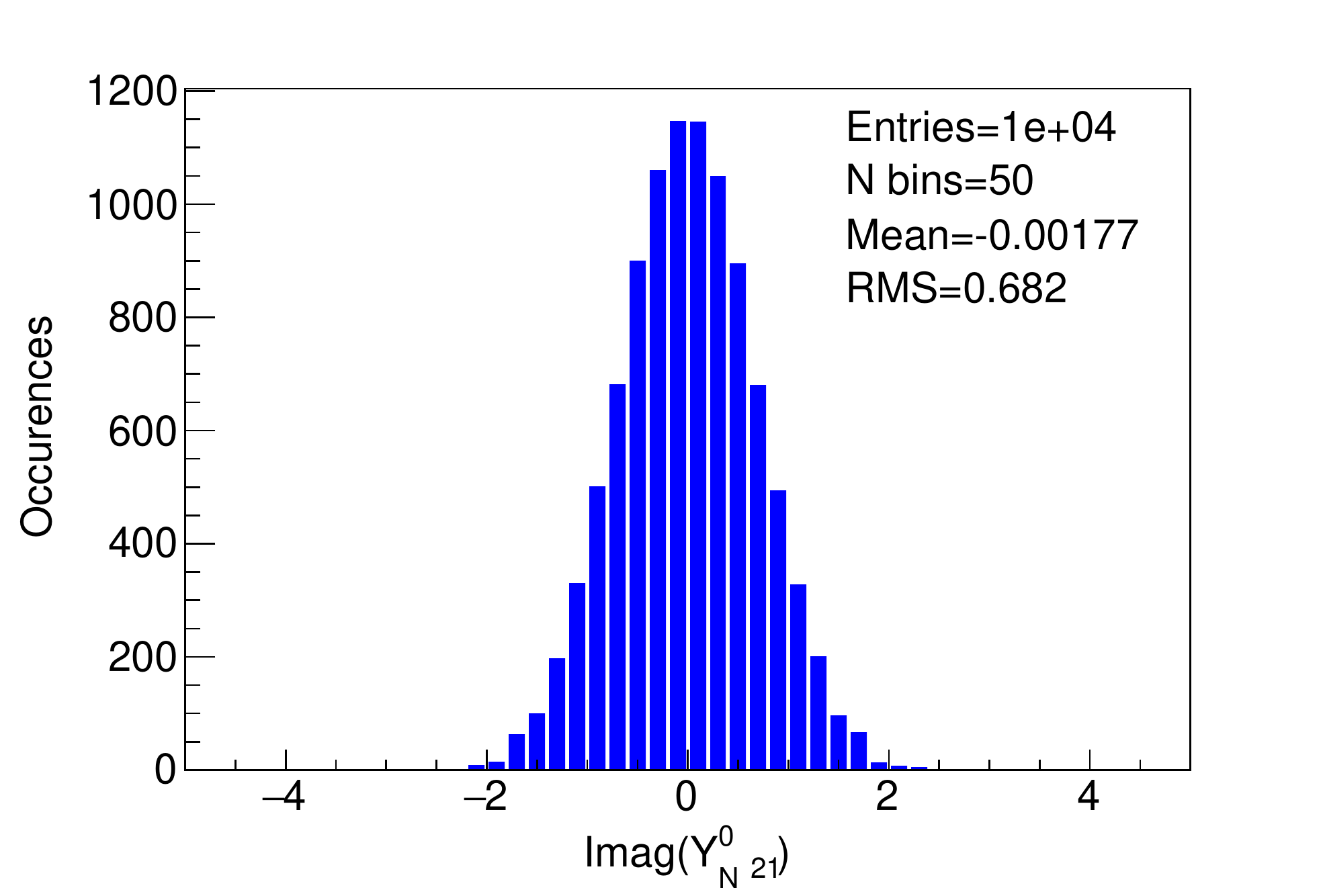}
\includegraphics[width=0.3\linewidth]{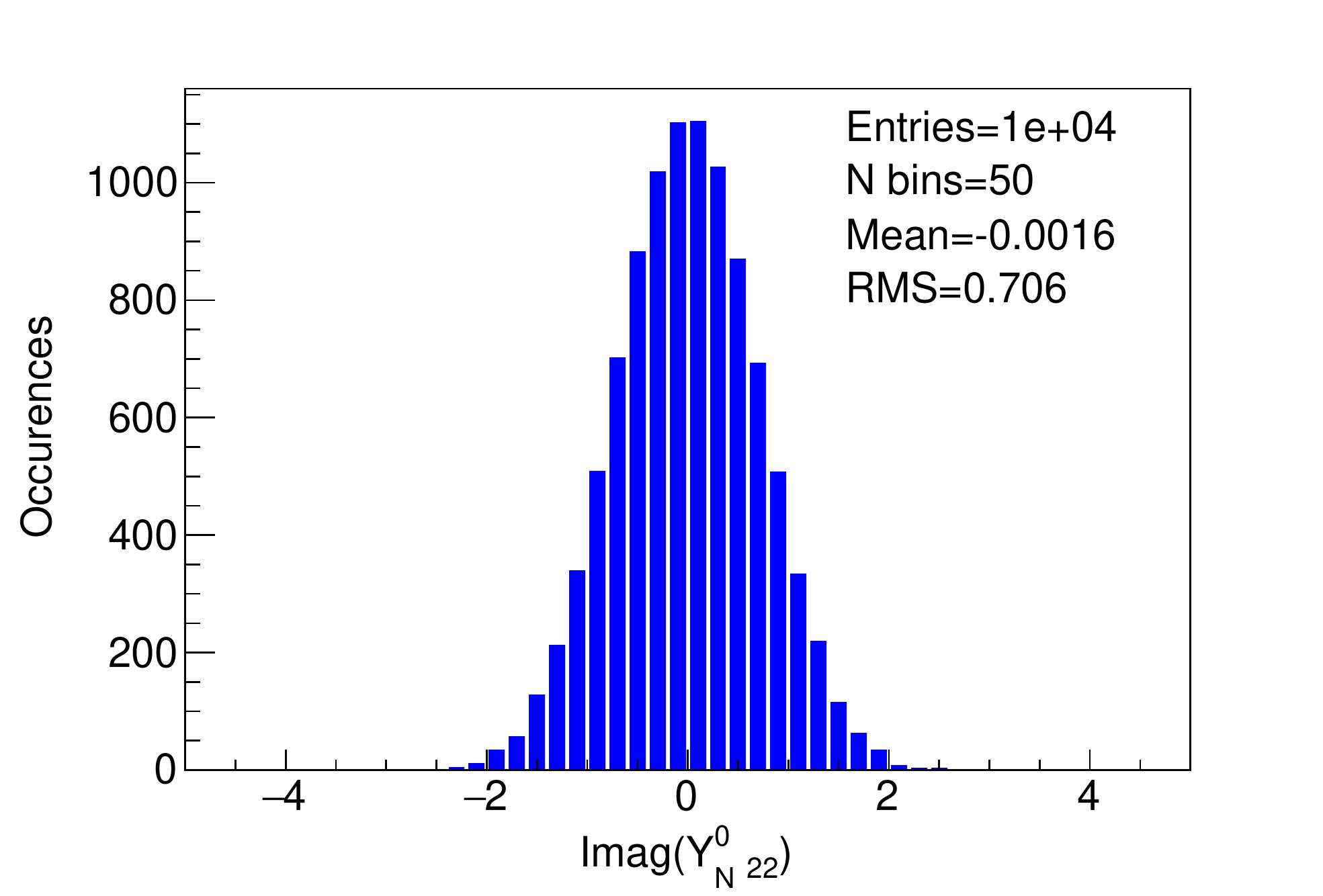}
\includegraphics[width=0.3\linewidth]{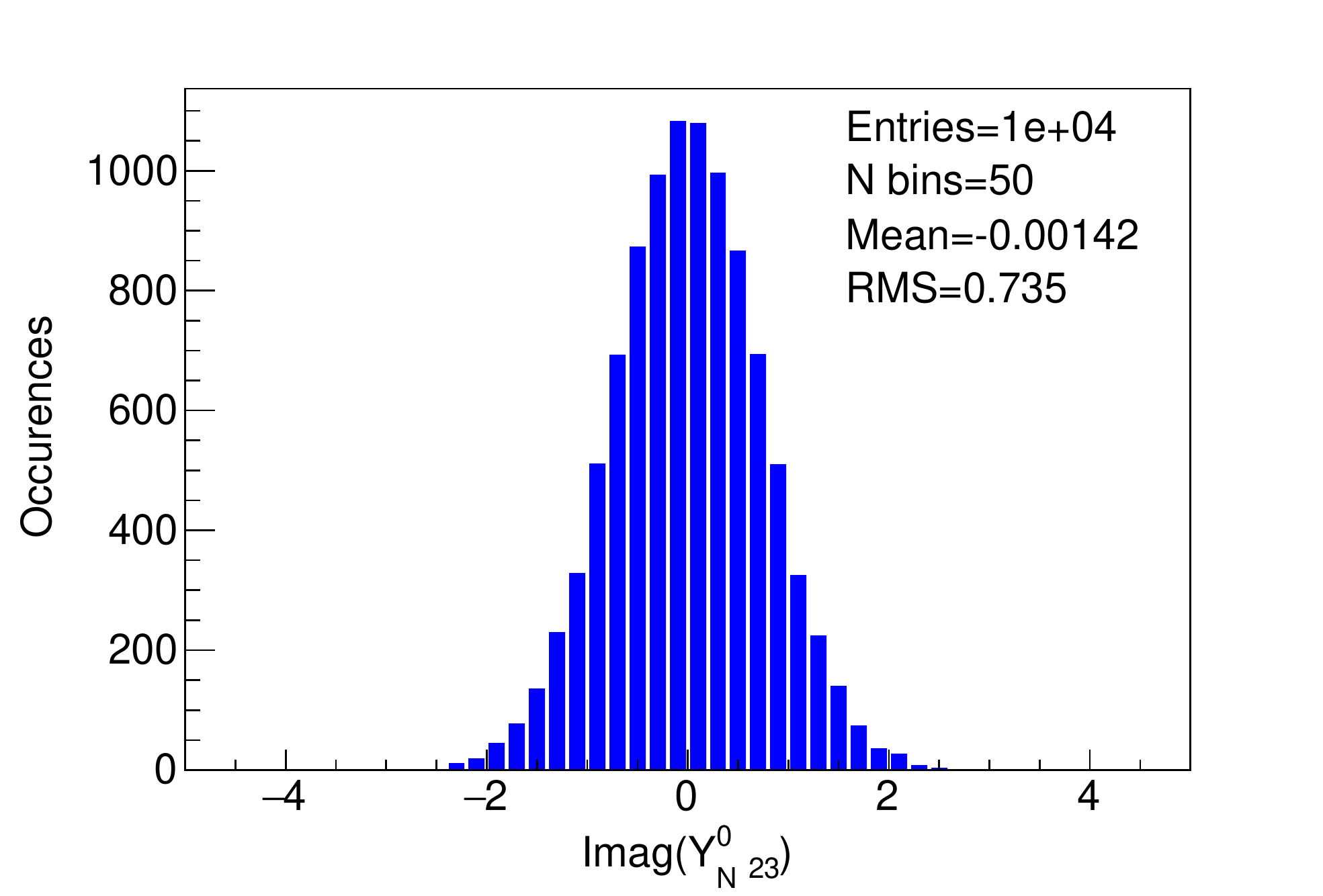}
\includegraphics[width=0.3\linewidth]{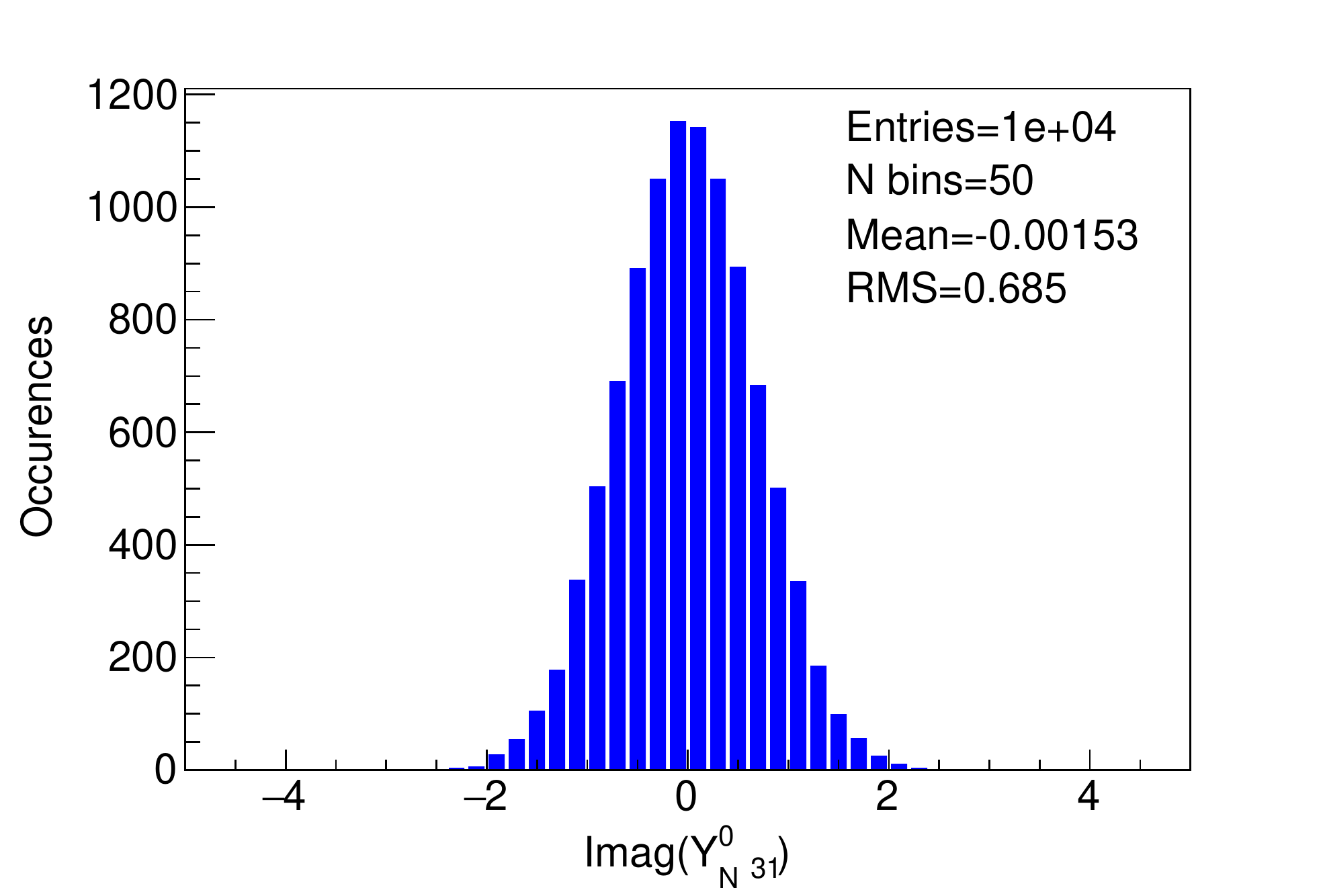}
\includegraphics[width=0.3\linewidth]{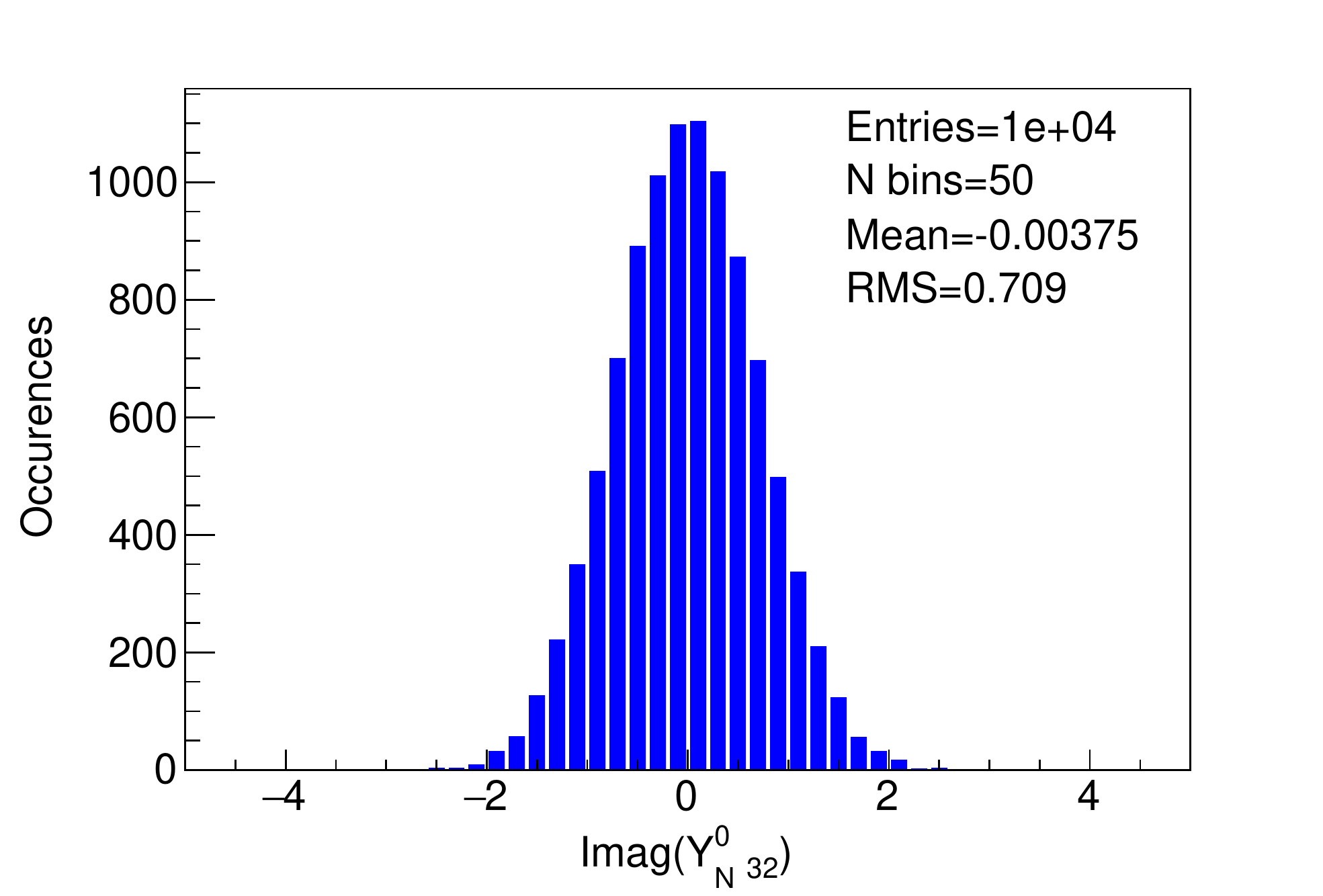}
\includegraphics[width=0.3\linewidth]{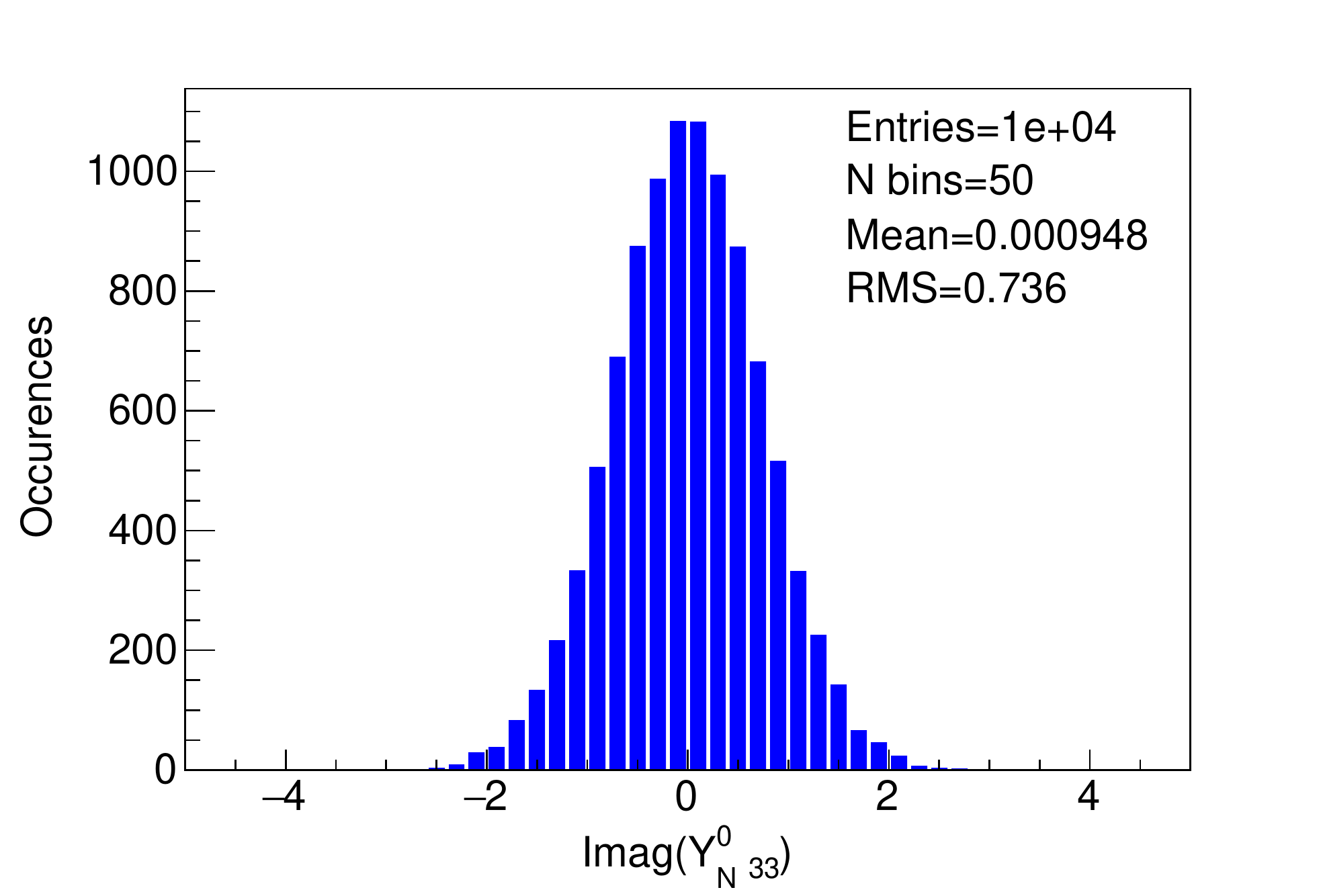}
\caption{ Distributions of the $O(1)$ random entries in the matrix $Y^{0}_N$ from the modified Monte Carlo approach  that produce the observables in Fig.\ref{fig:004N}. }\label{fig:005D3}
\end{figure}

\begin{figure}[th!]
\centering
\includegraphics[width=0.3\linewidth]{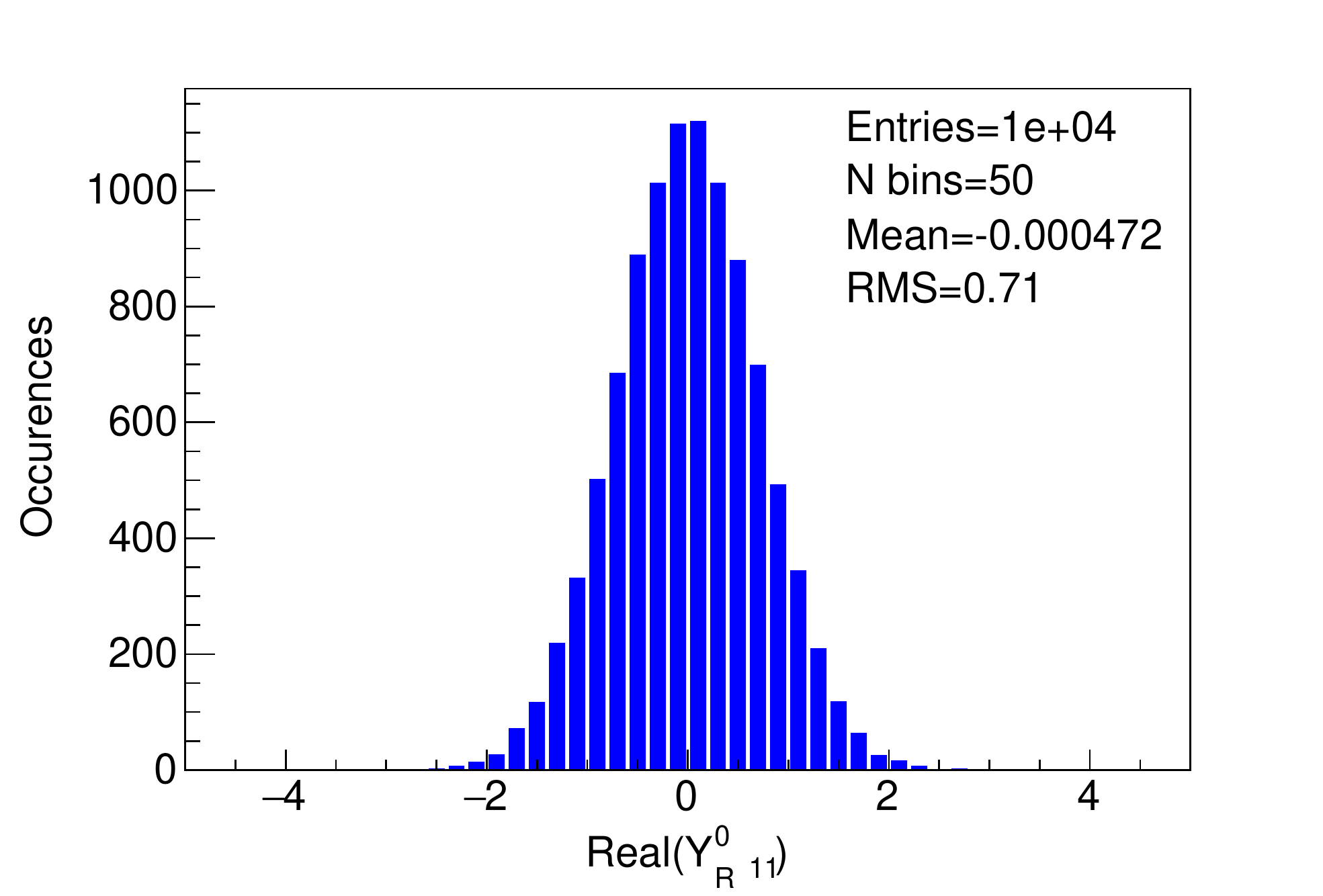}
\includegraphics[width=0.3\linewidth]{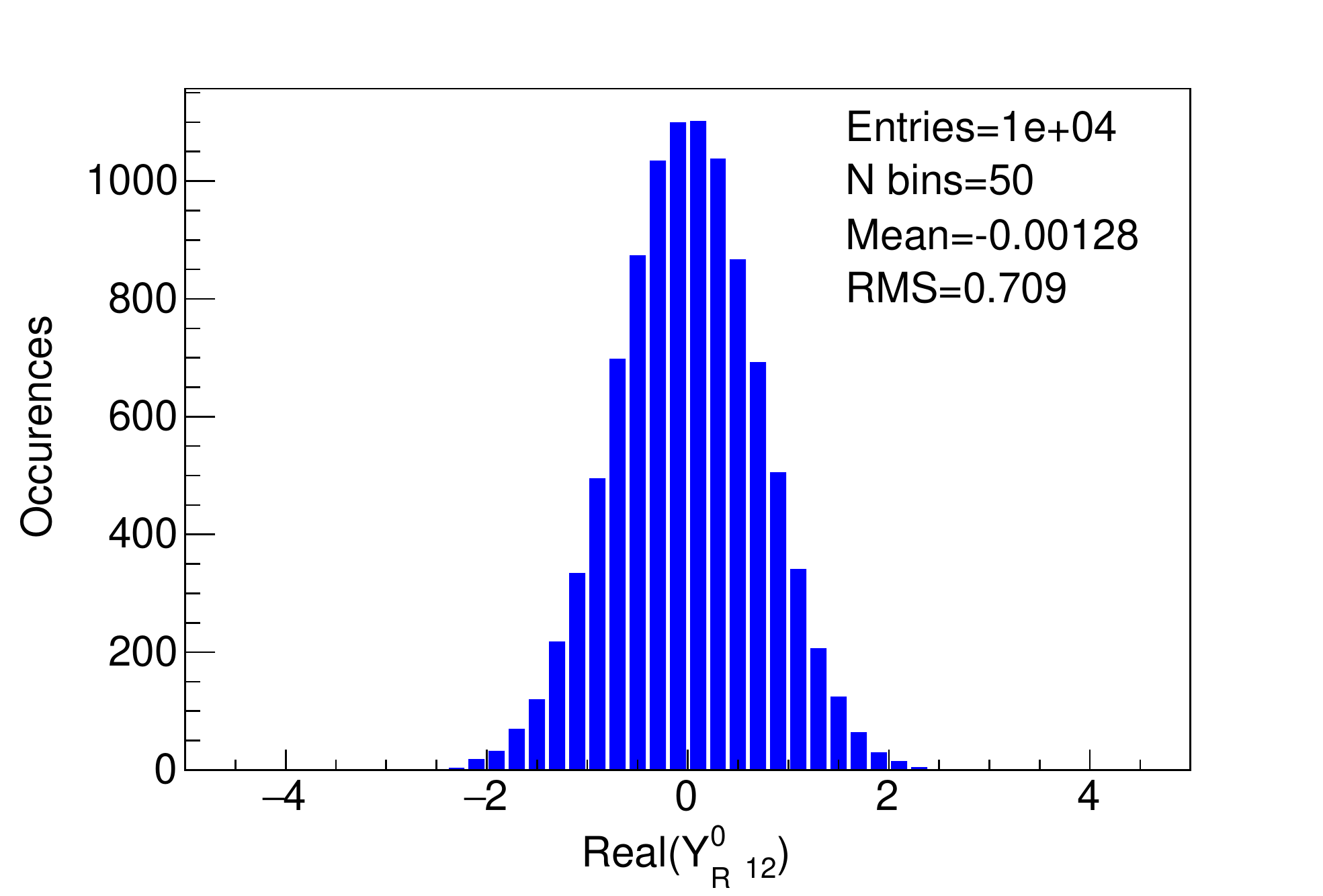}
\includegraphics[width=0.3\linewidth]{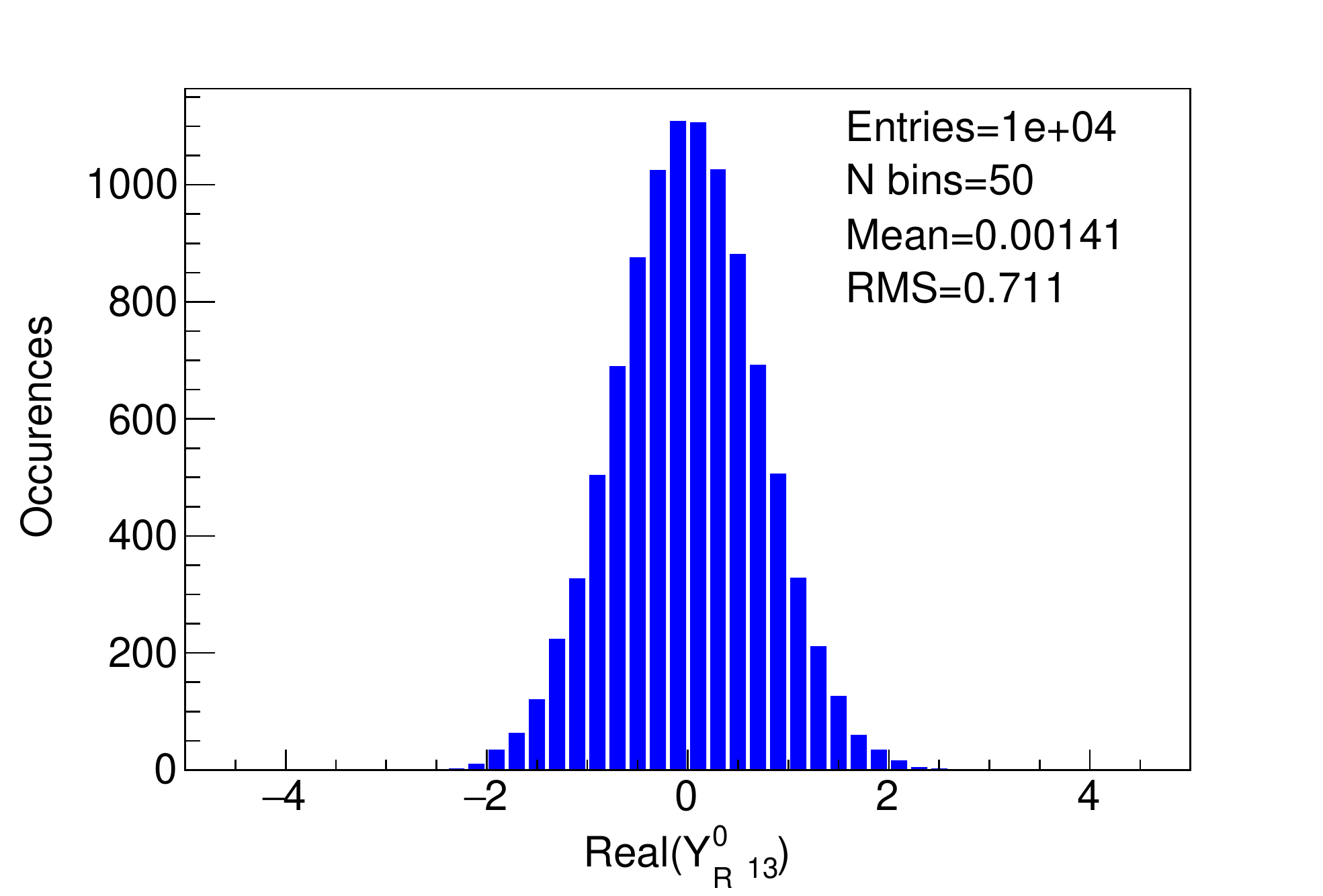}
\includegraphics[width=0.3\linewidth]{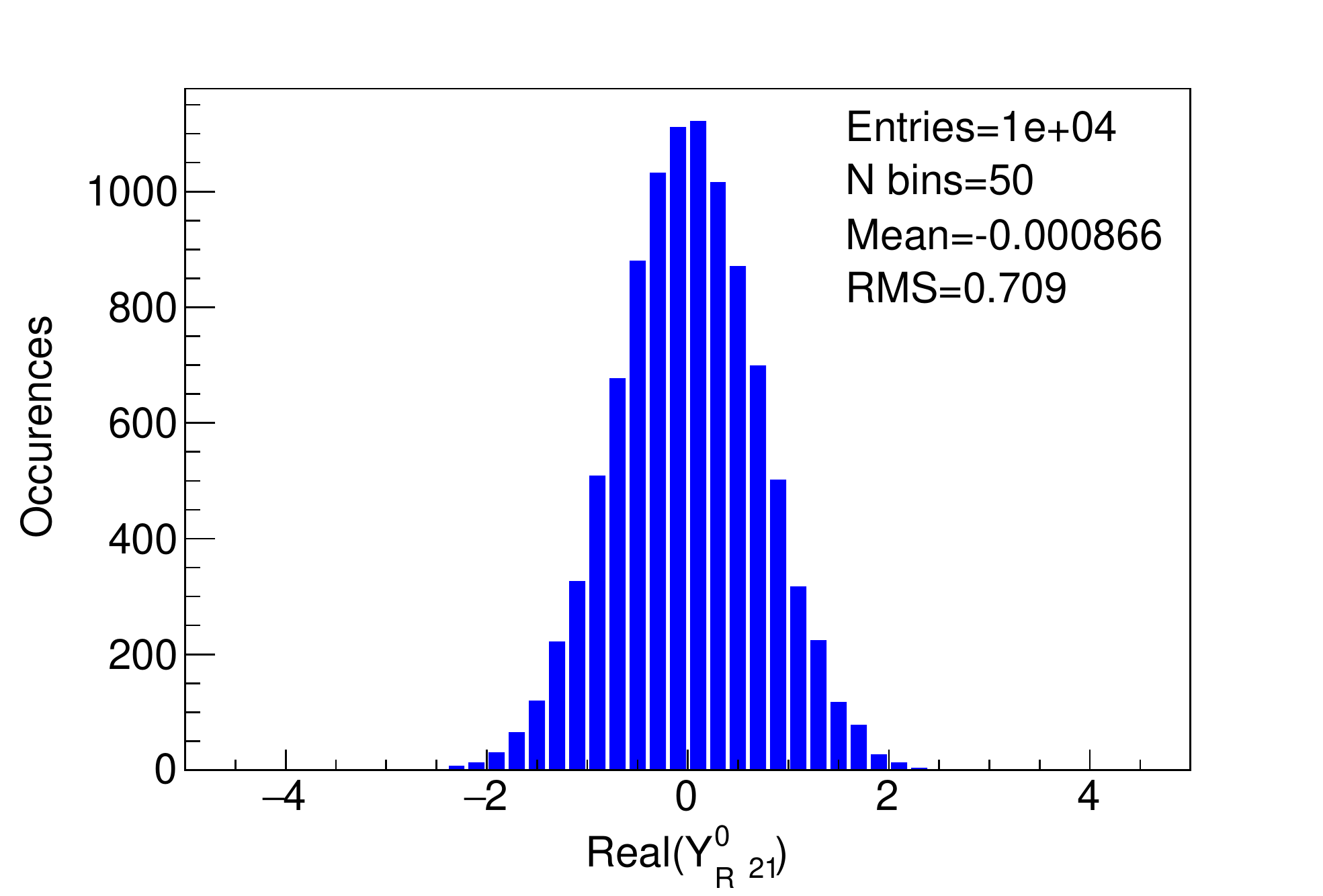}
\includegraphics[width=0.3\linewidth]{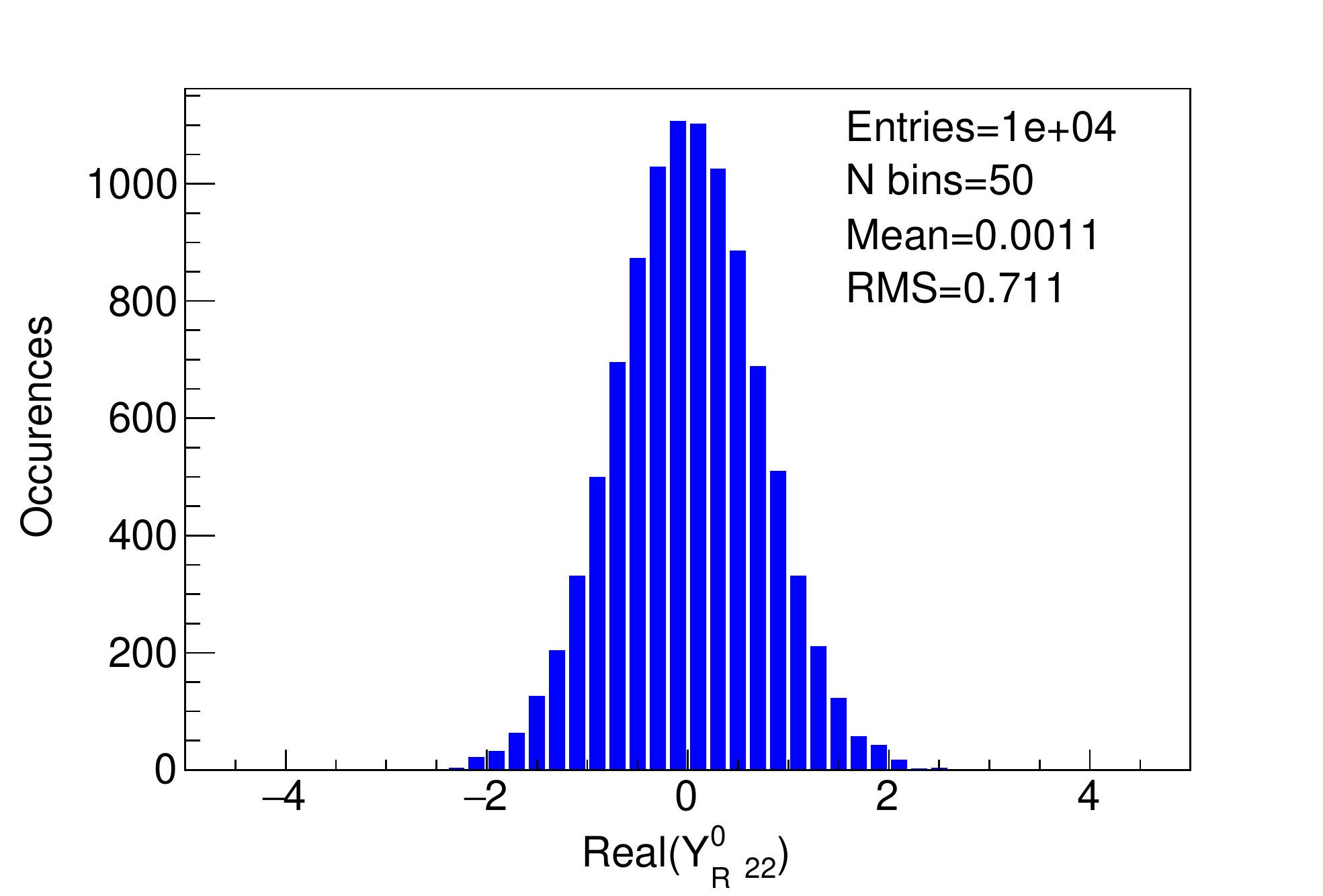}
\includegraphics[width=0.3\linewidth]{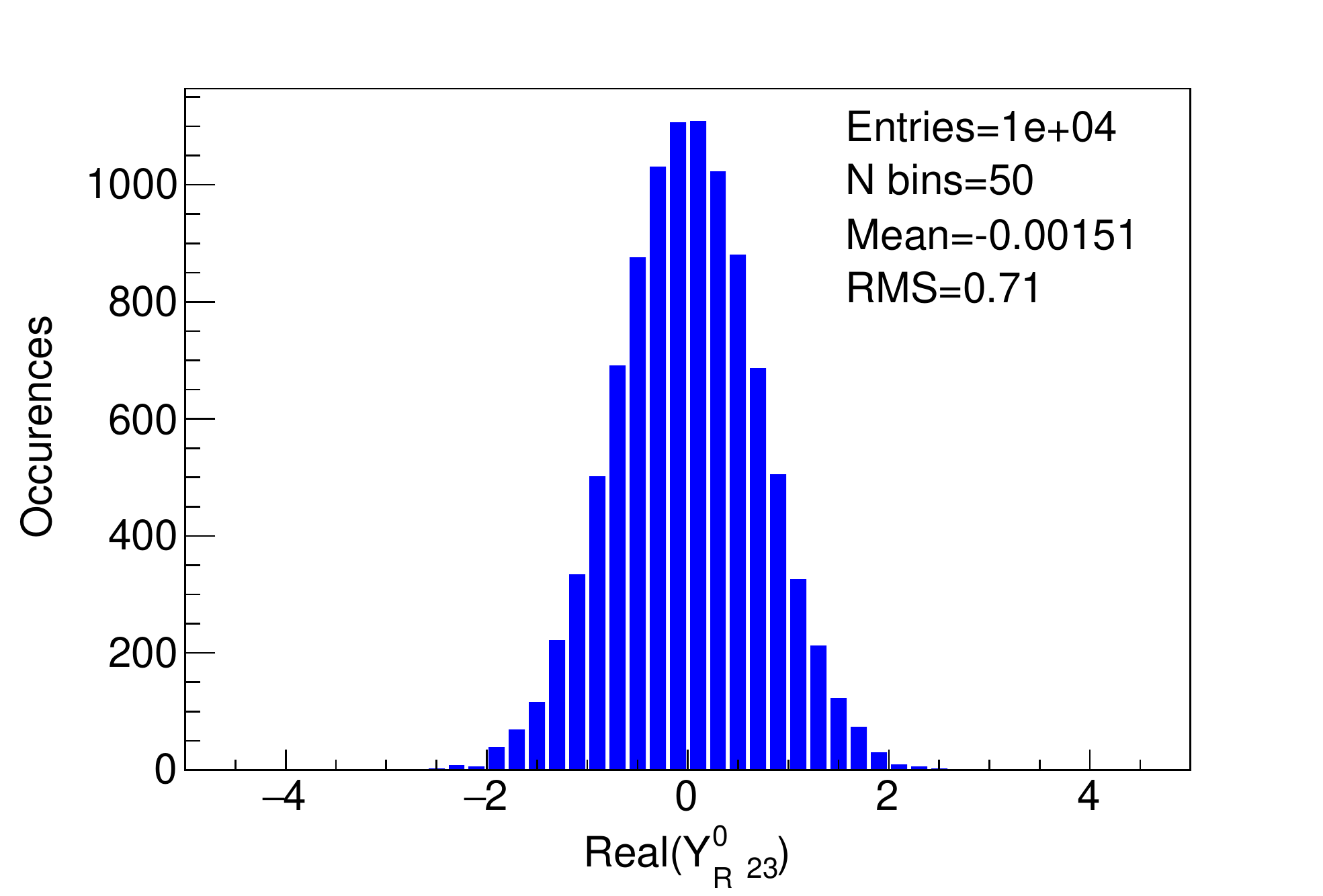}
\includegraphics[width=0.3\linewidth]{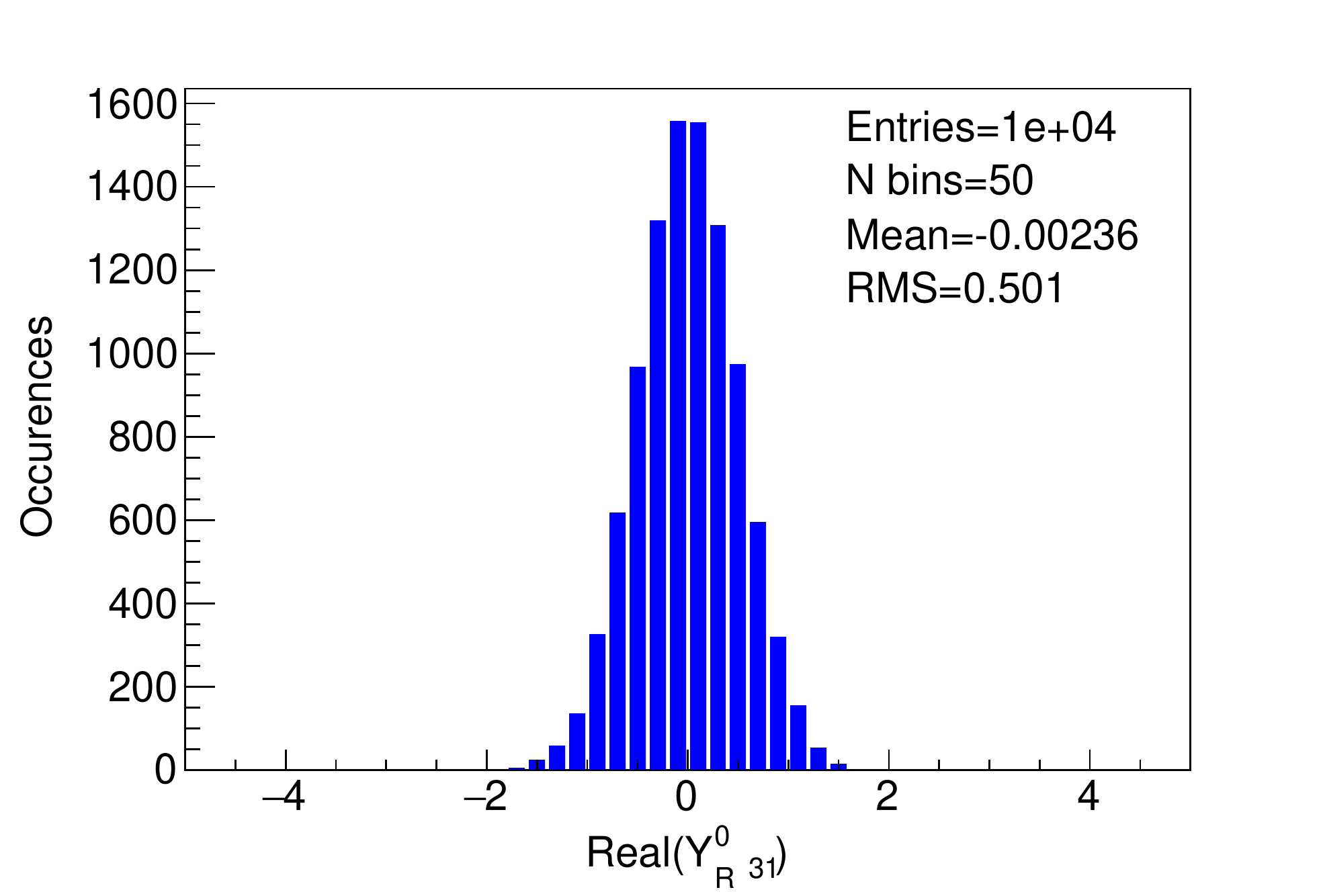}
\includegraphics[width=0.3\linewidth]{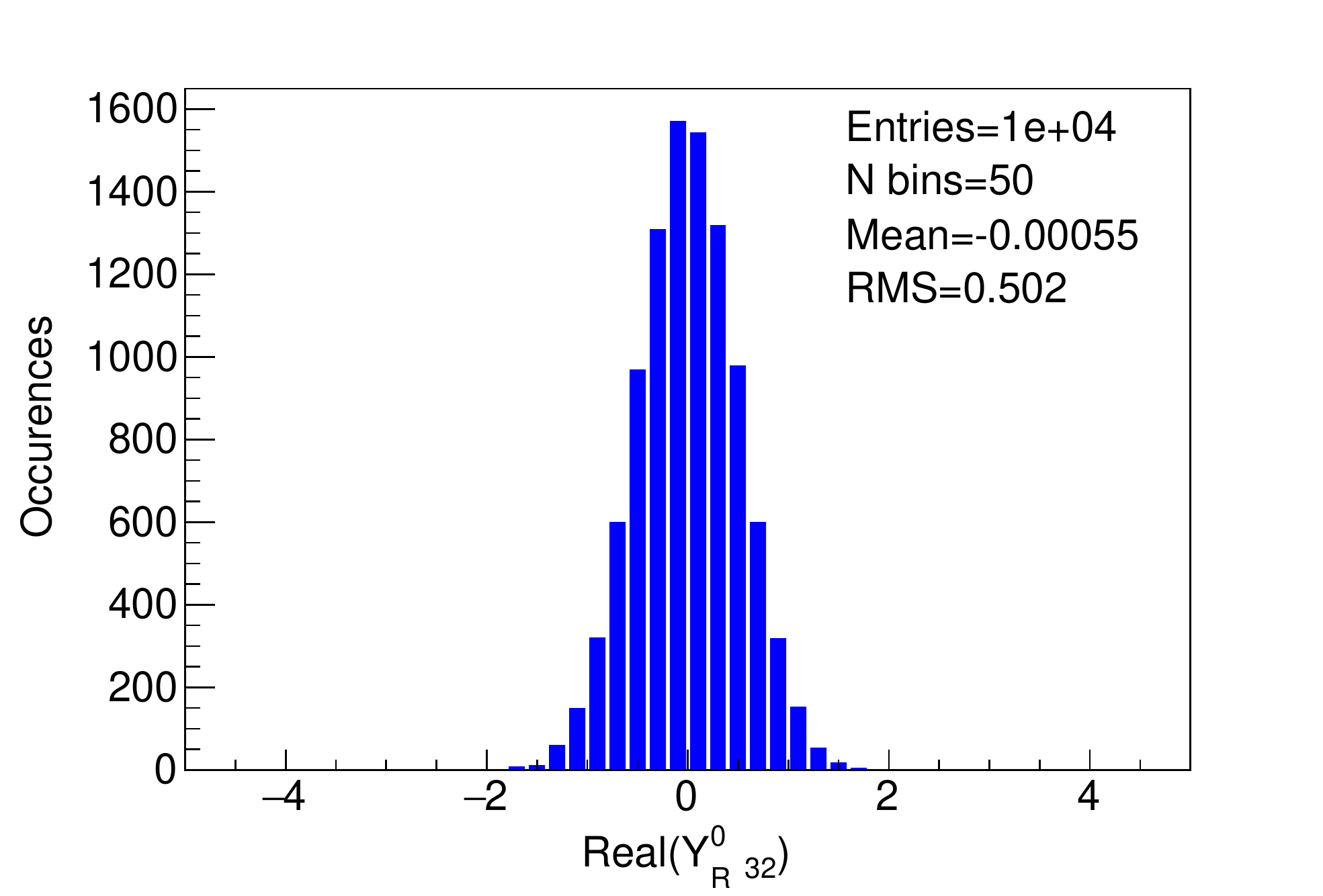}
\includegraphics[width=0.3\linewidth]{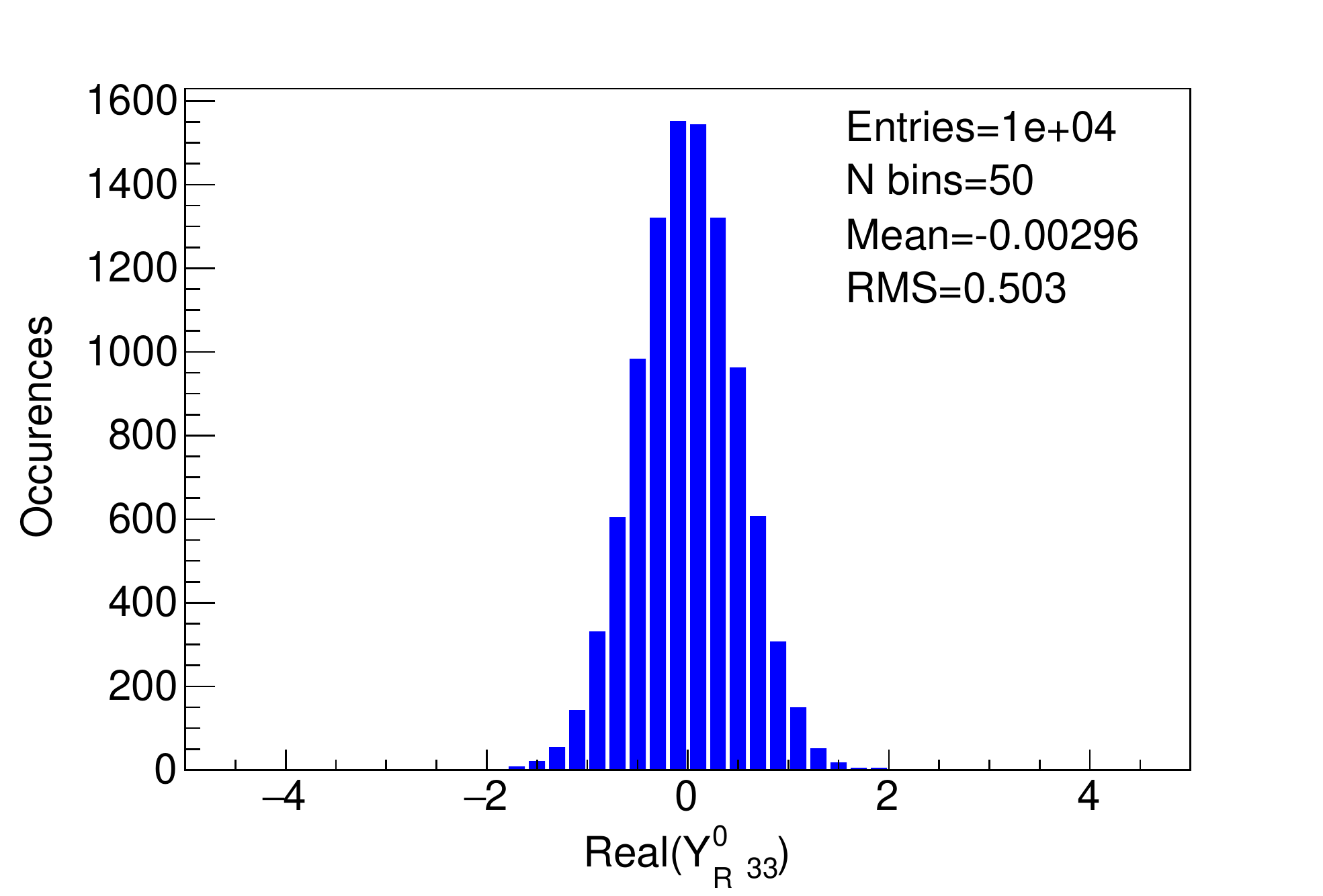}
\includegraphics[width=0.3\linewidth]{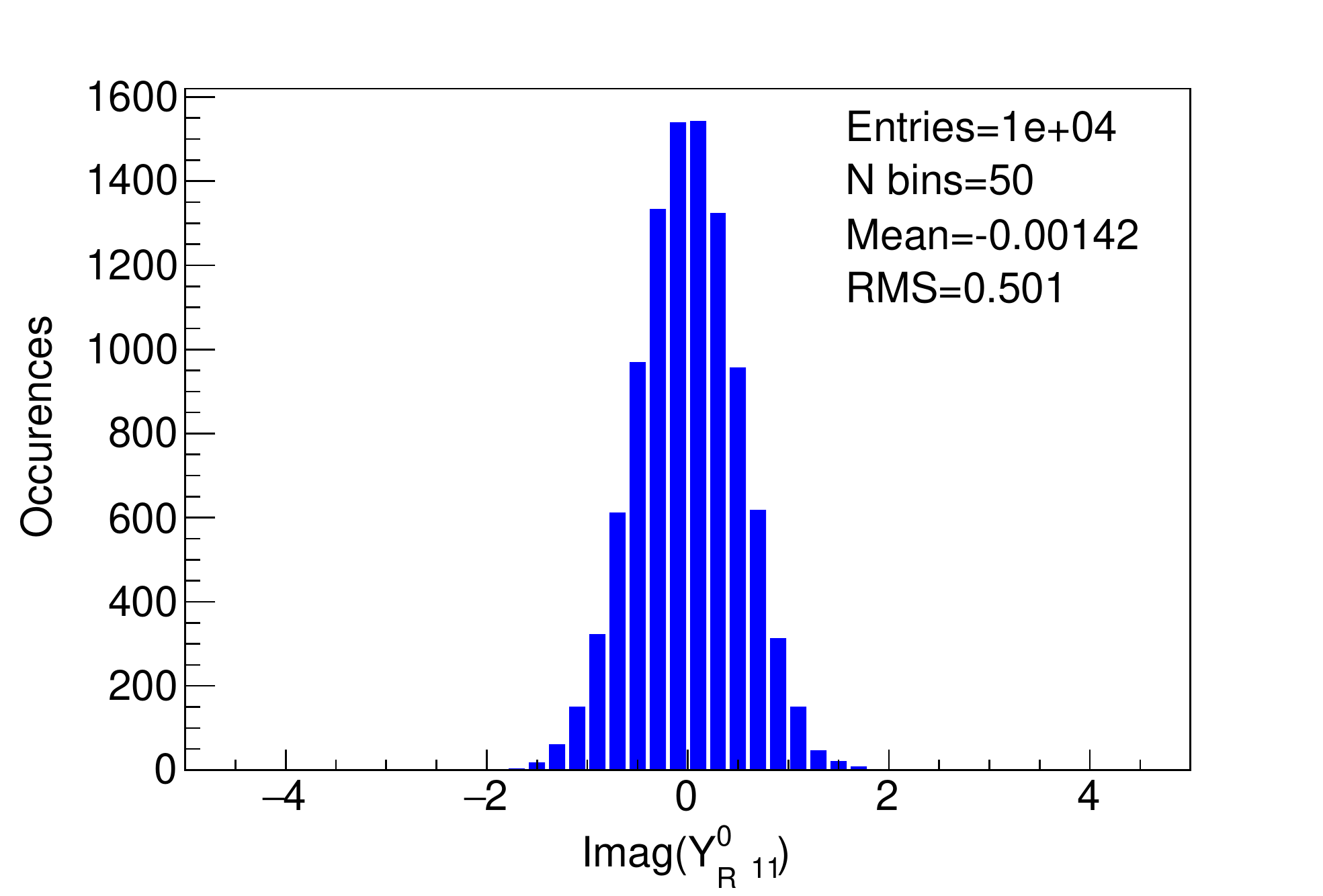}
\includegraphics[width=0.3\linewidth]{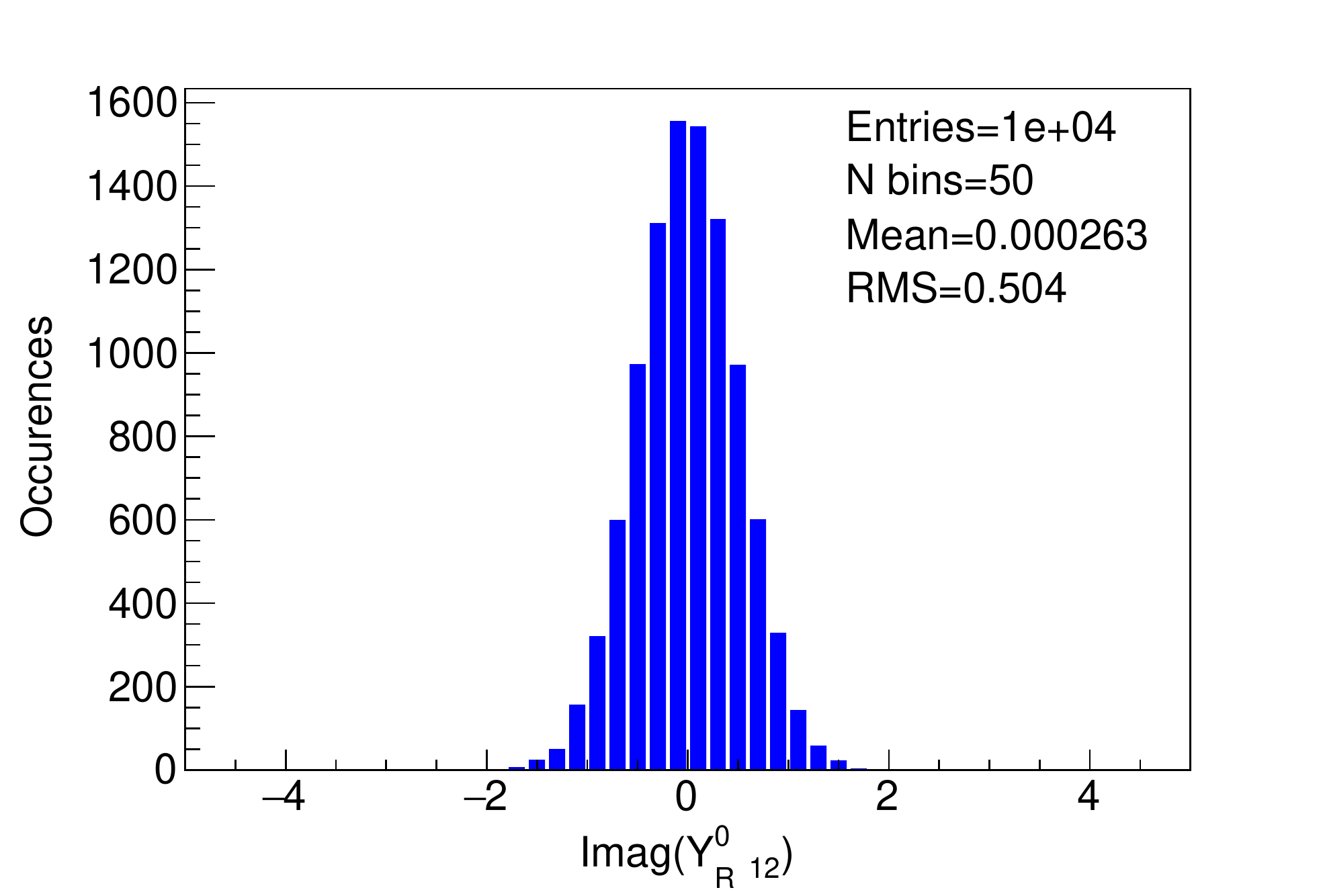}
\includegraphics[width=0.3\linewidth]{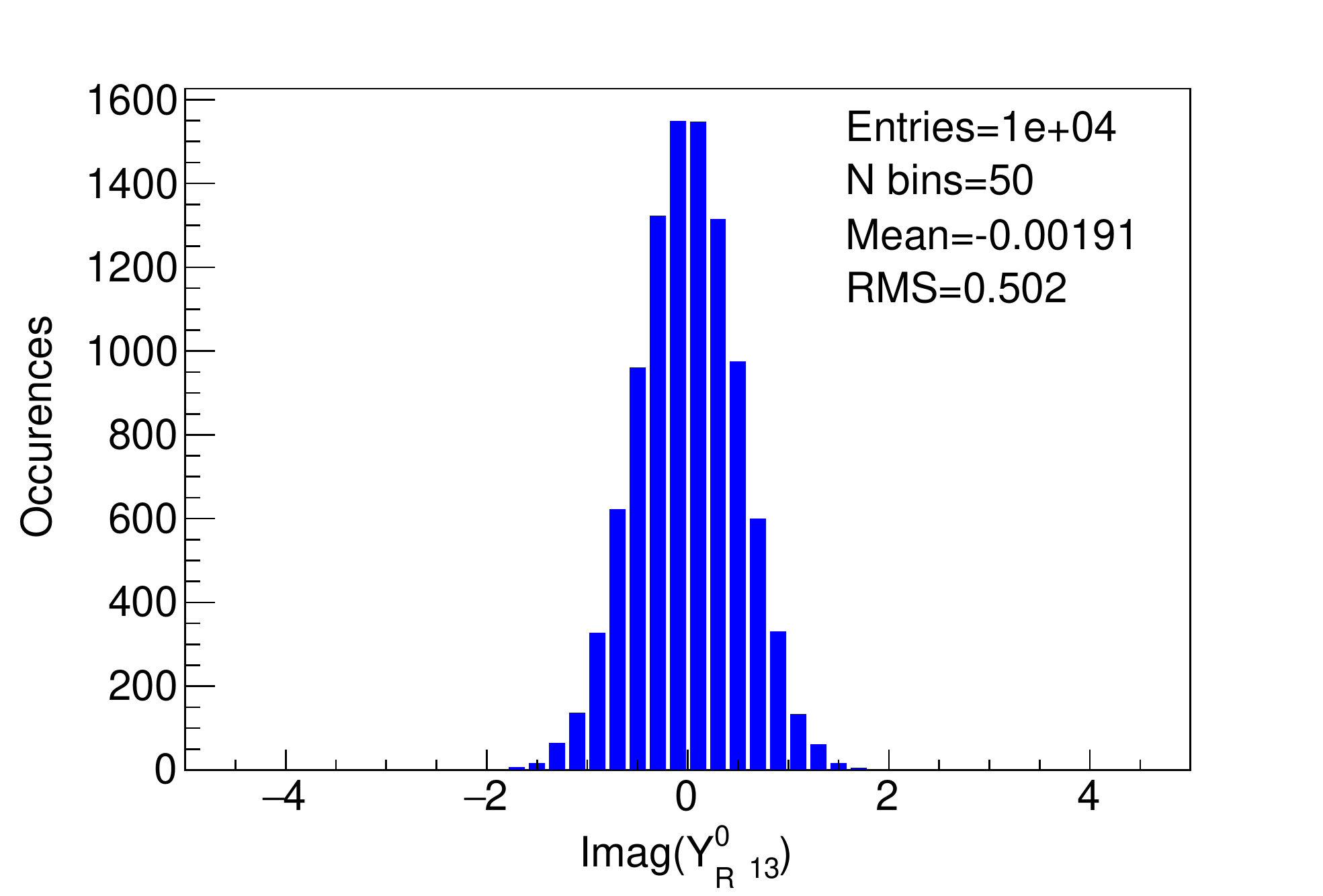}
\caption{ Distributions of the $O(1)$ random entries in the matrix $Y^{0}_R$ from the modified Monte Carlo approach that produce the observables in Fig.\ref{fig:004N}. }\label{fig:005E3}
\end{figure}

Here we present the distributions  of the biased random entries in the neutrino sector. These random entries are the result by employing the modified Monte Carlo analysis.  These random entries produce the theoretical distributions of the neutrino observables that are presented in Fig. \ref{fig:004N}. Modified random entries in the Dirac Yukawa coupling matrix are presented in Fig \ref{fig:005D3} and in Fig. \ref{fig:005E3} for the entries in the right-handed Yukawa couplings. All these entries get barely modified from the unbiased pattern.

\FloatBarrier

\FloatBarrier

\end{appendices}

\newpage
\FloatBarrier

\end{document}